\documentclass[10pt,journal,compsoc]{IEEEtran}
\usepackage{amsmath,amsfonts}
\usepackage{array}
\usepackage{amsthm}
\usepackage{color}
\usepackage{textcomp}
\usepackage{stfloats}
\usepackage{url}
\usepackage{verbatim}
\usepackage{graphicx}
\usepackage[tight,small]{subfigure}
\usepackage{cite}
\usepackage{bm}   
\usepackage{bbm}
\usepackage{amsthm,amsmath,amssymb,amsfonts}	
\usepackage{mathrsfs}
\usepackage{newunicodechar}
\usepackage[linesnumbered,ruled,vlined]{algorithm2e}
\usepackage{algpseudocode}
\usepackage{hyperref}
\usepackage{enumitem}
\usepackage{ragged2e} 
\usepackage{cuted}
\usepackage{xcolor}
\usepackage{colortbl}
\usepackage{setspace}
\usepackage{makecell}
\usepackage{booktabs,makecell, multirow, tabularx}
\newunicodechar{≤}{\ensuremath{\leq}}
\graphicspath{{}}
\newtheorem{Definition}{Definition}
\makeatletter

\newcommand{\Rmnum}[1]{\expandafter\@slowromancap\romannumeral #1@}
\makeatother
\definecolor{mycolor}{rgb}{0.9,0.9,0.9}

\begin{document}
	
	\title{Seamless Graph Task Scheduling over Dynamic Vehicular
		Clouds: A Hybrid Methodology for Integrating Pilot and
		Instantaneous Decisions}
	
	\author{Bingshuo Guo*, Minghui Liwang*, \IEEEmembership{Member}, \IEEEmembership{IEEE}, Xiaoyu Xia, \IEEEmembership{Member}, \IEEEmembership{IEEE}, Li Li, \IEEEmembership{Member}, \IEEEmembership{IEEE}, Zhenzhen Jiao, \IEEEmembership{Member}, \IEEEmembership{IEEE}, Seyyedali Hosseinalipour, \IEEEmembership{Member}, \IEEEmembership{IEEE}, Xianbin Wang,~\IEEEmembership{Fellow},  \IEEEmembership{IEEE} \thanks{B. Guo (guobingshuo@stu.xmu.edu.cn) and M. Liwang (minghuilw@xmu.edu.cn) are with the School of Informatics, Xiamen University, Fujian, China. X. Xia (xiaoyu.xia@rmit.edu.au) is with the School of Computing Technologies, RMIT University, Melbourne, Victoria, Australia. L. Li (lili@tongji.edu.cn) is with the Department of Control Science and Engineering, Tongji University, Shanghai, China. Z. Jiao (jiaozhenzhen@teleinfo.cn) is with the iF-Labs, Beijing Teleinfo Technology Co., Ltd., CAICT, China. S. Hosseinalipour (alipour@buffalo.edu) is with Department of Electrical Engineering, University at Buffalo-SUNY, NY, USA. X. Wang (xianbin.wang@uwo.ca) is with the Department of Electrical and Computer Engineering, Western University, Ontario, Canada. Corresponding author: Minghui Liwang. *B. Guo and M. Liwang contributed equally to this work.}
	}
	
	
	\IEEEtitleabstractindextext{
		\justify
		\begin{abstract}
			Vehicular clouds (VCs) play a crucial role in the Internet-of-Vehicles (IoV) ecosystem by securing essential computing resources for a wide range of tasks. This paPertackles the intricacies of resource provisioning in dynamic VCs for computation-intensive tasks, represented by undirected graphs for parallel processing over multiple vehicles. We model the dynamics of VCs by considering multiple factors, including varying communication quality among vehicles, fluctuating computing capabilities of vehicles, uncertain contact duration among vehicles, and dynamic data exchange costs between vehicles. Our primary goal is to obtain feasible assignments between task components and nearby vehicles, called \textit{templates}, in a timely manner with minimized task completion time and data exchange overhead. To achieve this, we \textbf{p}ropose a \textbf{h}ybrid graph \textbf{t}ask \textbf{s}cheduling (P-HTS) methodology that combines offline and online decision-making modes. For the offline mode, we introduce an approach called risk-aware pilot isomorphic subgraph searching (RA-PilotISS), which predicts feasible solutions for task scheduling in advance based on historical information. Then, for the online mode, we propose time-efficient instantaneous isomorphic subgraph searching (TE-InstaISS), serving as a backup approach for quickly identifying new optimal scheduling template when the one identified by RA-PilotISS becomes invalid due to changing conditions. Through comprehensive experiments, we demonstrate the superiority of our proposed hybrid mechanism compared to state-of-the-art methods in terms of various evaluative metrics, e.g., time efficiency such as the delay caused by seeking for possible templates and task completion time, as well as cost function, upon considering different VC scales and graph task topologies.
		\end{abstract}
		
		\begin{IEEEkeywords}
			Distributed task scheduling, undirected weighted graph, subgraph isomorphism, pilot and instantaneous decision-making, vehicular clouds
	\end{IEEEkeywords}}
	\vspace{-0.2cm}
	\maketitle

	\section{Introduction}
	\IEEEPARstart{T}{HE} Internet-of-Vehicles (IoV) has undergone remarkable development, driven by cutting-edge communication and computing technologies, as well as the growing popularity of intelligent vehicles. This progress has yielded numerous benefits for contemporary vehicular tasks and applications, such as autonomous driving, car games, and simultaneous localization \cite{liwang2019allocation,luo2021minimizing}. However, these tasks are often resource-hungry and computation-intensive, presenting challenges to the limited on-board computing capabilities of passenger vehicles. {While transferring these tasks to remote cloud servers offers a possible solution, this often suffers from a set of drawbacks,} including excessive network latency, additional energy consumption, and potential congestion over backhaul networks. Mobile edge computing (MEC) technology has helped alleviate some of these issues by bringing computing and storage resources closer to the network edge through the use of edge servers, providing low-cost and timely services \cite{hu2021efficient,mach2017mobile,rodrigues2016hybrid,samie2016computation,liu2021task}. {However, edge servers may face the risk of inadequate computing resources to manage concurrent tasks within the network, particularly when handling a large number of vehicles. Additionally, in areas where traditional MEC infrastructures may be less developed or constrained (such as rural or underdeveloped regions), collaborative vehicular computing technology has emerged as a promising solution. In these environments, vehicles can act as mobile servers, forming a vehicular cloud (VC) to support the flexible scheduling and smooth execution of computation-intensive tasks\cite{su2024reliable, liu2023rfid}. Although the number of vehicles may vary across different regions over time, VCs can leverage their growing computing and sensing capabilities to form a scalable computing platform, even in regions with lower vehicle density\cite{waheed2022comprehensive}. Thus, by combining resources from distributed smart vehicles with advanced sensors and processors, a VC enables tasks to be scheduled and processed over multiple vehicles in parallel, which in turn facilitates responsive, flexible, and cost-effective service provisioning over dynamic IoV.}\\	
	\indent A vehicular computation-intensive task can be modularized to provide a clear representation of its internal processes' interdependencies. This can be achieved through the use of an undirected weighted graph structure that consists of multiple subtasks (called components) and weighted edges \cite{liwang2023graph,liwang2020multi,gao2021truthful}. In such a representation (see the left-most subplot in Fig. \ref{fig1}), an edge between two subtasks describes the corresponding interrelation, such as intermediate data exchange; while the edge weight represents attributes such as the required data exchange duration between subtasks \cite{hosseinalipour2019power}\cite{ghaderi2016scheduling}. Moreover, a VC can also be modularized and modeled as a graph, where the edges among nodes (i.e., vehicles) represent the existence of vehicle-to-vehicle (V2V) links \cite{liwang2020multi,liwang2019allocation,gao2021truthful,liwang2023graph}. In this work, by leveraging these graph-based representations, \textit{we aim to investigate how task components can be mapped to mobile vehicles in a reliable and time-efficient manner}, which draws resemblance to a \textit{joint dynamic isomorphic subgraph search and network optimization} problem. To tackle this problem, we first model the dynamics of VCs, such as fluctuant on-board resource supply of vehicles, vehicular mobility, and varying communication qualities. We then introduce a novel hybrid graph task scheduling mechanism that effectively accounts for the dynamic nature of VCs during task scheduling. This mechanism incorporates both offline (i.e., pilot) and online (i.e., instantaneous) decision-making modes to identify an optimal template that minimizes graph tasks' completion time and execution cost. Notably, a \textit{template} refers to a mapping between task components and vehicles in a VC, which specifies the vehicle on which each task component will be processed.
	\vspace{-0.2 cm}
	\subsection{Background and Motivations}
	Allocating graph tasks over VCs is a nuanced aspect of future IoV networks. However, there exist certain challenges that should be addressed to guarantee its seamless realization. The following outlines these challenges.\\
	\noindent
	$\bullet$ Finding feasible mappings, called \textit{templates}, between task components and vehicles in a VC requires solving the \textit{subgraph isomorphism problem}, which is NP-complete \cite{Conte2004}. As a result, the complexity of obtaining these mappings can significantly hinder timely scheduling of graph tasks over large-scale VCs. This is a major concern and should be addressed to ensure smooth and efficient task processing.\\
	$\bullet$ To achieve optimal task scheduling, it is essential to take into account multiple evaluation indicators such as task completion time, energy efficiency, resource usage, among others. However, dealing with a mixture of multiple objectives can make the scheduling problem even more complex, in the form of \textit{a non-linear integer programming}, that is generally non-convex and NP-hard \cite{liwang2023graph,li2022bi}.\\
	$\bullet$ Scheduling graph tasks over real-world VCs further requires consideration of numerous network-related complex constraints, particularly those related to the dynamics of the VCs. For instance, the connectivity between vehicles in a VC may not always be stable. Therefore, to support the successful execution of connected task components, it is crucial to ensure that the contact duration between two vehicles in a VC is at least equal to or greater than the time duration required to complete at least one of the task components, to guarantee a reliable communication link throughout the task execution process.
		\begin{figure*}[htb]
		\centering
		\includegraphics[width=1\linewidth]{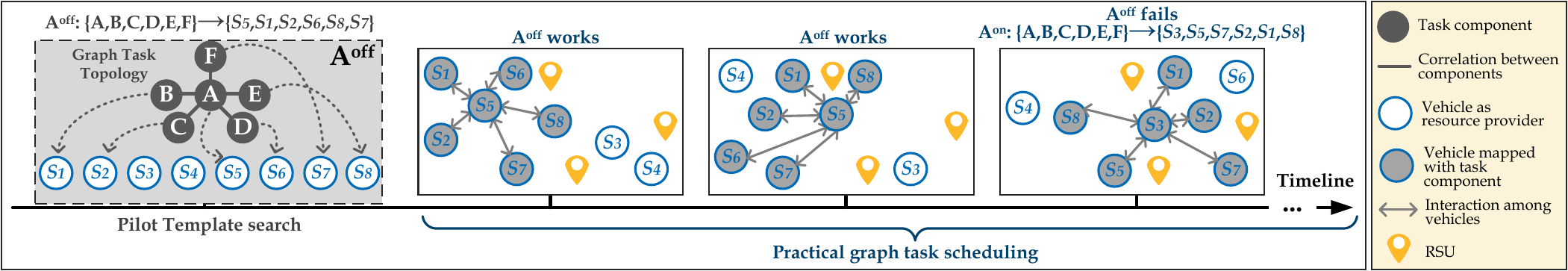}
		\caption{{A schematic of our hybrid graph task scheduling methodology in terms of a timeline.}}
		\label{fig1}
		\vspace{-0.5 cm}	
	\end{figure*}	
	
	To schedule graph tasks over VCs, existing studies mainly concentrate on online methodologies \cite{liu2020dependency,liu2023rfid,lv2022tbtoa,al2020task,abdisarabshali2023decomposition,liwang2020multi}, which allocate task components to vehicles at practical/actual task scheduling events based solely on the current network condition. However, considering the aforementioned challenges, such methods can suffer from the following drawbacks.
	
	\noindent
	$\bullet$ \textit{Low time/energy efficiency:} Obtaining the templates and making graph task scheduling decisions online requires a certain amount of time to analyze the current network conditions that can lead to increased latency. For instance, if searching for feasible mappings between a graph task and vehicles in a VC takes 3 seconds, the actual task execution can only start after 3 seconds. During this period, the topology of VC can change, which may result in inadequate contact duration among the vehicles to support the intermediate data exchange required by task components. Moreover, a prolonged decision-making process can lead to excessive energy consumption on the computing units in charge of task scheduling, which negatively impacts the sustainability of networks. Therefore, it is imperative to optimize the decision-making process while considering environmental sustainability to ensure efficient network performance.
	
	\noindent
	$\bullet$ \textit{Potential scheduling failures:} Vehicles and tasks may encounter potential risks, such as a vehicle that has reserved resources for tasks but fails to be assigned to any task component after a certain period of decision-making time. Similarly, a task may face the risk of not receiving its required computing resources if the topology of a VC can no longer support the task structure after a prolonged decision-making time. These cases can lead to a low quality of service (QoS) and quality of experience (QoE) for both task owners and vehicles, discouraging their involvement in VCs.
		
{In this work, our key goal is to tackle the aforementioned challenges while also addressing the limitations identified in existing studies. Also, we aim to capture and evaluate VC dynamics, taking into account factors such as uncertain VC topology, changing wireless channel quality, and fluctuating vehicular resource supply. To tackle the limitations of conventional online scheduling, we propose a \textit{hybrid graph task scheduling methodology} (P-HTS), combining the online scheduling mode with an offline/pilot one, which moves the decision-making process ahead of practical task scheduling events, by obtaining a template denoted by $\mathbf{A^{off}}$. We also conduct an unique view of addressing the potential risks during this offline mode}, such as inaccurate predictions of network statistics that can negatively affect task completion performance and make the use of the template obtained in the offline mode impractical. To mitigate these risks, our methodology then devises an efficient online task scheduling strategy that serves as a backup plan, which will only be executed when template $\mathbf{A^{off}}$ fails to be used for the current network conditions. 

{Fig. \ref{fig1} illustrates the overall procedure of our hybrid task scheduling methodology, showing how the designed two-stage approach operates over time and how they can impact each other. }{As depicted in this figure, the leftmost diagram provides an example using an undirected graph, where six task components (A, B, C, D, E, F) are interconnected and allow for parallel execution. In this example, eight vehicular servers ($S1$ to $S8$) are available to process these task components. Our goal is to efficiently identify a feasible mapping (i.e., a template) between task components from the task graph and computing nodes from the vehicular service topology while minimizing overall task execution costs (such as latency and energy consumption). The first box in Fig. \ref{fig1} represents the offline scheduling stage, with the key principle of pre-deciding a feasible scheduling solution called $\mathbf{A^{off}}$ under historical network statistics, helping with the rapid decision-making when dealing with real-time execution. 
The second to fourth boxes in Fig. \ref{fig1} illustrate the online stage in which real-time task scheduling occurs. Each time a new task scheduling request arrives, the system first checks whether $\mathbf{A^{off}}$ remains valid under the current network state (i.e., whether it still satisfies scheduling constraints); if yes, it can directly be used as the scheduling template for executing the task (e.g., the second and third box of Fig. \ref{fig1}), significantly improving time efficiency for computing service delivery. Nevertheless, if $\mathbf{A^{off}}$
becomes invalid due to dynamic changes in network conditions or vehicular topology (e.g., a vehicle involved in the original template moves out of the network), the system switches to online backup approach to search a new optimal template $\mathbf{A^{on}}$ (the fourth box of Fig. \ref{fig1}).}

\vspace{-0.2 cm}
\subsection{Related Studies}
There are generally three main categories of research on task scheduling. The first category involves analyzing tasks as bit streams, without taking into account any inherent task dependencies/topology. The second category focuses on tasks represented as directed acyclic graphs, which require sequential processing of task components. Finally, the third category considers tasks represented as undirected graphs, allowing for the parallel processing of all task components (which represents our emphasis in this paper). In the following, we provide a summary of the studies regarding each category.

\noindent
\textit{\textbf{Tasks represented as bit streams. }}Bit stream-based task representations have been a common practice, where the tasks of interest lack any internal processing complexities and interdependencies. Considering such tasks, efforts have been made to achieve a trade-off between delay and task execution cost \cite{li2022trade}, improve the task processing time efficiency \cite{alameddine2019dynamic}, and enhance task offloading under delay constraints \cite{yue2021todg,tutuncuouglu2022online}. However, our work migrates from bit-stream task representations and focuses on a different scheduling approach suitable for computation-intensive tasks with internal processing topologies, represented as graphs.

\noindent
\textit{\textbf{Tasks represented as directed acyclic graphs (DAGs). }}{DAG tasks, with their directed edges, facilitate sequential execution between adjacent task components and, consequently, the servers handling them. In DAG task scheduling, the primary focus is to determine the proPerexecution order of task components and the mapping of components to the appropriate servers, respecting the directed edges. This consideration does not impose stringent requirements on keeping inter-server communications, as some components will not be processed in parallel. Instead, ensuring the correct handling of inputs and outputs between task components becomes the primary concern.} Several researchers have proposed innovative scheduling schemes for DAG task scheduling to improve system stability, time efficiency, and resource utilization. For instance, Sun \textit{et al.} \cite{sun2018cooperative} used a genetic algorithm-based approach, Zhang \textit{et al.} \cite{zhang2022dag} leveraged advanced reinforcement learning and graph neural networks to minimize task completion time, and Liu \textit{et al.} \cite{liu2020dependency} proposed a priority-aware scheduling algorithm to speed up the task completion process. To map DAG tasks to virtual computing clusters, Liu \textit{et al.} \cite{liu2023rfid} proposed a dynamic scheduling scheme based on sorting and prediction techniques. Lv \textit{et al.} \cite{lv2022tbtoa} introduced a heuristic algorithm that considers DAG tasks and service cache constraints to predict the impact of offloading decisions on task completion time and energy consumption. In general, DAG task components are processed sequentially according to the directions of edges in their DAG representation (i.e., the scheduling of DAG tasks does not require addressing the subgraph isomorphism problem). However, in this work, we focus on computation-intensive applications with an undirected graph representation, where all the task components can be executed in parallel, and thus our developed methodology is significantly different from the above literature. 
	
\noindent{
\textit{\textbf{Tasks represented as undirected graphs (UGs).}} {In contrast to DAG tasks, the primary focus in scheduling UG tasks is to identify parallel computation opportunities between connected task components, which necessitates continuous communication among computing servers when handling connected components. Generally, this requires solving the isomorphic subgraph problem to optimize task execution. In this paper, we focus on the scheduling of UG tasks in dynamic networks, an area that is not yet well-explored in the literature.} Although various studies have explored task scheduling across both static and dynamic networks, their scope is constrained by the specific conditions of the network or the status of the servers. For instance, Ghaderi \textit{et al.} \cite{ghaderi2016scheduling} proposed a randomized task scheduling algorithm for static networks. Al-Habob \textit{et al.} \cite{al2020task} introduced two scheduling algorithms based on the genetic algorithm and conflict graph model for dynamic networks. Hosseinalipour \textit{et al.} \cite{hosseinalipour2019power} studied the graph task allocation over geo-distributed cloud networks with various scales. Shi \textit{et al.} \cite{shi2016energy} investigated task scheduling aiming to minimize the task completion time while considering energy consumption. Abdisarabshali \textit{et al.} \cite{abdisarabshali2023decomposition} studied redundancy-aware task processing over dynamic vehicular networks.
As compared to the works discussed above, this paPerexplores the scheduling of graph tasks in the presence of uncertainties in resource supply and vehicle mobility, which further complicates the scheduling problem and calls for nuanced solutions. We have made some early efforts towards this direction \cite{liwang2020multi,gao2021truthful,liwang2020energy,liwang2019allocation,liwang2023graph}, which include proposing a randomized task allocation mechanism under task concurrency \cite{liwang2020multi}, auction-promoted task scheduling under resource reutilization \cite{gao2021truthful}, randomized task scheduling under hierarchical tree decomposition \cite{liwang2019allocation}, and simultaneous resource allocation and task scheduling over air-ground integrated networks \cite{liwang2023graph}. Moreover, existing studies generally focus on online decision-making for graph task scheduling according to the current network conditions, which sometimes leads to unacceptable prolonged task scheduling decisions, especially over complicated task/VC topologies. Thus, our proposed hybrid methodology opens a new degree of freedom (i.e., offline/pilot scheduling mode) to the above literature. Moreover, the majority of these studies consider predetermined system information, e.g., a fixed and known computing capability of vehicles. In contrast, this paPeraddresses practical challenges posed by network uncertainties. Table \ref{relate work comparison} presents a summary of related studies, highlighting the key differences of our paper. 
\begin{table}[htb]
	\centering
	\scriptsize
	\vspace{-0.2 cm}
	\caption{A summary of related works}
	\setlength{\tabcolsep}{3.0 pt} 
	\centering
	\vspace{-0.2 cm} 
	\begin{tabular}{|c|c|c|c|c|c|c|c|}
		\hline
		\multirow{2}{*}{\textbf{Reference}} & \multicolumn{2}{c|}{\makecell{\textbf{Environmental}\\ \textbf{attributes}}} & \multicolumn{2}{c|}{\textbf{Decision mode}} & \multicolumn{3}{c|}{\textbf{Task model}} \\ \cline{2-8} 
		& Stable & Dynamic & Online & Offline & Bit streams & DAG & UG \\
		\hline
		[20][21] &  & $\checkmark$ & $\checkmark$ &  & $\checkmark$ &  &  \\
		\hline
		[22][23] & $\checkmark$ &  & $\checkmark$ &  & $\checkmark$ &  &  \\
		\hline
		[2][16] &  & $\checkmark$ & $\checkmark$ &  &  & $\checkmark$ &  \\
		\hline
		[17][24] & $\checkmark$ &  & $\checkmark$ &  &  & $\checkmark$ &  \\
		\hline
		[19] &  & $\checkmark$ & $\checkmark$ &  &  & $\checkmark$ & $\checkmark$ \\
		\hline
		[11][12][18][25] & $\checkmark$ &  & $\checkmark$ &  &  &  & $\checkmark$ \\
		\hline
		[1][8][9][10][26] &  & $\checkmark$ & $\checkmark$ &  &  &  & $\checkmark$ \\
		\hline
		Our paPer&  & $\checkmark$ & $\checkmark$ & $\checkmark$ &  &  & $\checkmark$ \\
		\hline
	\end{tabular}
	\label{relate work comparison}
\end{table}

\vspace{-0.3 cm}
\subsection{Contributions}
	In this paper, we are interested in a novel graph task scheduling mechanism that effectively combines offline/pilot and online/instantaneous decision-making modes into a hybrid setting, over a dynamic vehicular computing environment. We also carefully consider multiple uncertainties within the network to capture the dynamic nature of networks and enhance the effectiveness of our P-HTS. To the best of our knowledge, this is the first attempt in the literature towards such a stagewise decision-making direction. Our key contributions are summarized as follows:
	\begin{enumerate}[leftmargin=4mm] 
		\item This work examines the scheduling and processing of graph tasks over a dynamic VC amid various uncertainties, e.g., fluctuating vehicular computing capability, uncertain contact duration among vehicles, varying channel quality, and dynamic data exchange cost, to capture the dynamic nature of the network. We devise a stagewise and hybrid methodology that seamlessly integrates both offline and online decision-making modes to ensure a reliable and time-efficient task scheduling process.
		
		\item Our proposed offline task scheduling mode pre-determines a template for mapping task components and vehicles within a VC, from a long-term view. In doing so, it aims to minimize the expected value of the cost function that takes both task completion time and data exchange cost into consideration. To achieve this goal, we develop a novel approach called \textbf{r}isk-\textbf{a}ware \textbf{p}ilot \textbf{i}somorphic \textbf{s}ubgraph \textbf{s}earching (RA-PilotISS). This technique utilizes historical statistical analysis and risk management to enhance the probability of executing future tasks with high performance. Consequently, it facilitates timely resource provisioning.
		
		\item We then design an online task scheduling mode to provide a reliable backup solution in case the offline template cannot be used for task execution given the current network condition at practical/actual task scheduling events. For the online mode, we investigate a \textbf{t}ime-\textbf{e}fficient \textbf{insta}ntaneous \textbf{i}somorphic \textbf{s}ubgraph \textbf{s}earching (TE-InstaISS) approach, which enables us to quickly find the optimal template according to the current network conditions.
		
		\item We conduct comparative experiments to compare the performance of our proposed mechanism against existing methods across various problem scales. The results reveal the superiority of our methodology in terms of critical evaluation indicators such as running time (scheduling efficiency), data exchange cost, and task completion time.
	\end{enumerate}
\begin{figure}[t!]
\vspace{-0.3 cm}
\centering
\includegraphics[width=1\linewidth]{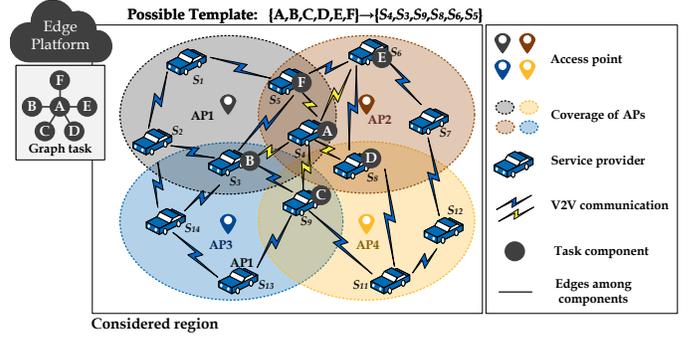}
\caption{{A schematic of graph task scheduling over dynamic VCs.}}
\vspace{-0.1 cm}
\label{fig2}
\vspace{-0.2 cm}
\end{figure}
\vspace{-0.1 cm}
\section{System Overview and Models}
\vspace{0.1 cm}
\subsection{Overview}
In this paper, we propose a novel methodology for assigning graph tasks to mobile vehicles in a specific region, such as Google Park or DisneyLand. {As depicted in Fig. \ref{fig2}, our model comprises three key entities, namely, an edge platform (EP), multiple access points (APs), and several vehicles (service providers, SPs). Particularly, the EP\footnote{{Typically, the EP is positioned centrally relative to multiple APs, enabling efficient task distribution to available SPs via nearby APs. 
The EP and APs are connected to each other via wired links\cite{li2022trade,zhang2022task,hu2019task}.}} acts as a coordinator that periodically generates, collects, and announces task requirements through APs (act as intermediaries, helping vehicles get access to the EP)\cite{Josilo2020}, while ensuring that the task scheduling procedure runs smoothly. Moreover, SPs can get access to the EP through APs and form a vehicular cloud (VC) to process the assigned tasks, where they can interact with APs through Vehicle-to-Infrastructure (V2I) communications. Communication among vehicles can be supported by Vehicle-to-Vehicle (V2V) links \cite{dey2016vehicle}. In our model, we consider a fixed number of vehicles in a given region, where the uncertain contact duration among them reflects their mobility inside the region (similar considerations can be found in \cite{gao2021truthful,liwang2023graph}).}

We represent tasks and the VC topology by utilizing a graph-based model where they are both represented as undirected weighted graphs. We will further consider the dynamic and ever-changing nature of the network environment, which is influenced by factors such as the availability of computing resources of vehicles, the duration of contact between them, and the quality of V2I and V2V communication channels. By taking these factors into account, we can create a more comprehensive and constructive representation of the network environment.

Our overall goal is to \textit{find the best template that describes assigning task components to SPs in a VC under a responsive and cost-effective manner, while minimizing task completion time and data exchange cost.} {In order to strike a balance between decision-making overhead on task scheduling (measured in terms of the time spent on obtaining the templates and identifying the best one for task execution) and task execution performance (measured in terms of delay and processing cost), we have developed a hybrid mechanism that utilizes both offline and online modes.} The offline task scheduling analyzes past data to identify a near-optimal template (which we refer to as $\mathbf{A^{off}}$) for assigning task components and SPs. This is achieved through our designed \textbf{r}isk-\textbf{a}ware \textbf{pilot} \textbf{i}somorphic \textbf{s}ubgraph \textbf{s}earching (RA-PilotISS) approach, which aims to minimize the expected weighted sum of task completion time and data exchange cost. If $\mathbf{A^{off}}$ is not successful during practical task scheduling events, the online task scheduling seeks the optimal template ($\mathbf{A^{on}}$) using our proposed \textbf{t}ime-\textbf{e}fficient \textbf{insta}ntaneous \textbf{i}somorphic \textbf{s}ubgraph \textbf{s}earching (TE-InstaISS) approach. {Major notations in this paPerare summarized in Table \ref{table3}. }For the tractability of analysis, similar to \cite{hou2020reliable,deb2020deft,pham2019coalitional}, we focus on scheduling a single graph task\footnote{Our proposed hybrid methodology can also be applied when considering multiple tasks, as these tasks can be regarded as a large undirected (disconnected) weighted graph.}. In the following, we mathematically formalize different elements of our system of interest.

Throughout the presentation of the paper, we will introduce a set of parameters, noting their adherence to certain distributions. These variables serve to encapsulate the uncertainties inherent in the network, with our approach treating them as generic random variables. It is only during simulations that we actualize their values, providing a practical manifestation of their impact on our methodology.

\vspace{-0.1 cm}
\subsection{Modeling of VC Graph}
We represent the VC as an undirected weighted graph $\bm{G}^{\bm{serv}}=\left(\bm{V}^{\bm{serv}},\bm{E}^{\bm{serv}},\bm{W}^{\bm{serv}}\right)$, where $\bm{V}^{\bm{serv}}=\left\{s_m|m\in{\left\{1,2,\ldots,\left|\bm{V}^{\bm{serv}}\right|\right\}}\right\}$ represents the set of SPs\footnote{We consider that each service provider can only handle one task component simultaneously, for analytical simplicity \cite{liu2023rfid,liu2020dependency}.}. Each SP $s_m\in\bm{V^{serv}}$ has an attribute $f_m$ that describes its computing capability (i.e., its CPU frequency measured in Hz), which is considered as a random variable that follows a certain distribution (detailed in Sec. 4.1), capturing the dynamic resource supply. The set of edges is given by $\bm{E}^{\bm{serv}}=\left\{e_{m,m^{\prime}}^{serv}|s_m,s_{m^\prime}\in\bm{V}^{\bm{serv}},m\neq m^{\prime}\right\}$, where $e_{m,m^{\prime}}^{serv}$ implies the existence of a connection between SPs $s_m$ and $s_{m^\prime}$, i.e., a V2V link. 
{Each edge $e_{m,m^{\prime}}^{serv}\in\bm{E^{serv}}$ has a weight $t_{m,m^{\prime}}^{conn}$, which helps capture the contact duration between two SPs. For convenience of expression, we use matrix $\bm{W}^{\bm{serv}}=[t_{m,m^{\prime}}^{conn}]_{e_{m,m^\prime}^{serv}\in\bm{E}^{\bm{serv}}}$ to represent the set of $t_{m,m^{\prime}}^{conn}$.} Specifically, $t_{m,m^\prime}^{conn}$ is considered as a random variable following a certain distribution (discussed in Sec. 4.1), as the vehicle mobility can make the contact duration between them fluctuate. We use $\bm{f} = \left[f_m\right]_{1 \leqslant m\leqslant |\bm{V}^{\bm{serv}}|}$ to denote the profile of computing capabilities of SPs. 
\setlength{\tabcolsep}{0.5mm}{
	\begin{table}[h!t]
		{\small
            \vspace{-0.2 cm}
			{\setlength{\extrarowheight}{2.3pt}
				{
					\smaller
					\caption{{Major Notations}}
					\renewcommand\arraystretch{1}
					\vspace{-3mm}
					\begin{tabular}{|ll|}
						\hline\\[-3.5mm]
						\multicolumn{1}{|c|}{{\textbf{Notation}}} & \multicolumn{1}{c|}{{\textbf{Description}}} \\ 
						\hline
						\multicolumn{1}{|l|}{{$\bm{G}^{\bm{serv}}$, $\bm{V}^{\bm{serv}}$}} & {VC graph and the set of SPs} \\
						\hline
						\multicolumn{1}{|l|}{{$\bm{E}^{\bm{serv}}$, $\bm{W}^{\bm{serv}}$}} & {Set of edges, and edges weights in VC graph}\\
						\hline
						\multicolumn{1}{|l|}{{$s_m$, $f_m$}} & {$m^\text{th}$ SP and corresponding computing capacity} \\
						\hline
						\multicolumn{1}{|l|}{{$e_{m,m^\prime}^{serv}$, $t_{m,m^\prime}^{conn}$}} & {Edge and edge weight between SPs $s_m$ and $s_{m^\prime}$} \\
						\hline
						\multicolumn{1}{|l|}{{$\bm{G}^{\bm{task}}$, $\bm{V}^{\bm{task}}$}} & {Task graph and the set of task components} \\
						\hline
						\multicolumn{1}{|l|}{{$\bm{E}^{\bm{task}}$, $\bm{W}^{\bm{task}}$}} & {Set of edges and edges weights in task graph} \\
						\hline
						\multicolumn{1}{|l|}{{$v_n$, $t_n^{max}$}} & {$n^\text{th}$ task component and its tolerable completion time} \\
						\hline
						\multicolumn{1}{|l|}{{$d_n$, $q_n$}} & {Data size and required computing resources of $v_n$} \\
						\hline
						\multicolumn{1}{|l|}{{$e_{n,n^\prime}^{task}$, $w_{n,n^\prime}^{task}$}} & {Edge and edge weight between components $v_n$ and $v_{n^\prime}$} \\
						\hline
						\multicolumn{1}{|l|}{{$c_{m,m^\prime}^{exch}$}} & {Data exchange cost between SPs $s_m$ and $s_{m^\prime}$} \\
						\hline
						\multicolumn{1}{|l|}{{$r_{m}$}} & {Data transmission rate from an AP to SP $s_m$} \\
						\hline
						\multicolumn{1}{|l|}{{$\alpha_{n,m}$}} & {Assignment between component $v_n$ and SP $s_m$} \\
						\hline
						\multicolumn{1}{|l|}{{$\beta_{m,m^\prime}$}} & {Existence of data exchange between SPs $s_m$ and $s_{m^\prime}$} \\
						\hline
						\multicolumn{1}{|l|}{{$\mathbbm{t}_{n,m}^{comp}$}} & {Execution time of $v_n$ on $s_m$} \\
						\hline
						\multicolumn{1}{|l|}{{$\mathbbm{t}_{n,m}^{comm}$}} & {Time for transmitting data of $v_n$ to $s_m$} \\
						\hline
						\multicolumn{1}{|l|}{{$\mathbbm{t}_{n,m}^{sum}$}} & {Completion time of $v_n$ on $s_m$} \\
						\hline
						\multicolumn{1}{|l|}{{$\mathbf{A}$, $\mathbf{B}$}} & {Profiles of $\alpha_{n,m}$ and $\beta_{m,m^\prime}$} \\
						\hline
						\multicolumn{1}{|l|}{{$\mathbbm{T}\left(\mathbf{A}\right)$}} & {Completion time of the graph task} \\
						\hline
						\multicolumn{1}{|l|}{{$\mathbbm{C}\left(\mathbf{B}\right)$}} & {Data exchange cost for completing the graph task} \\
						\hline
						\multicolumn{1}{|l|}{{$\lambda_t$, $\lambda_c$}} & {Non-negative weight coefficients} \\
						\hline
						\multicolumn{1}{|l|}{{$\xi$, $\xi^\prime$}} & {Threshold for risks} \\
						\hline
						\multicolumn{1}{|l|}{{$\mathbf{A^{off}}$, $\mathbf{A^{on}}$}} & {Template from offline and online mode} \\
						\hline
					\end{tabular}
					\label{table3}
		}}}
        \vspace{-0.3 cm}
\end{table}}

\vspace{-0.1 cm}
\subsection{Modeling of Graph Task}
A computation-intensive task is represented as an undirected weighted graph $\bm{G}^{\bm{task}}=(\bm{V}^{\bm{task}},\bm{E}^{\bm{task}},\bm{W}^{\bm{task}})$, where the vertex set $\bm{V}^{\bm{task}}=\left\{v_n|n\in\left\{1,2,\ldots,\left|\bm{V}^{\bm{task}}\right|\right\}\right\}$ collects the set of task components, each of which requires certain computing resources and can be assigned to a SP for processing. Each task component ${v_n}$ is further characterized by a triple $(t_n^{max},q_n,d_n)$, where $t_n^{max}$ denotes its tolerable completion time, $q_n$ represents its required computing resources (measured by the number of CPU cycles), and $d_n$ describes its data size. The set of edges among components is given by $\bm{E}^{\bm{task}}=\left\{e_{n,n^\prime}^{task}|v_n,v_{n^\prime}\in\bm{V}^{\bm{task}},n\neq n^\prime\right\}$, where the existence of an edge $e_{n,n^\prime}^{task}$ between components $v_n$ and $v_{n^\prime}$ reflects the interdependency among these components, i.e., they need to exchange data during their execution. {Further, the weight matrix $\bm{W}^{\bm{task}}=[w_{n,n^\prime}^{task}]_{e_{n,n^\prime}^{task}\in\bm{E}^{\bm{task}}}$  collects edge weights, where $w_{n,n^\prime}^{task}$ indicates the required time duration for data exchange.} As a result, for successful task execution, the contact duration between two SPs $v_m$ and $v_{m'}$ (i.e., $t_{m,m'}^{conn}$) that handle connected components $v_n$ and $v_{n^\prime}$ should exceed (or at least be equal to) weight $w_{n,n^\prime}^{task}$.

\vspace{-0.2 cm}
\subsection{Contact and Communication Model}
Considering two vehicles denoted by $s_m$ and $s_{m'}$, a V2V contact event between $s_m$ and $s_{m'}$ within time interval $\tau\in\left(\tau_1,\tau_2\right)$ occurs when the following conditions are jointly satisfied: $L_{m,m'}\left(\tau_1\right)>R$, $L_{m,m'}\left(\tau\right)\le R$, and $L_{m,m'}\left(\tau_2\right)>R$, where $L_{m,m'}\left(\tau\right)$ denotes the Euclidean distance between the vehicles at time $\tau$, and $R$ represents the communication radius of vehicles. For V2I communications, each vehicle can interact with an AP when it is located inside of the AP's coverage\footnote{Despite considering APs in the studied region, we refrain from personalizing them—meaning, we do not introduce specific mathematical notations for individual APs. This decision is rooted in the vehicular mobility model assumed in our paper. To maintain generality, we lack precise knowledge of SPs exact locations during each practical task scheduling event in the designed offline mode. Consequently, it becomes challenging to determine the specific AP accessible to each SP. In our approach, SPs interact with the EP by connecting to a nearby AP, which also elucidates why the V2I channel quality is represented as a random variable.}.
{Specifically,
the data transmission model mainly considers V2I communication links\footnote{Since it is challenging to capture the size of intermediate exchanged data between SPs when handling connected components, we thus do not consider the data transmission rate associated with V2V links\cite{liwang2019allocation,liwang2020multi} with dynamic nature. Instead, the cost (e.g., delay for transmitting intermediate data) incurred by data sharing among these SP can be estimated by a random variable detailed in Sec. 2.6, without loss of generality.}. To get access to uncertain time-varying communication qualities caused by factors such as vehicular mobility, we model the downlink data transmission rate $r_{m}$ from a nearby AP to a SP $s_m$\ as a random variable with a certain distribution (discussed in Sec. 4.1), describing the changing channel quality between APs and SPs. Let $\bm{r}=[r_{m}]_{1≤m≤|\bm{V}^{\bm{serv}}|}$ denote the profile of $r_m$ for notational simplicity. Also, we omit the communication cost (e.g., delay) between the {EP} and APs due to the high-speed wired links (also as supported by \cite{hu2019task})}.

\vspace{-0.3 cm}
\subsection{Task Completion Time}
Upon executing task component $v_n$ on SP $s_m$, the execution time is given by $\mathbbm{t}_{n,m}^{comp}=\frac{{q}_n}{f_m}$, while the corresponding data transmission time from the AP to the SP is given by $\mathbbm{t}_{n,m}^{comm}=\frac{{d}_n}{{r_{m}}}$.
Therefore, the overall completion time\footnote{The time for the resulting feedback obtained by processing of the task is omitted in our model since the result often possesses a small size, as also presumed in \cite{huang2012dynamic,liu2020dependency,jovsilo2018joint}.} of $v_n$ upon being processed on $s_m$ is given by
\begin{equation} 
\setlength{\abovedisplayskip}{3pt}
\setlength{\belowdisplayskip}{3pt}
\mathbbm{t}_{n,m}^{sum}=\mathbbm{t}_{n,m}^{comp}+\mathbbm{t}_{n,m}^{comm}= \frac{q_n}{f_m} + \frac{d_n}{r_m}. 
\label{eq1} 
\small
\end{equation} 
Also, let binary indicator $\alpha_{n,m}$ denote the assignment between component $v_n$ and SP $s_m$, where $\alpha_{n,m}=1$ indicates that $v_n$ is processed on $s_m$; $\alpha_{n,m}=0$, otherwise. 
Let $\mathbf{A}=\left[\alpha_{n,m}\right]_{1\le n\le\left|\bm{V}^{\bm{task}}\right|,\ 1\le m\le\left|\bm{V}^{\bm{serv}}\right|}$ denote task allocation profile. The overall completion time of graph task $\bm{G^{task}}$ denoted by $\mathbbm{T}\left(\mathbf{A}\right)$ is determined by its component with the longest completion time. Mathematically, we have
\begin{equation}
\setlength{\abovedisplayskip}{3pt}
\setlength{\belowdisplayskip}{3pt}
\mathbbm{T}\left(\mathbf{A}\right)=\mathrm{max}{\left[\alpha_{n,m}\mathbbm{t}_{n,m}^{sum}\right]}_{1\le n\le\left|\bm{V}^{\bm{task}}\right|,\ 1\le m\le|\bm{V}^{\bm{serv}}|}
\label{eq2}  
\small
\end{equation} 
\vspace{-0.6 cm}
\subsection{Data Exchange Cost}
Upon executing two connected task components on two separate SPs, the data exchange requirement among the components can lead to a data exchange cost. To model this cost, let $c_{m,m^\prime}^{exch}$ denote the cost\footnote{As the task structure may require vehicles to maintain communications during component processing, costs can be incurred during this time. For example, the energy consumption for transmitting intermediate data, the financial cost on data traffic (e.g., data bills, electricity bills), etc. Thus, in this paper, we use data exchange cost ($c^{exch}_{m,m^\prime}$) to reflect such a consideration\cite{liwang2023graph}.} incurred by data sharing between SPs $s_m$ and $s_{m^\prime}$. To better capture the varying V2V channel qualities, $c_{m,m^\prime}^{exch}$ is considered as a random variable which obeys a certain distribution (will be specified in Sec. 4.1). Also, let $\bm{c}=[c_{m,m^\prime}^{exch}]_{1 \leqslant m,m^{\prime}\leqslant |\bm{V}^{\bm{serv}}|}$ denote the data exchange cost profile among vehicles.

To describe the cost of processing two components on $s_m$ and $s_{m'}$, we introduce a binary indicator $\beta_{m,m^\prime}$, where $\beta_{m,m^\prime}=1$ denotes that data transmission cost is incurred; while $\beta_{m,m^\prime}=0$, otherwise. Accordingly, $\beta_{m,m^\prime}$ is a piecewise function of component-to-vehicle allocation indicators $\alpha_{n,m}$ and $\alpha_{n^\prime,m^\prime}$ as follows:
\begin{equation}
	\setlength{\abovedisplayskip}{3pt}
	\setlength{\belowdisplayskip}{3pt}
	\small
	\beta_{m,m^\prime} =
	\begin{cases}
		1, & \forall e_{n,n^\prime}^{task}\in\bm{E}^{\bm{task}},\ n\neq{n}^\prime,m\neq {m^\prime} \\
		& \text{and} \enspace \alpha_{n,m}\alpha_{n^\prime,m^\prime}=1\\
		0, & \text{otherwise} 
	\end{cases}.
	\label{eq3}
	\small
\end{equation}
Let $\mathbf{B}=[\beta_{m,m^\prime}]_{1\le m,m^\prime\le\bm{|V^{serv}|}}$ collect all the $\beta_{m,m^\prime}$. The overall data exchange cost for task processing can be calculated as   
\begin{equation}
	\setlength{\abovedisplayskip}{3pt}
	\setlength{\belowdisplayskip}{3pt}
	\mathbbm{C}\left(\mathbf{B}\right)=\frac{1}{2}\sum_{m=1}^{|\bm{V}^{\bm{serv}}|}\sum_{m^{\prime}=1}^{|\bm{V}^{\bm{serv}}|}{\beta_{m,m^\prime}c_{m,m^\prime}^{exch}}, 
	\label{eq4}
	\small
\end{equation} 
where the normalization by 1/2 is applied to prevent redundant computation of data exchange costs caused by undirected edges: this adjustment ensures an accurate reflection of the actual costs without double-counting.

\vspace{-0.2 cm}
\section{Hybrid Graph Task Scheduling over Dynamic VCs}
In this section, we describe our proposed hybrid graph task scheduling methodology over dynamic vehicles. 
\vspace{-0.2 cm}
\subsection{Cost Function}
We evaluate the performance and effectiveness of task execution through measuring the overall task completion time and data exchange cost. Our approach involves defining a cost function that takes into account these two key indicators/metrics, represented as a weighted sum between them, as follows:
\begin{equation} 
	\setlength{\abovedisplayskip}{3pt}
	\setlength{\belowdisplayskip}{3pt}
	{\mathcal{F}(\mathbf{A,B})= \lambda_t{\mathbbm{T}\left(\mathbf{A}\right)}+\lambda_c{\mathbbm{C}\left(\mathbf{B}\right)}},
	\label{eq5}
	\small
\end{equation}
where $\lambda_t$ and $\lambda_c$ are non-negative weighting coefficients that weigh the importance of the two metrics. Given the uncertainties in the system, it is imperative that we focus on the expected value of the cost function, taking into account any potential uncertainties. Mathematically, the expected value of the cost function (the expectation is taken with respect to all the uncertainties in the network, such as vehicle contact duration) is given by
\begin{equation}
	\setlength{\abovedisplayskip}{3pt}
	\setlength{\belowdisplayskip}{3pt}
	{\overline{\mathcal{F}}(\mathbf{A,B})= \lambda_t{\overline{\mathbbm{T}}\left(\mathbf{A}\right)}+\lambda_c{\overline{\mathbbm{C}}\left(\mathbf{B}\right)}}, 
	\label{eq6} 
	\small
\end{equation}
where $\overline{\mathbbm{T}}\left(\mathbf{A}\right)$ and $\overline{\mathbbm{C}}\left(\mathbf{B}\right)$ denote the expectation\footnote{Unless otherwise state, an overline above a notation (e.g., $\overline{\mathbbm{T}}\left(\mathbf{A}\right),\overline{\mathbbm{t}_{n,m}^{sum}},\overline{c_{m,m^\prime}^{exch}}$) denotes the corresponding expectation.} of task completion time and data exchange cost, respectively. Specifically, $\overline{\mathbbm{T}}\left(\mathbf{A}\right)$ can be approximated as follows, where the corresponding derivation is given in Appendix A:
\begin{equation}
	\setlength{\abovedisplayskip}{3pt}
	\setlength{\belowdisplayskip}{3pt}
	\small
	\begin{aligned}
			\overline{\mathbbm{T}}\left(\mathbf{A}\right)\approx\mathrm{max}{\left[\alpha_{n,m}\overline{\mathbbm{t}_{n,m}^{sum},}\right]}_{1\le n\le\left|\bm{V}^{\bm{task}}\right|,1\le m\le\left|\bm{V}^{\bm{serv}}\right|},
	\end{aligned}
	\label{eq7} 
\end{equation}
where $\overline{\mathbbm{t}^{sum}_{n,m}}$ is given by
\begin{equation}
	\setlength{\abovedisplayskip}{3pt}
	\setlength{\belowdisplayskip}{3pt} 	
	\overline{\mathbbm{t}_{n,m}^{sum}} =\ \overline{\mathbbm{t}_{n,m}^{comp}}+\overline{\mathbbm{t}_{n,m}^{comm}}=\left(\overline{\tau_m^{comp}}q_n+\overline{\tau_{m}^{comm}}d_n\right).
	\label{eq8} 
	\small
\end{equation} 
In \eqref{eq8}, $\overline{\tau_m^{comp}} = \mathrm{E}[\frac{1}{f_m}]$ and $\overline{\tau_{m}^{comm}}=\mathrm{E}[\frac{1}{r_m}]$ denote the expected values of unit execution time (e.g., PerCPU cycle) and data transmission time (e.g., Perbit), respectively, which can be obtained numerically given the distribution of $f_m$ and $r_m$. Moreover, $\overline{\mathbbm{C}}\left(\mathbf{B}\right)$ in \eqref{eq6} is given by
\begin{equation}
	\setlength{\abovedisplayskip}{3pt}
	\setlength{\belowdisplayskip}{3pt}
	\overline{\mathbbm{C}}\left(\mathbf{B}\right)=\frac{1}{2}\sum_{m=1}^{|\bm{V}^{\bm{serv}}|}\sum_{m^{\prime}=1}^{|\bm{V}^{\bm{serv}}|}{\beta_{m,m^\prime}\overline{c_{m,m^\prime}^{exch}}},
	\label{eq9} 
	\small
\end{equation} 
where $\overline{c_{m,m^\prime}^{exch}}$ represents the expectation of $c^{exch}_{m,m'}$, which can be obtained numerically given the distribution of $c_{m,m^\prime}^{exch}$. 

%

\vspace{-0.2 cm}
\subsection{Offline Approach Design: Risk-Aware Pilot Isomorphic Subgraph Searching (RA-PilotISS)}
{Building on the cost function $\mathcal{F}(\mathbf{A,B})$ presented in Section 3.1, the goal of this section is to identify the task scheduling template $\mathbf{A^{off}}$ that minimizes \textbf{the expected value} of $\mathcal{F}(\mathbf{A,B})$. This requires addressing a subgraph isomorphism and network optimization formulations, upon considering the uncertainties within the network. We next introduce \textbf{r}isk-\textbf{a}ware \textbf{pilot} \textbf{i}somorphic \textbf{s}ubgraph \textbf{s}earching (RA-PilotISS) to obtain $\mathbf{A^{off}}$ based on the historical statistics of uncertainties (i.e., $f_m, t_{m,m^\prime}^{conn}, c_{m,m^\prime}^{exch}, r_{m}$) in advance to practical task scheduling events\footnote{This paPerdoes not delve into the design of period of validity for $\mathbf{A^{off}}$. A heuristic approach is to renew $\mathbf{A^{off}}$ after a specific duration or a predefined number of failures. We leave further investigations on effective validation techniques for $\mathbf{A^{off}}$ as future work.}.} 
To simplify the notations, we use $\overline{\bm{f}}, \overline{\bm{W^{serv}}}, \overline{\bm{r}}$, and $\overline{\bm{c}}$ to denote the collection of statistics obtained from historical data. For example, $\overline{\bm{f}}=\left[\overline{f_m}\right]_{1 \leqslant m\leqslant |\bm{V}^{\bm{serv}}|}$ represents the profile of the expected values of $f_m$. 

\vspace{-0.2 cm}
\subsubsection{Risk analysis}
In general, uncertainties in the network can introduce potential risks to task scheduling, leading to the case where $\mathbf{A^{off}}$ template fails to support task completion in practical scheduling events. To make $\mathbf{A^{off}}$ robust against such risks, we take into account two key risk factors: completion time and graph task structure preservation. Firstly, a task component $v_n \in \bm{V}^{\bm{task}}$ faces the risk of completing overtime on SP $s_m \in \bm{V}^{\bm{serv}}$. Such a risk is measured by the probability of the corresponding completion time exceeding its tolerable time $t_n^{max}$ as follows:
\begin{equation}
	\setlength{\abovedisplayskip}{3pt}
	\setlength{\belowdisplayskip}{3pt}
	R^{time}_{n,m} =\mathrm{Pr}\left({\mathbbm{t}_{n,m}^{sum}}>t_n^{max}\right). 
	\label{eq10} 
	\small
\end{equation}
Secondly, for a graph task to be completed smoothly, there must be ongoing communications between its interconnected components. This necessitates V2V connections between vehicles that process connected task components to maintain the task's structure. Thus, the likelihood of any vehicle pair (such as SPs $s_m$ and $s_{m^\prime}$) failing to facilitate the transmission of crucial data for their assigned components (e.g., components $v_n$ and $v_{n^\prime}$) can be calculated as another probabilistic risk factor:
\begin{equation}
	\setlength{\abovedisplayskip}{3pt}
	\setlength{\belowdisplayskip}{3pt}
	R^{struc}_{m,m^\prime} =\mathrm{Pr}\left({t_{m,m^\prime}^{conn}}< w_{n,n^\prime}^{task}\right),
	\label{eq11}
	\small
\end{equation}
given that $\beta_{m,m^\prime}=1$. Derivations of \eqref{eq10} and \eqref{eq11} are provided in Appendix B and C.
\vspace{-0.2 cm}
\subsubsection{Design of RA-PilotISS}
Our designed offline task scheduling aims to acquire template {$\mathbf{A^{off}}$ that can efficiently minimize the value of $\overline{\mathcal{F}}(\mathbf{A,B})$, for which we formulate this goal as the following optimization problem $\bm{\mathcal{P}}$:}
\begin{equation} 
	\setlength{\abovedisplayskip}{3pt}
	\setlength{\belowdisplayskip}{3pt}
	\bm{\mathcal{P}}: \mathbf{A^{off}}={\mathop{\arg\min}_{\mathbf{A}}\overline{\mathcal{F}}(\mathbf{A,B})} 
	\label{eq12}
	\small 
\end{equation}
\vspace{-0.3 cm}
\begin{flalign}
	\tag{C1}
	\label{C1}
	&\ \text{s.t.}~~~~~~~~~~~~~~~~	\sum_{m=1}^{|\bm{V}^{\bm{serv}}|}{\alpha_{n,m}= 1}, \forall{ v_n \in \bm{V}^{\bm{task}}}&
	\small
\end{flalign}
\begin{equation}
	R^{time}_{n,m}\le{\xi}, \ \text{if} \ \alpha_{n,m}=1
	\tag{C2}
	\label{C2} 
	\small
\end{equation}
\begin{equation} 
	R^{struc}_{m,m^\prime}\le{\xi^\prime},\ \text{if} \ \beta_{m,m^\prime}=1
	\tag{C3}
	\label{C3} 
	\small
\end{equation} 
In $\bm{\mathcal{P}}$, constraint \eqref{C1} ensures that each task component is assigned to only one vehicle. Constraints \eqref{C2} and \eqref{C3} are probabilistic constraints that limit the possibility of risks falling in certain ranges. For example, \eqref{C2} constrains the likelihood of $v_n$'s completion time exceeding its tolerance, where $\xi\in(0,1]$. Similarly, \eqref{C3} controls the probability of the contact duration between two SPs ($s_m$ and $s_{m^\prime}$) that handle connected components ($v_n$ and $v_{n^\prime}$) failing to meet the required time $w_{n,n^\prime}^{task}$ within a certain range, where $\xi^\prime\in(0,1]$. These risk-related constraints ensure that $\mathbf{A^{off}}$ can be effectively used during future/practical task scheduling events with a high probability, reducing the time spent on online decision-making.

It is straightforward to verify that the optimization problem $\bm{\mathcal{P}}$ is a non-linear integer programming (NIP) problem, which is generally considered to be NP-hard\cite{liwang2023graph}. Moreover, due to the need for the SPs to maintain the same structure as the graph task during its execution, constraints associated with $\bm{\mathcal{P}}$ require solving the subgraph isomorphism problem, which is known to be NP-complete\cite{Conte2004}. Obtaining the optimal template in a timely manner is therefore a daunting challenge, particularly as the time required for solving $\bm{\mathcal{P}}$ increases exponentially with growing vehicular density (e.g., an increasing number of SPs) and the complexity of VC topology (i.e., intricate connections among SPs). To address this challenge, we propose RA-PilotISS, which identifies the template $\mathbf{A^{off}}$ in advance to future task scheduling processes with acceptable risks. In the following, we first introduce the key definitions in RA-PilotISS.
\vspace{-0.1 cm}
\begin{Definition}
Neighborhood $\left(\rm{NHood(.)}\right)$: A neighborhood of component $v_n\in\bm{V^{task}}$ covers components with an edge connecting them to $v_n$. We denote the neighborhood of $v_n$ by $\mathrm{NHood}(v_n)$, where  $\mathrm{NHood}(v_n)=\left\{{v_{n^\prime}|e_{n,n^\prime}^{task}\in{\bm{E^{task}}}}\right\}$. Similar definition can be applied to SPs, where the neighborhood of SP $s_m \in \bm{V^{serv}}$ is given by $\mathrm{NHood}(s_m)=\left\{{s_{m^\prime}|e_{m,m^\prime}^{serv}\in{\bm{E^{serv}}}}\right\}$.
\end{Definition}
\vspace{-0.4 cm}
\begin{Definition}
	Degree $\left(\rm{Deg(.)}\right)$: The degree of component $v_n \in \bm{V^{task}}$ is equal to the number of nodes in its neighborhood, where $\mathrm{Deg}(v_n)=\left|\mathrm{NHood}(v_n)\right|$. Similar definition can be applied to SPs, where the degree of SP $s_m \in \bm{V^{serv}}$ is given by $\mathrm{Deg}(s_m)=\left|\mathrm{NHood}(s_m)\right|$.
\end{Definition}
\vspace{-0.4 cm}
\begin{Definition}
	Maximum Neighborhood-Degree $\left(\mathrm{(MNdeg(.)}\right)$: The maximum neighborhood-degree of component $v_n\in{\bm{V^{task}}}$ is denoted by 
	${\mathrm{MNdeg}{\left(v_n\right)}}$, which is equal to the maximum degree of the nodes in its neighborhood, i.e., ${\mathrm{MNdeg}{\left(v_n\right)}}$ = $\mathrm{max}\left[\mathrm{Deg}{\left(v_{n^\prime}\right)}\right]_{v_{n^\prime}\in{\mathrm{NHood}(v_n)}}$. Similarly, for SP $s_m \in \bm{V^{serv}}$, we have ${\mathrm{MNdeg}{\left(s_m\right)}}$ = $\mathrm{max}\left[\mathrm{Deg}{\left(s_{m^\prime}\right)}\right]_{s_{m^\prime}\in{\mathrm{NHood}(s_m)}}$.
\end{Definition}
\vspace{-0.4 cm}
\begin{Definition}
	Distance $\left(\rm{Dst(.,.)}\right)$: The distance between components $v_n$ and $v_{n^\prime}$ is denoted by $\mathrm{Dst}(v_{n},v_{n^\prime})$, which is equal to the length of the shortest path \cite{west2001introduction} from $v_n$ to $v_{n^\prime}$ in the task graph topology. Similarly, $\mathrm{Dst}(s_{m},s_{m^\prime})$ describes the length of the shortest path from $s_m$ to $s_{m^\prime}$ in the VC graph topology.
\end{Definition}
\vspace{-0.4 cm}
\begin{Definition}
	Eccentricity $\left(\mathrm{Ecc(.)}\right)$: The eccentricity of component $v_n\in{\bm{V^{task}}}$ is defined as the maximum distance between $v_n$ and other components in graph task $\bm{G^{task}}$ and is denoted by $\mathrm{Ecc}\left(v_n\right)$. Mathematically, $\mathrm{Ecc}\left(v_n\right) = \mathrm{max}\left[\mathrm{Dst}(v_{n},v_{n^\prime})\right]$, where $n\neq n^\prime, v_{n'}\in{\bm{V^{task}}}$.
\end{Definition}
\vspace{-0.4 cm}
\begin{Definition}
	Alternative SPs $\left(\mathrm{ASP(.)}\right)$: The alternative SP set of component $v_n\in{\bm{V^{task}}}$ denotes the set of SPs that can be mapped to $v_n$, represented as $\mathrm{ASP}\left(v_n\right)$.
\end{Definition}
\vspace{-0.4 cm}
\begin{Definition}
	Pivot component: A pivot component in a graph task topology indicates the component $v_n$ with the minimum value of ${\left|\mathrm{ASP}\left(v_n\right)\right|\times \mathrm{Ecc}\left(v_n\right)}$.
\end{Definition}
Based on the above definitions, we develop RA-PilotISS (Algorithm 1), which involves multiple steps (Algorithms 2-5). Among them, we first search for all isomorphic subgraphs that meet constraints \eqref{C1}-\eqref{C3} through steps 1-2, and then determine the template with the lowest expectation of cost among all isomorphic subgraphs by step 3.

\begin{algorithm}[htb!]
	\small
	{\setstretch{0.6} 
		\caption{Risk-aware advance isomorphic subgraph searching (RA-PilotISS)}\label{algorithm} \label{algorithm}
		\KwIn{$\bm{G^{task}}$, $\bm{G^{serv}}$, $\overline{\bm{r}}$, $\overline{\bm{f}}$, $\overline{\bm{W^{serv}}}$, $\overline{\bm{c}}$}
		\KwOut{Template $\mathbf{A^{off}}$}
		$\widetilde{{v}_n},\mathrm{ASP}\left(\widetilde{v_n} \right)\leftarrow{{\rm{PivotCSelectOffline}}}$ \% Algorithm 2 \\
		${SG}_{all}\gets\emptyset$ \\
		\ForEach{$\widetilde{s_m}\ \in\ {\mathrm{ASP}\left(\widetilde{v_n}\right)}$}
		{$Candi\gets {\rm{RegionExploreOffline}}$ \% Algorithm 3 \\
			\If{$Candi\neq\emptyset$}
			{
				${SG}_m\gets {\rm{SubGSearchOffline}}$ \% Algorithm 4 \\
			}
			${SG}_{all}\gets{SG}_{all}\cup{SG}_m$
		}
		$\mathbf{A^{off}}\gets {\rm{OptTSelectOffline}}$ \% Algorithm 5 \\
	}
	\label{Alg1}
\end{algorithm}
\begin{algorithm}[h!]
	\small
	\setstretch{0.6} 
	\caption{PivotCSelectOffline}\label{algorithm}
	\KwIn{$\bm{G^{task}}$, $\bm{G^{serv}}$, $\overline{\bm{r}}$, $\overline{\bm{f}}$ , $\overline{\bm{W^{serv}}}$}
	\KwOut{The pivot component $\widetilde{v_n}$, its alternative SPs $\mathrm{ASP}\left(\widetilde{v_n}\right)$, and eccentricity $\mathrm{Ecc}\left(\widetilde{v_n}\right)$}
	$\mathcal{L} \gets $  List of ordered pairs $\left< v_n,\left|\mathrm{ASP}\left(v_n\right)\right|\right>$ \hspace{6em} 
	\% $\left|\mathrm{ASP}\left(v_n\right)\right|$ is the number of alternative SPs\\
	\ForEach{$v_n \in \bm{V^{task}}$}
	{\ForEach{$s_m \in \bm{V^{serv}}$}
		{
			\If
			{
				$R^{time}_{n,m}\le\xi \ \mathrm{and} \  \mathrm{Deg}\left(v_n\right)\le \mathrm{Deg}\left(s_m\right) \   \mathrm{and} \  
				{\mathrm{MNdeg}{\left(v_n\right)}} \le {\mathrm{MNdeg}{\left(s_m\right)}} $
			}
			{
				$\mathrm{ASP}\left(v_n\right)\gets\mathrm{ASP}\left(v_n\right)\cup s_m$
			}
		}
		$\mathcal{L}(n) \gets \left<v_n,\left|\mathrm{ASP}\left(v_n\right)\right|\right>$ \\
		$\mathrm{Ecc}\left({v_n}\right) \gets$ Calculate the eccentricity of ${v_n}$ \\
	}
	$\widetilde{v_n}\gets v_n \ \text{with the minimum value of} ~ {\left|\mathrm{ASP}\left(v_n\right)\right|\times \mathrm{Ecc}\left(v_n\right)}$ \\
	return $\widetilde{v_n}, \mathrm{ASP}\left(\widetilde{v_n}\right), \mathrm{Ecc}\left(\widetilde{v_n}\right)$
	\label{Alg2}
\end{algorithm}

\noindent$\bullet$ \textbf{Step 1.} Pivot component selection: In RA-PilotISS, we start with selecting a pivot component ($\widetilde{{v}_n}$) through the PivotCSelectOffline function (Algorithm 2), along with its corresponding alternative SPs $\mathrm{ASP}\left(\widetilde{v_n}\right)$. {In Algorithm 2, we employ constraint \eqref{C2}, and consider both the degree and neighborhood information of every component $v_n$, to determine the alternative SP set denoted by $\mathrm{ASP}\left(v_n\right)$ (lines 1-6).} Additionally, the function calculates the eccentricity $\mathrm{Ecc}\left(v_n\right)$ for all components (line 7). Finally, the algorithm selects the pivot component $\widetilde{v_n}$ by applying line 8, which is the component with the minimum value of ${\left|\mathrm{ASP}\left(v_n\right)\right|}\times{\mathrm{Ecc}\left(v_n\right)}$.

\begin{algorithm}[h!]
	\setstretch{0.6} 
	\small
	\caption{RegionExploreOffline}
	\KwIn{$\bm{G^{task}}$, $\bm{G^{serv}}$, $\widetilde{v_n}$, $\widetilde{s_m}$, $\overline{\bm{r}}$, $\overline{\bm{f}}$, $\overline{\bm{W^{serv}}}$}
	\KwOut{$Candi$: SPs in $\bm{G^{sub}}$ that can be mapped to task components}
	$\bm{G^{sub}}$\ $\leftarrow$\ {Subgraph with $\widetilde{s_m}$ as center and $\mathrm{Ecc}\left(\widetilde{v_n}\right)$ as the radius in VC graph $\bm{G^{serv}}$} \\
	\If{$\left|\bm{V^{sub}}\right|<\left|\bm{V^{task}}\right|$}
	{return $\emptyset$}
	$Candi \gets \emptyset$ \\
	\ForEach{$v_n\in \bm{V^{task}}$}
	{$\mathrm{ASP}\left(v_n\right) \gets \emptyset$ \\
		\ForEach{$s_m\in\bm{V^{sub}}$}
		{
			\If
			{
				$\mathrm{Dst}(v_n,\widetilde{v_n}) \geq \mathrm{Dst}(s_m,\widetilde{s_m})$
			}
			{
				\If
				{
					$R^{time}_{n,m}\le\xi ~ \mathrm{and} ~  \mathrm{Deg}\left(v_n\right)\le \mathrm{Deg}\left(s_m\right) ~ \mathrm{and} ~
					{\mathrm{MNdeg}{\left(v_n\right)}} \le {\mathrm{MNdeg}{\left(s_m\right)}}$
				}
				{
					$\mathrm{ASP}\left(v_n\right)\gets\mathrm{ASP}\left(v_n\right) \cup s_m$
				}
			}
			
		}
		
	}
	$Candi \gets Candi \ \cup \left<v_n,\mathrm{ASP}\left(v_n\right)\right>$ \\
	return $Candi$
	\label{Alg3}
\end{algorithm}
\begin{algorithm}[htb]
	\setstretch{0.6} 
	\small
	\caption{SubGSearchOffline} \label{algorithm}
	\KwIn{$\bm{G^{task}}$, $\bm{G^{serv}}$, $Candi$, $\overline{\bm{r}}$, $\overline{\bm{f}}$, $\overline{\bm{W^{serv}}}$}
	\KwOut{${{SG}_m}$: The set of isomorphic subgraphs from\ \ $Candi$}
	${{SG}_m}$\ $\leftarrow$\ {All combinations of candidate SPs for task components from\ $Candi$} \\
	\ForEach{${G}_i\in{{SG}_m}$}
	{
		$\mathbf{B}_i$ $\gets$ Profile of $\beta_{m,m^\prime}$ associated with ${G}_i$\\
		\ForEach{$e_{n,n^\prime}^{task}\in\bm{E^{task}} \ \mathrm{and} \ \beta_{m,m^\prime}=1$
		}
		{
			\If
			{
				${e_{m,m^\prime}^{serv} \notin \bm{E^{serv}}} \ \mathrm{or} \  R^{struc}_{m,m^\prime}>\xi^{\prime}$ 
			}
			{
				Delete\ ${{G}_i}$\ from ${{SG}_m}$
				
			}
		}
		
	}
	return ${{SG}_m}$
	\label{Alg4}
\end{algorithm}

\noindent$\bullet$ \textbf{Step 2.} Subgraph search: By utilizing functions RegionExploreOffline (Algorithm 3) and SubGSearchOffline (Algorithm 4), as well as thoroughly examining the SPs in set $\mathrm{ASP}\left(\widetilde{v_n}\right)$, we can obtain all feasible isomorphic subgraphs. Note that the SPs within $\mathrm{ASP}\left(\widetilde{v_n}\right)$ are appropriately labeled as $\widetilde{s_m}$ to distinguish them. Particularly, in Algorithm 3, each SP $\widetilde{s_m}\in\mathrm{ASP}\left(\widetilde{v_n}\right)$ undergoes a process where a subgraph, denoted as $\bm{G^{sub}}$, is extracted from $\bm{G^{serv}}$. In this subgraph, $\widetilde{s_m}$ is regarded as the center point and $\mathrm{Ecc}\left(\widetilde{v_n}\right)$ as the radius. The radius indicates the longest path from $\widetilde{s_m}$ to other SPs (line 1). For instance, if $\mathrm{Ecc}\left(\widetilde{v_n}\right)$ is equal to 3, $\bm{G^{sub}}$ will encompass all the SPs within 3 hops from $\widetilde{s_m}$ in $\bm{G^{serv}}$. If the SPs in $\bm{G^{sub}}$ are insufficient to support component processing, an empty set is returned (lines 2-3), indicating that no isomorphic subgraph can be found in $\bm{G^{sub}}$. If there are enough SPs, Algorithm 3 then examines the degree and neighborhood details of all the task components and searches for alternative SPs for them under $\bm{G^{sub}}$. These alternative SPs are saved in set $Candi$ (lines 4-12). Afterwards, Algorithm 4 first identifies all isomorphic subgraphs in $\bm{G^{sub}}$ that have the same structure as task graph $\bm{G^{task}}$. This operation can be implemented by exploring alternative SPs for different components in $Candi$ (line 1), to find all the possible isomorphic subgraphs. Next, lines 2-6 eliminate subgraphs that fail to meet constraint \eqref{C3}. Finally, Algorithm 4 saves isomorphic subgraphs in set $SG_m$.

\noindent$\bullet$ \textbf{Step 3.} Determination of template $\mathbf{A^{off}}$: Once all isomorphic subgraphs have been obtained, referred to as set ${{SG}_{all}}$, RA-PilotISS utilizes OptTSelectOffline function (as outlined in Algorithm 5) to identify the optimal template. This involves calculating the value of $\overline{\mathcal{F}}(\mathbf{A,B})$ for various templates, with the optimal template $\mathbf{A^{off}}$ being selected based on the minimum expected value of the cost function.
\begin{algorithm}[htb]
	\setstretch{0.6} 
	\small
	\caption{OptTSelectOffline} 
	\label{algorithm}
	\KwIn{$\bm{G^{task}}$, $\bm{G^{serv}}$, ${{SG}_{all}}$, $\overline{\bm{r}}$, $\overline{\bm{f}}$, $\overline{\bm{W^{serv}}}$, $\overline{\bm{c}}$}
	\KwOut{Template $\mathbf{A^{off}}$}
	$\overline{\mathcal{F}}_{min} \gets $ Initialize the minimum expected value of cost function with a large value\\
	\ForEach{${G}_i\in {{SG}_{all}}$}
	{
		$\mathbf{A}_i,\mathbf{B}_i$ $\gets$ Scheduling template from subgraph ${G}_i$\\
		$\overline{\mathcal{F}}(\mathbf{A}_i,\mathbf{B}_i)$ $\gets$ The expected value of cost function on template $\mathbf{A}_i,\mathbf{B}_i$\\
		\If{$\overline{\mathcal{F}}(\mathbf{A}_i,\mathbf{B}_i) < \overline{\mathcal{F}}_{min}$}
		{
			$\mathbf{A^{off}} \gets {\mathbf{A}_i}$, $\overline{\mathcal{F}}_{min} \gets \overline{\mathcal{F}}(\mathbf{A}_i,\mathbf{B}_i) $
		}
	}
	return $\mathbf{A^{off}}$
	\label{Alg5}
\end{algorithm}

\vspace{-0.2 cm}
\subsection{Online Approach Design: Time-Efficient Instantaneous Isomorphic Subgraph Searching (TE-InstaISS)}
To ensure efficient and dependable scheduling of tasks, we next develop a scheduling backup plan called \textbf{t}ime-\textbf{e}fficient \textbf{insta}ntaneous \textbf{i}somorphic \textbf{s}ubgraph \textbf{s}earching (TE-InstaISS). This plan is designed to be activated when our previously obtained template $\mathbf{A^{off}}$ fails to work at practical/actual task scheduling events due to current network conditions. For instance, some SPs in $\mathbf{A^{off}}$ may not be able to support the necessary data exchange because of the limited contact duration between them during the practical/actual task scheduling. At each practical task scheduling event, TE-InstaISS is initiated to check the availability of template $\mathbf{A^{off}}$. If it is unavailable, TE-InstaISS will solve an optimization problem $\bm{\mathcal{P}^{\prime}}$ based on the current network conditions such as practical values of $\bm{r},\bm{f},\bm{W^{serv}}$, and $\bm{c}$. The ultimate goal of TE-InstaISS is to achieve the optimal template $\mathbf{A^{on}}$ that minimizes the practical value of the cost function $\mathcal{F}(\mathbf{A,B})$. We formulate $\bm{\mathcal{P}^{\prime}}$ as follows:
\begin{equation}
	\bm{\mathcal{P}^{\prime}}:
	{\mathbf{A^{on}}={\mathop{\arg\min}_{\mathbf{A}}{\mathcal{F}(\mathbf{A,B})}}} 
	\label{eq13}
	\small
\end{equation}
\begin{center}
	s.t. (C1),
\end{center}
\begin{equation}
	\mathbbm{t}_{n,m}^{sum} \leq t_n^{max},\ \forall \ \alpha_{n,m} = 1 
	\label{C4}
	\tag{C4}
	\small
\end{equation}
\begin{equation} 
	t_{m,m^\prime}^{conn} \geq w_{n,n^\prime}^{task},\ \forall \ \beta_{m,m^\prime} =1
	\label{C5} 
	\tag{C5}
	\small
\end{equation}
Note that the focus of $\bm{\mathcal{P}^{\prime}}$ is no longer on the expected value of the cost function. This is because $\bm{\mathcal{P}^{\prime}}$ operates in real-time with known information of the network. Additionally, to ensure reliable task completion, we have introduced constraints \eqref{C4} and \eqref{C5}. Constraint \eqref{C4} ensures the timely completion of task components, while constraint \eqref{C5} guarantees the stability of vehicular connections during component processing. For brevity, we have moved the details of TE-InstaISS (Algorithms 7-11) to Appendix D, as some steps designed for addressing problem $\bm{\mathcal{P}^{\prime}}$ are similar to those of problem $\bm{\mathcal{P}}$ (i.e., Algorithms 1-5).


\begin{algorithm}[htb]
	\small
	\setstretch{0.6} 
	\caption{Proposed hybrid graph task scheduling mechanism} \label{algorithm6}
	\KwIn{$\bm{G^{task}}$, $\bm{G^{serv}}$, $\bm{r}$, $\bm{f}$, $\bm{W^{serv}}$, $\bm{c}$}
	\KwOut{Template $\mathbf{A^{on}}$}
	$\mathbf{A^{off}} \ \gets$ RA-PilotISS \% Algorithm 1 \\
	$\bm{G^{off}},\mathbf{B}^*\ \gets$ SP graph and profile of $\beta_{m,m^\prime}$ associated with $\mathbf{A^{off}}$ \\
	\If{graph $\bm{G^{task}}$ and $\bm{G^{off}}$ are isomorphic}
	{
		\ForEach{$\beta_{m,m^\prime}=1$}
		{
			\If{$\mathbbm{t}_{n,m}^{sum} \leq t_n^{max}$ \ \rm{and} \ $\mathbbm{t}_{n^\prime,m^\prime}^{sum} \leq t_{n^\prime}^{max}$ \ \rm{and} \ $t_{m,m^\prime}^{conn} \geq w_{n,n^\prime}^{task}$} 
			{
				$\mathbf{A^{on}}\gets \mathbf{A^{off}}$
			}
			\Else
			{
				$\mathbf{A^{on}}\ \gets $ TE-InstaISS \% Algorithm 7
			}
		}
	}
	\Else{$\mathbf{A^{on}}\ \gets$ TE-InstaISS \% Algorithm 7}
	\label{Alg6}
\end{algorithm}

\vspace{-0.2 cm}
\subsection{Hybrid Graph Task Scheduling Methodology}
Our hybrid methodology presents a systematic approach that combines the two complementary strategies discussed above: RA-PilotISS for offline and TE-InstaISS for online scheduling, as detailed in Algorithm 6. {In particular, RA-PilotISS aims to obtain the mapping {$\mathbf{A^{off}}$} before the future task processing demands by analyzing historical statistics, which is expected to perform good from a long-term view.} Thus, Algorithm 6 first explores $\mathbf{A^{off}}$ through RA-PilotISS (line 1). This template is risk-aware and can be directly used during practical task scheduling to enhance responsive computing services. Subsequently, at each practical task scheduling event, our mechanism assesses the availability of $\mathbf{A^{off}}$ for practical tasks in a prompt manner. If $\mathbf{A^{off}}$ meets the task structure (line 3) and constraints \eqref{C4}-\eqref{C5} (lines 4-5), it becomes the definitive template for guiding task scheduling (line 6). Otherwise, TE-InstaISS searches for the optimal template $\mathbf{A^{on}}$ based on the current network conditions (lines 7-10).	

\noindent \textbf{Optimality analysis.} In the offline mode of our proposed methodology, we use the RA-PilotISS approach to identify all the feasible templates, while determining the best one(i.e., $\mathbf{A^{off}}$)) with the minimum expectation of cost function (i.e., $\overline{\mathcal{F}}(\mathbf{A,B})$) among them, based on historical data. Thus, $\mathbf{A^{off}}$ can serve as an optimal solution for task scheduling from a long-term perspective, i.e., in the time-average sense as time extends to infinity. {However, during each practical task scheduling event, $\mathbf{A^{off}}$ may be sub-optimal upon considering the current network condition, due to the dynamic and uncertain nature of the network of our interest. In other words, $\mathbf{A^{off}}$ helps reach the minimum of $\overline{\mathcal{F}}(\mathbf{A,B})$, but cannot guarantee the optimal value of $\mathcal{F}(\mathbf{A,B})$.} Then, during practical task scheduling events, the online mode first confirms the availability of $\mathbf{A^{off}}$, and deploys TE-InstaISS if $\mathbf{A^{off}}$ fails. During this time, our TE-InstaISS approach enables achieving the optimal template by minimizing the practical cost function (i.e., $\mathcal{F}(\mathbf{A,B})$) under given network conditions, in a timely manner.
	
\noindent\textbf{Efficiency analysis}. 
We delve into two perspectives for analyzing the efficiency of our proposed hybrid methodology, which are \textit{i)} computational complexity analysis from the view of algorithm design, and \textit{ii)} the advantages brought by the hybrid setting. For \textit{i)}, Since both RA-PilotISS and TE-InstaISS follow similar procedures, with key differences in constraint and the objective function design, their time complexities can be the same, namely $O(mn!)$. Here, $m$ depends on the size of $\mathrm{ASP}\left(\widetilde{v_n}\right)$ and $n$ is related to the size of $\mathrm{Ecc}\left(\widetilde{v_n}\right)$. For \textit{ii)}, Since our proposed methodology is stagewise, we offer an unique view of computational complexity in forms of the lower bound and upPerbound of P-HTS. In particular, the best complexity (lower bound) of P-HTS can be $O(1)$ when template $\mathbf{A^{off}}$ is valid, since $\mathbf{A^{off}}$ can be used directly during task scheduling without additional decisions. Then, when $\mathbf{A^{off}}$ is invalid, the backup approach TE-InstaISS can be triggered, which offers a complexity of $O(mn!)$ for P-HTS (upPerbound). Although our designed RA-PilotISS may introduce some latency in obtaining $\mathbf{A^{off}}$, this process occurs solely before practical task scheduling events. Ultimately, this facilitates an expedited practical task scheduling, leveraging the pre-established template $\mathbf{A^{off}}$. The synergy between RA-PilotISS and TE-InstaISS, operating as two complementary strategies within the hybrid mechanism, enables timely and seamless graph task scheduling across dynamic VCs.

\begin{figure}[t!]
	\centering
	\includegraphics[width=1\linewidth]{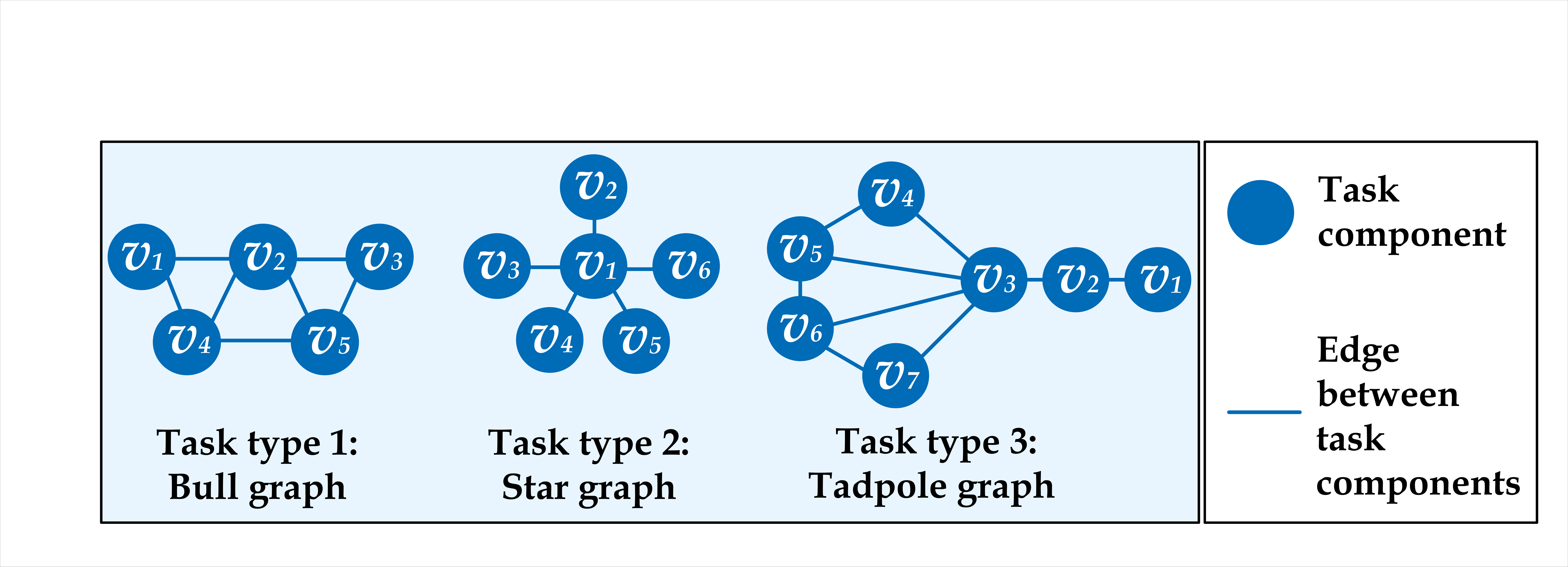}
	\vspace{-0.7 cm}
	\caption{Various graph task types.\cite{hosseinalipour2019power,gao2021truthful}}
	\label{fig3}
	\vspace{-0.2 cm}
\end{figure}

\vspace{-0.1cm}
\section{Performance Evaluation} 
In this section, We conduct comprehensive simulations to illustrate the performance of our methodology P-HTS.
\begin{figure*}[t!] \centering    
	\vspace{0 cm} 
	\subfigtopskip=2pt
	\subfigbottomskip=1pt
	\subfigcapskip=0 cm
	\setlength{\abovecaptionskip}{0 cm} 
	\subfigure[] {
		\label{fig4a }     
		\includegraphics[width=0.66 \columnwidth]{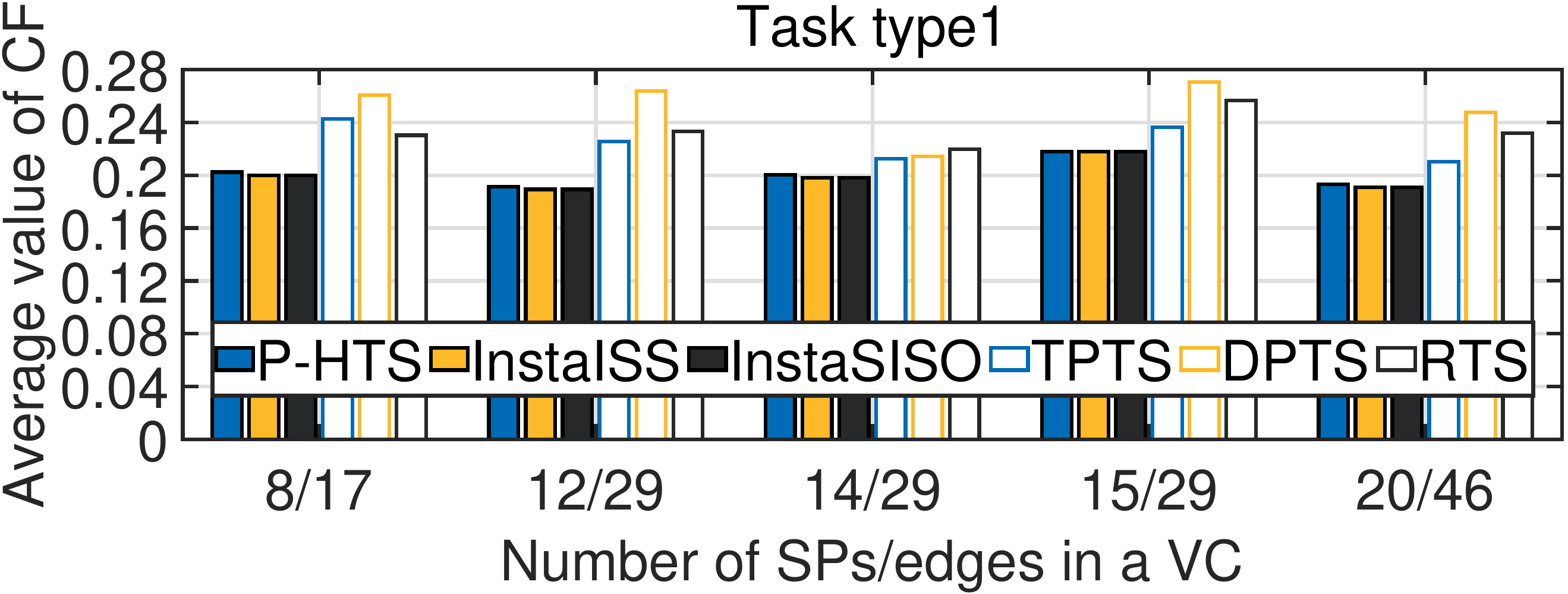}   
	}    \hspace{-3 mm}  
	\subfigure[] { 
		\label{fig4b}     
		\includegraphics[width=0.66\columnwidth]{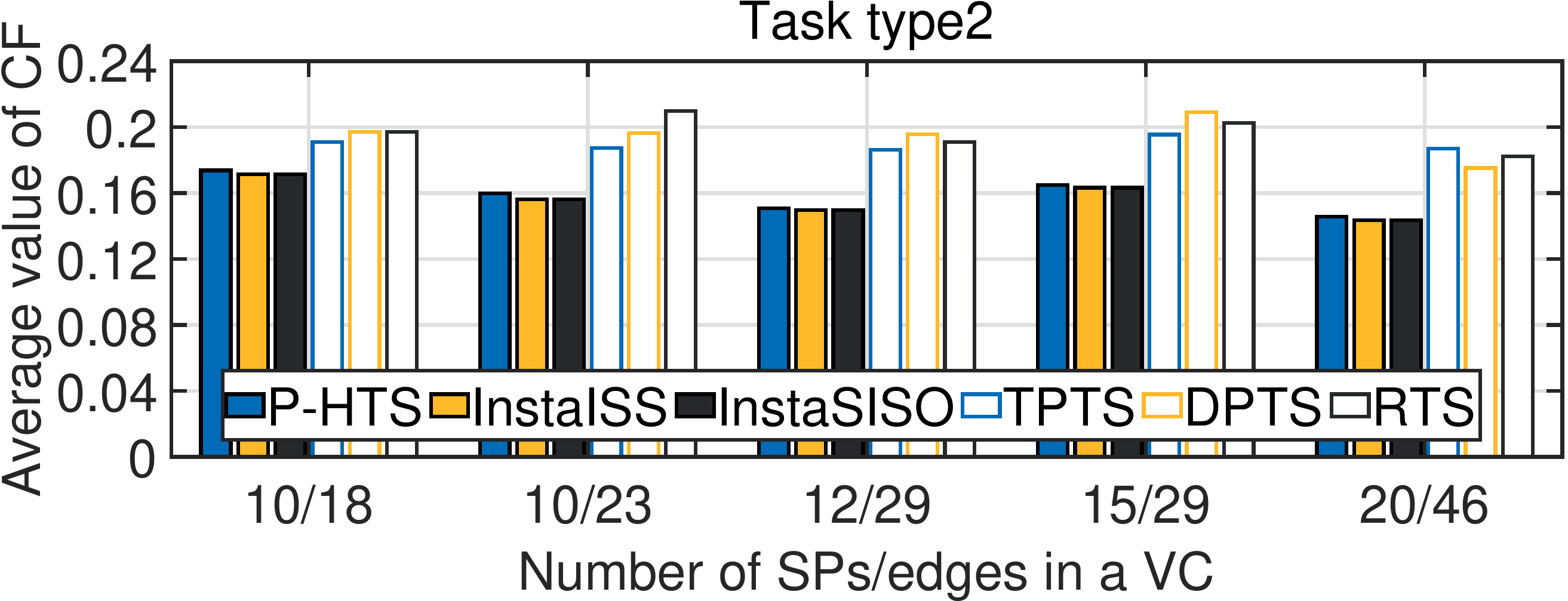}     
	}     \hspace{-3 mm} 
	\subfigure[] {
		\label{fig4c}     
		\includegraphics[width=0.66\columnwidth]{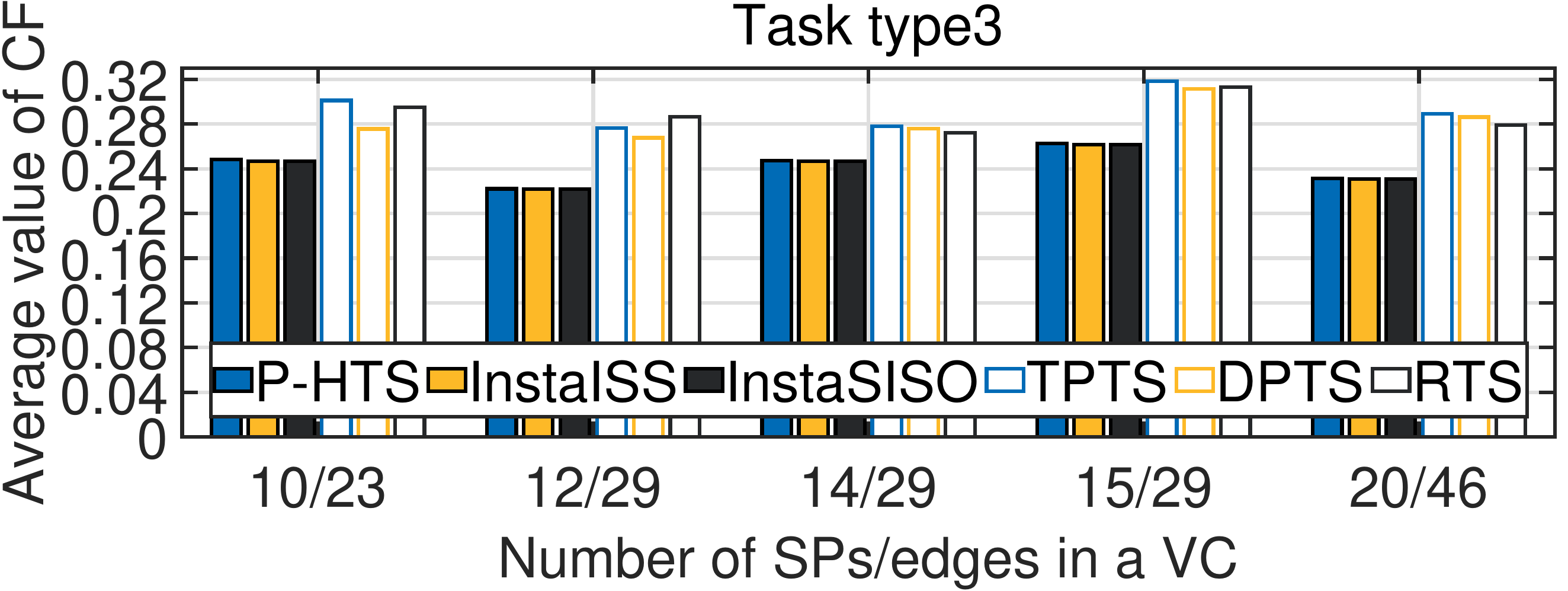}  
	}     		
	\caption{Performance comparison in terms of the average value of CF under various problem sizes: (a) Task type1; (b) Task type2; (c) Task type3.}      
	\label{fig4}    
\end{figure*}
\begin{figure*}[t!] \centering    
	\vspace{-0.3 cm} 
	\subfigtopskip=2pt
	\subfigbottomskip=1pt
	\subfigcapskip=0 cm
	\setlength{\abovecaptionskip}{0 cm} 
	\subfigure[] {
		\label{fig5a }     
		\includegraphics[width=0.66 \columnwidth]{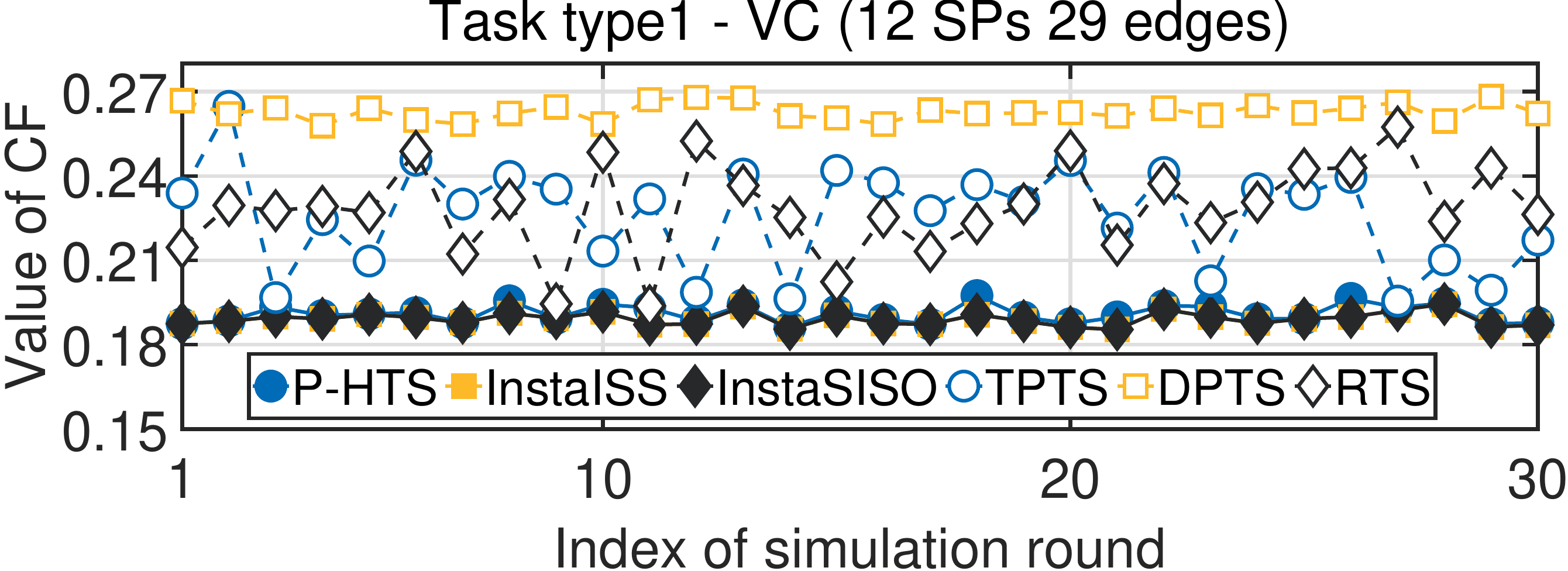}   
	}    \hspace{-3 mm}  
	\subfigure[] { 
		\label{fig5b}     
		\includegraphics[width=0.66 \columnwidth]{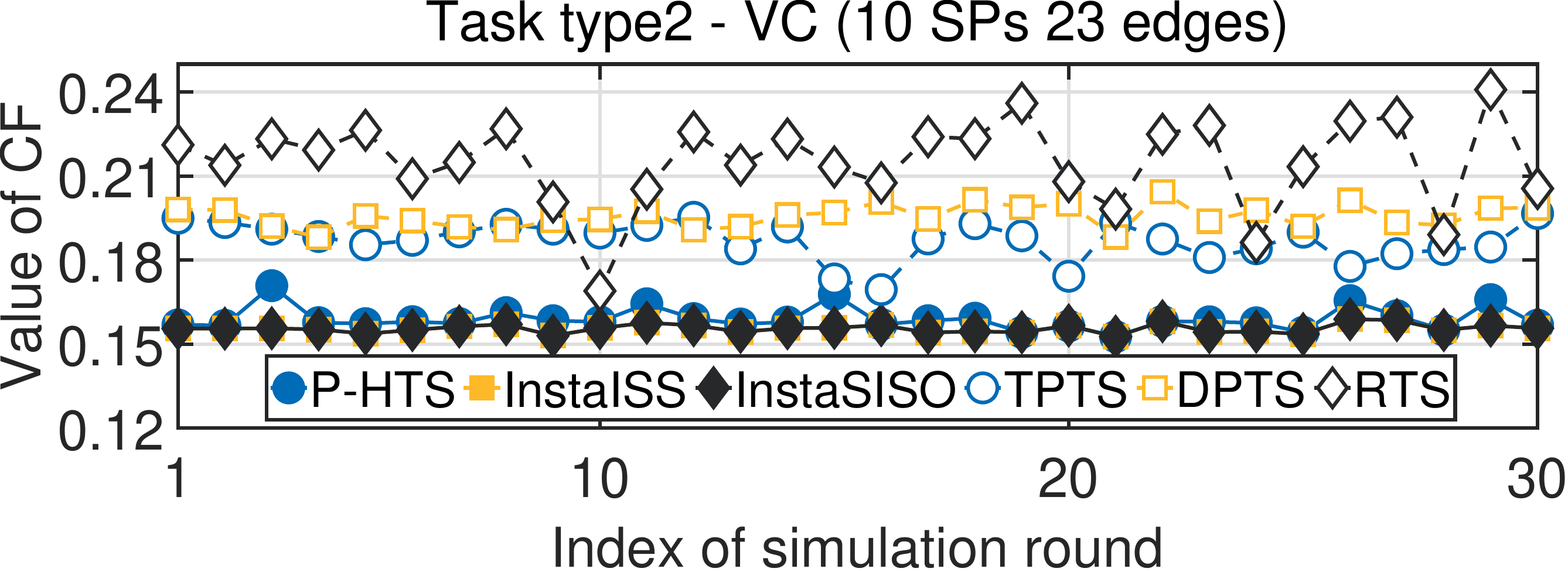}     
	}     \hspace{-3 mm} 
	\subfigure[] {
		\label{fig5c}     
		\includegraphics[width=0.66 \columnwidth]{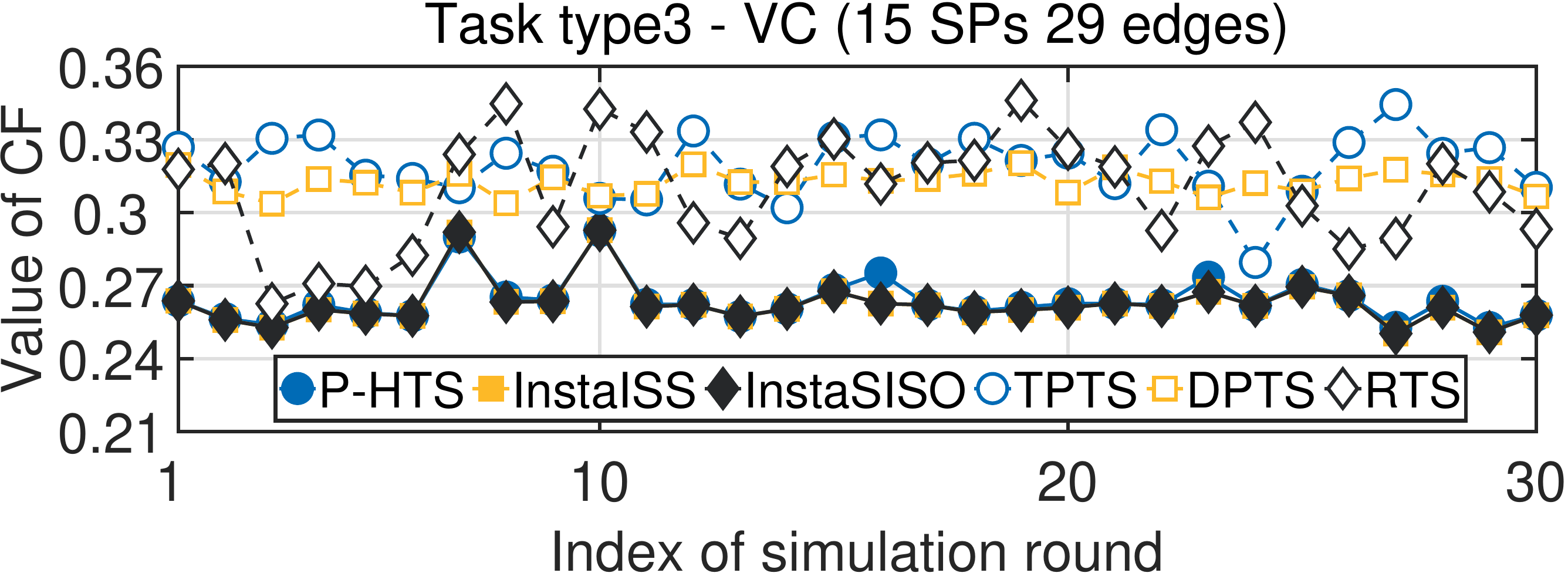}  
	}    \hspace{-2 mm} 		
	\caption{Performance Comparison of 30 simulations on the practical value of CF, considering various task types: (a) Task type1; (b) Task type2; (c) Task type3.}      
	\label{fig5}   
	\vspace{-0.2 cm}   
\end{figure*}

\begin{figure*}[htb] \centering    
	\vspace{-0.1 cm} 
	\subfigtopskip=2pt
	\subfigbottomskip=1pt
	\subfigcapskip=0 cm
	\setlength{\abovecaptionskip}{0 cm} 
	\subfigure[] {
		\label{fig6a }     
		\includegraphics[width=0.66 \columnwidth]{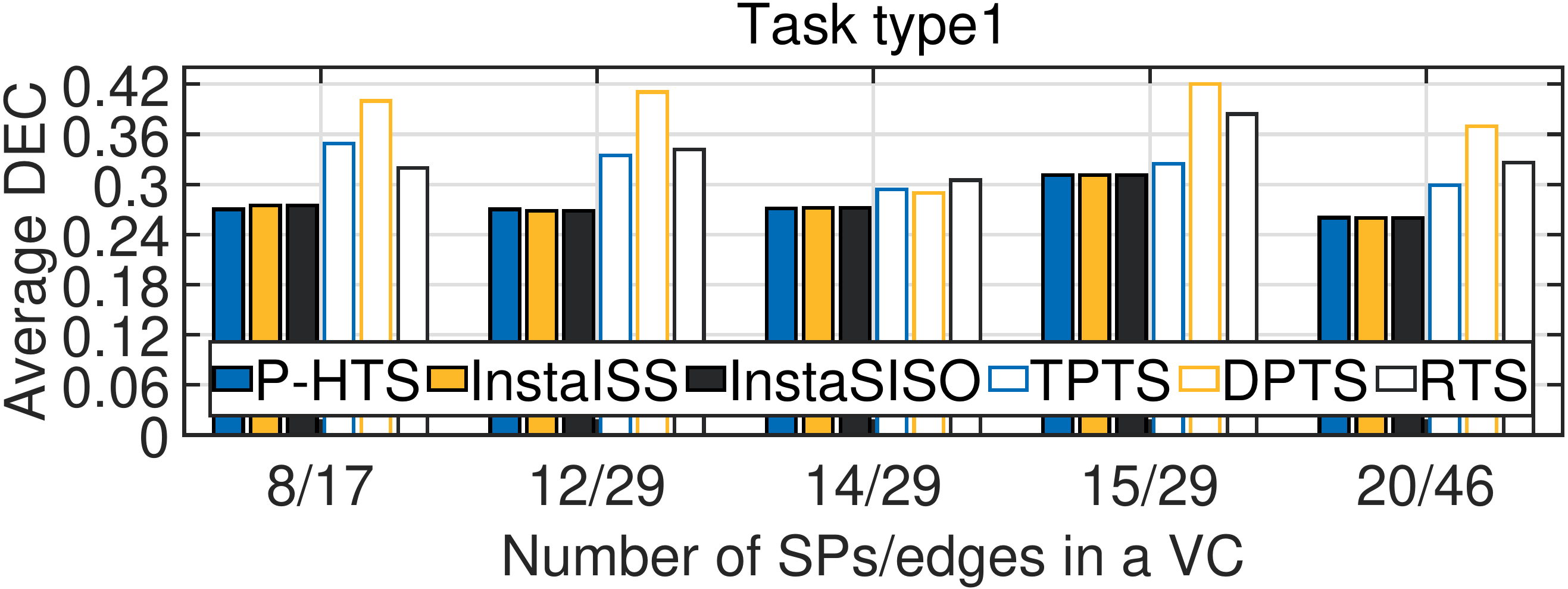}   
	}    \hspace{-3 mm}  
	\subfigure[] { 
		\label{fig6b}     
		\includegraphics[width=0.66\columnwidth]{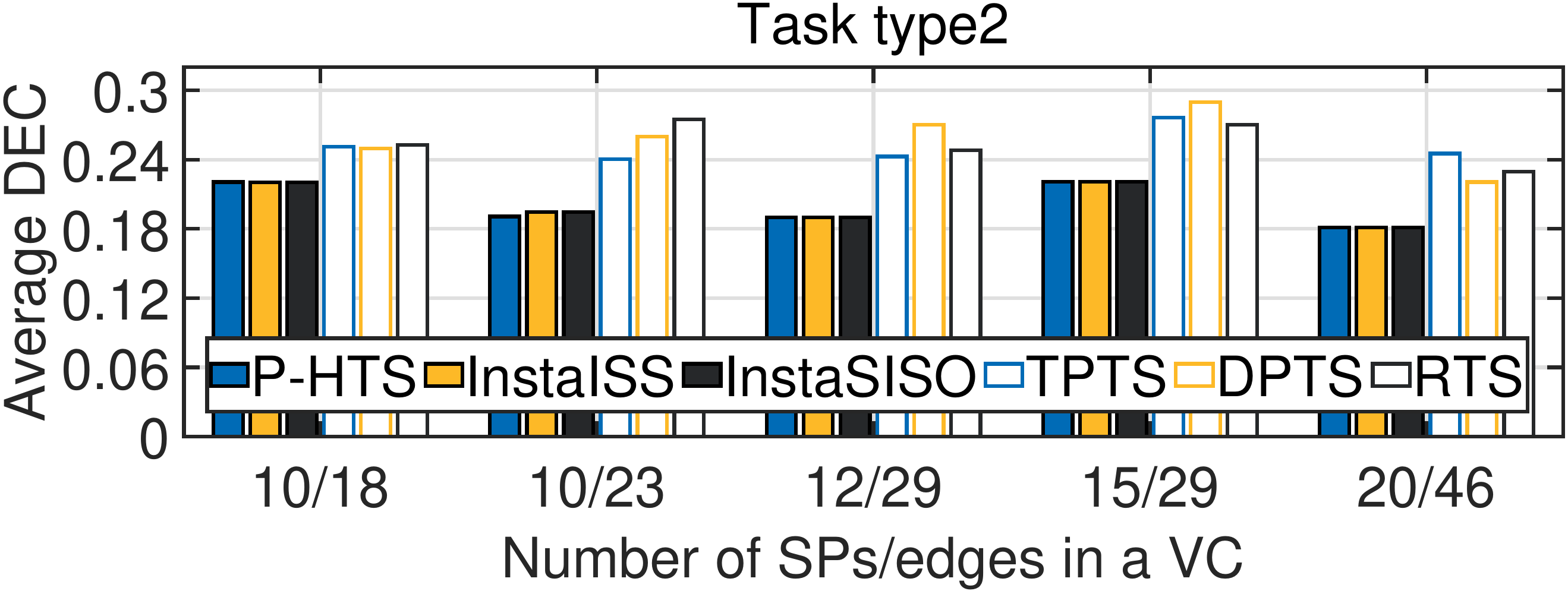}     
	}     \hspace{-3 mm} 
	\subfigure[] {
		\label{fig6c}     
		\includegraphics[width=0.66\columnwidth]{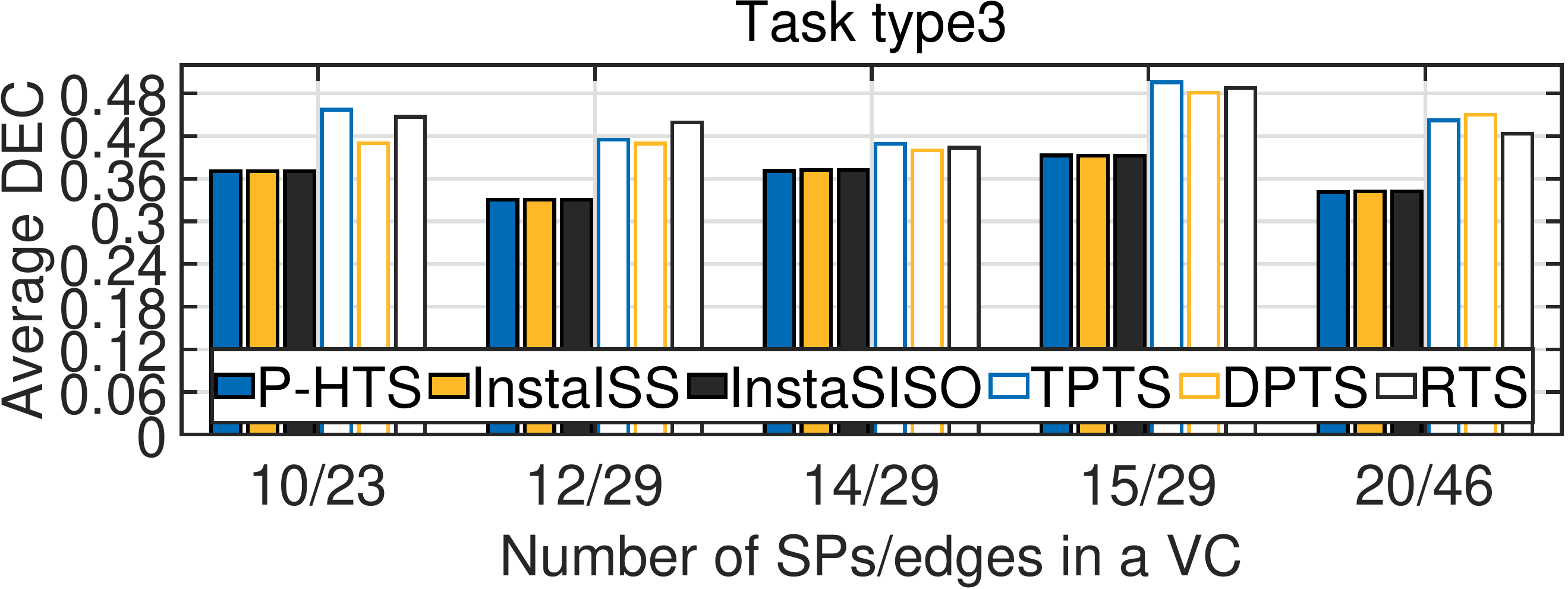}  
	}    
	\caption{Performance comparison in terms of the average DEC under various problem sizes: (a) Task type1; (b) Task type2; (c) Task type3.}      
	\label{fig6}     
\end{figure*}
\vspace{-0.2 cm}
\subsection{Simulation Setup}
\begin{table}[t!]
	\smaller
	\caption{Choice of uncertain parameters \cite{ghimire2021dynamic}\cite{Huang2021}}
	\centering
	\setlength{\tabcolsep}{2.5pt} 
	\vspace{-1em}
	\begin{tabular}{|c|c|c|c|c|}
		\hline
		Parameters & Distribution & Mean & Variance & Practical value \\
		\hline
		$t_{m,m^\prime}^{conn}$(seconds) & Exp. & $5\sim15$  & - & $0\sim60$ \\
		\hline
		$c_{m,m^\prime}^{exch}$ & Trunc. Gauss. & $0.03\sim0.07$ & 0.001 & $0.025\sim0.075$ \\
		\hline
		$r_m$(Mbps) & Trunc. Gauss. & $5\sim7$ & 0.2 & $4\sim8$ \\
		\hline
		$f_m$(GHz)  & Trunc. Gauss. & $2\sim4$ & 0.04-0.07 & $1.5\sim4.5$\\
		\hline
	\end{tabular}
	\vspace{-0.5 cm}
	\label{table1}
\end{table}

We assess performance evaluations upon having different network configurations, graph task types (refer to Fig. \ref{fig3}), and VC structures. To accurately reflect the uncertainties of networks, we account for factors such as contact duration ($t_{m,m^\prime}^{conn}$), data exchange cost ($c_{m,m^\prime}^{exch}$), data transmission rate ($r_m$), and computing capacity ($f_m$). These factors follow certain distributions as outlined in Table \ref{table1}, partially relying on 
a real dataset from ACT Government Open Data Portal dataACT and references \cite{ghimire2021dynamic}\cite{Huang2021}\footnote{The 
scenarios for task scheduling of our interest mainly focus on relatively detailed, micro-level view, e.g., each task component will be assigned to which vehicle, making it difficult to find real-world datasets that fully meet the requirements of our study. Without loss of generality, we set the parameters within our experiments to combine real-world data with simulation data. Similar considerations are also common in existing literature.}. Note that in real-world networks, practical values of these factors are often limited by vehicular hardware settings and communication standards, and therefore are constrained within specific ranges. Other key parameters are as follows\cite{zhan2020deep}\cite{Chen2022}: $d_n \in [200,400]$ Kb; $q_n \in [0.1,0.2]$ GHz; $t_n^{max} \in [0.5,2]$ seconds; $\xi , \xi^\prime \in (0,0.1]$; $\lambda_c = \lambda_t =0.5$.

\vspace{-0.2 cm}
\subsection{Performance Evaluation: Proposed Hybrid Scheduling vs. Online Scheduling Methods}
Comparative analysis of performance is presented in this section, taking into account various metrics such as cost function (CF), running time (RT), data exchange cost (DEC), and task completion time (TCT). The evaluation is compared against the baseline methods considered in \cite{liwang2023graph,luo2020hfel}:\\
\textbf{$\bullet$ Exhaustive template search (ETS):} The ETS method is one of the common approaches in solving 1-0 integer programming problems. By thoroughly evaluating the entire solution space, this method enables the selection of the optimal template with the lowest value of cost function \eqref{eq5}, while satisfying constraints \eqref{C4}-\eqref{C5}.\\
\textbf{$\bullet$ Time-preferred template search (TPTS):} TPTS algorithm uses a greedy-based approach, where task components are sorted through depth-first search. TPTS enables components to be seamlessly mapped onto available SPs with the lowest execution time, while adhering to all constraints \eqref{C4}-\eqref{C5}.\\
\textbf{$\bullet$ Degree-preferred template search (DPTS):} DPTS algorithm adopts a greedy-based approach, leveraging depth-first search to sort task components in a methodical manner. With this algorithm, components are mapped to available SPs with the highest degree (i.e., the largest number of V2V connections), while adhering to constraints \eqref{C4}-\eqref{C5}.\\
\textbf{$\bullet$ Random template search (RTS):} In RTS, task components are randomly mapped to available SPs, while satisfying constraints \eqref{C4}-\eqref{C5}.\\
\textbf{$\bullet$ Instantaneous isomorphic subgraph search (InstaISS):} InstaISS implements only a part of P-HTS, namely, the backup approach of our methodology.\\
\textbf{$\bullet$ Instantaneous SubISO (InstaSISO)\cite{abulaish2019subiso}:} InstaSISO deploys SubISO algorithm proposed in \cite{abulaish2019subiso} for isomorphic subgraph search.
\begin{figure*}[t!] \centering    
	\vspace{-0.1 cm} 
	\subfigtopskip=2pt
	\subfigbottomskip=1pt
	\subfigcapskip=0 cm
	\setlength{\abovecaptionskip}{0 cm} 
	\subfigure[] {
		\label{fig7a}     
		\includegraphics[width=0.66 \columnwidth]{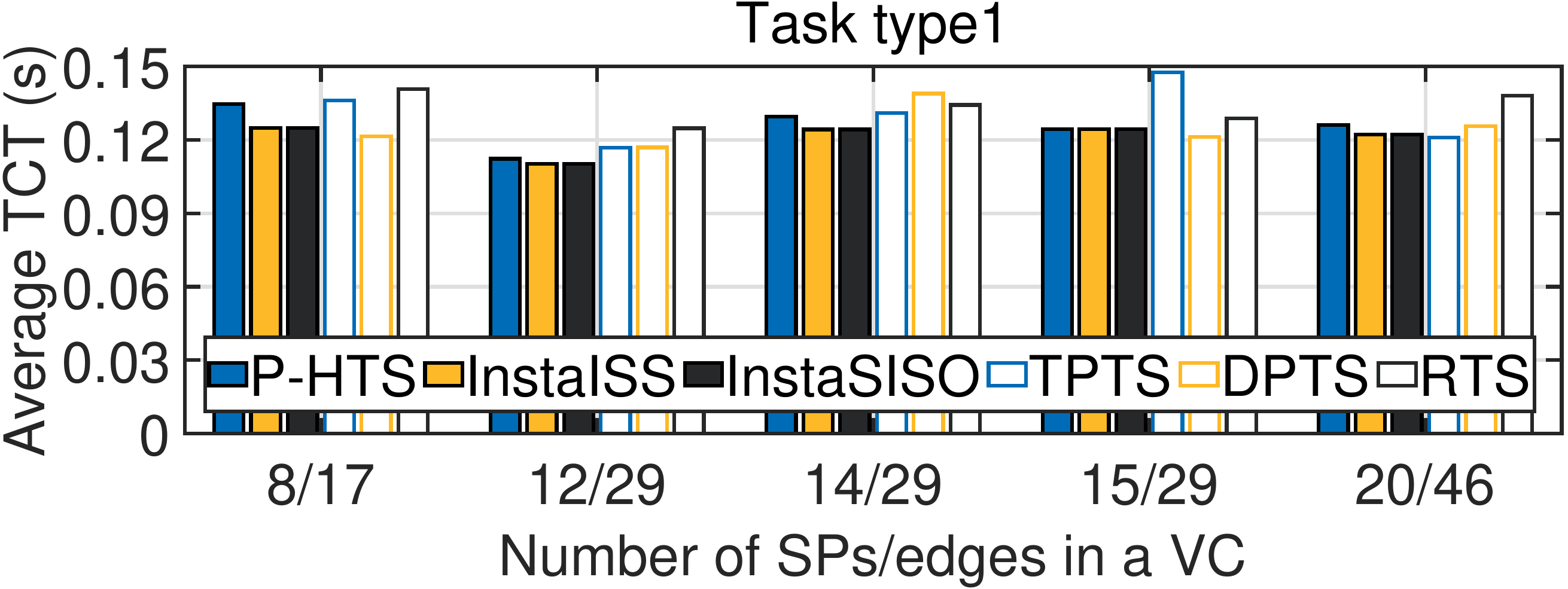}   
	}    \hspace{-3 mm}  
	\subfigure[] { 
		\label{fig7b}     
		\includegraphics[width=0.66 \columnwidth]{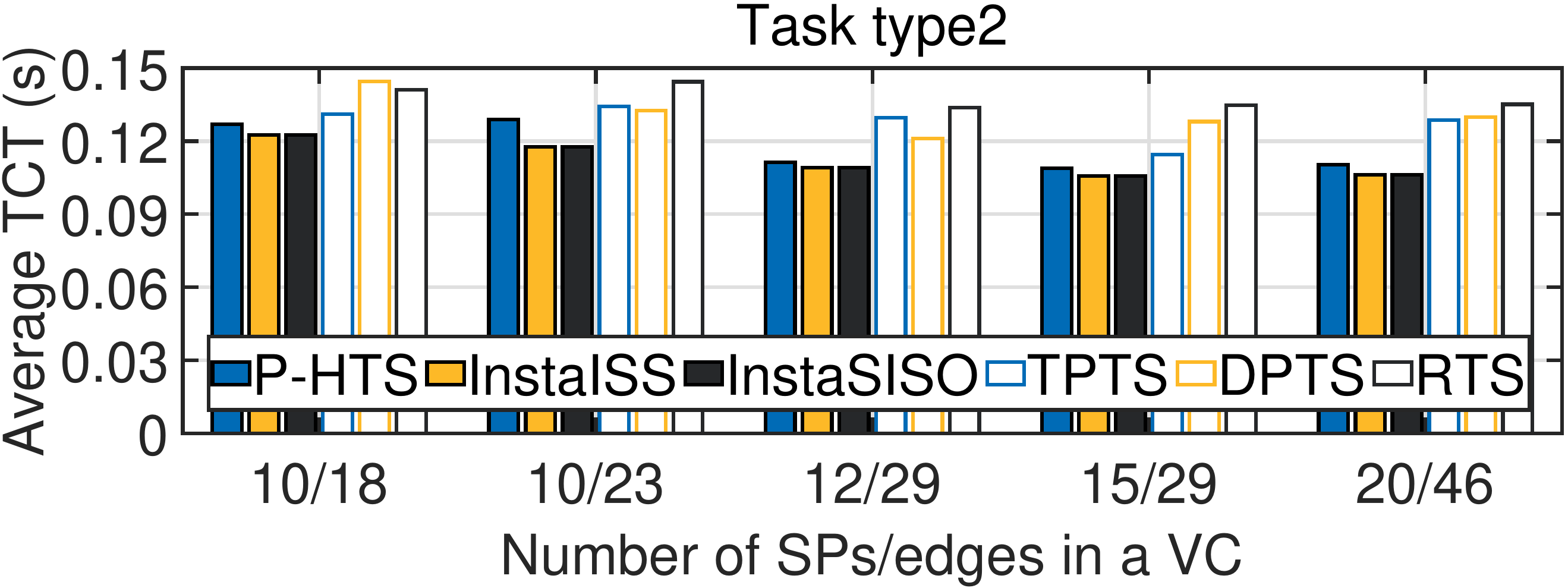}     
	}     \hspace{-3 mm} 
	\subfigure[] {
		\label{fig7c}     
		\includegraphics[width=0.66 \columnwidth]{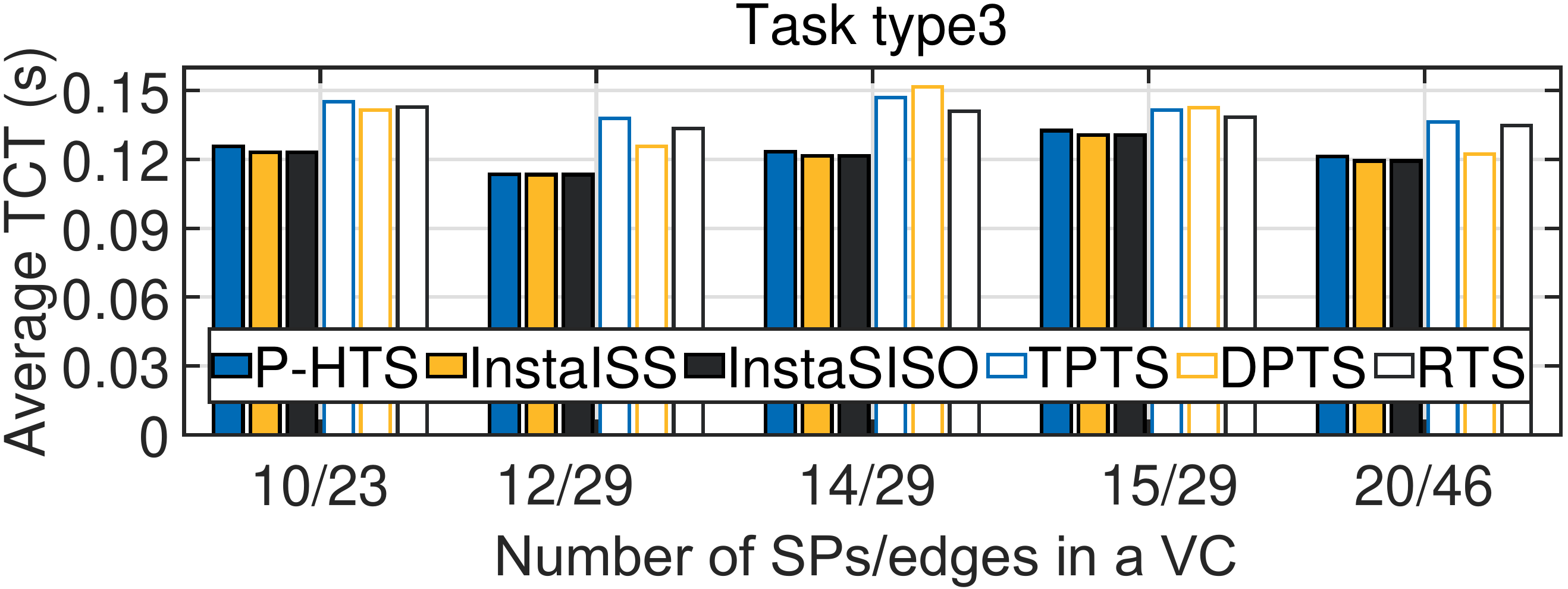}  
	}    
	\caption{Performance comparison in terms of the average TCT under various problem sizes: (a) Task type1; (b) Task type2; (c) Task type3.}      
	\label{fig7}     
	\vspace{-0.4 cm}
\end{figure*}

To conduct better performance comparisons, we have adopted the Monte-Carlo method and created 100 independent simulations for each unique problem setting. For instance, every value in Fig. \ref{fig4} was derived from averaging over 100 experimental simulations. Considering the significant gap in RT between ETS and P-HTS (as depicted in Fig. \ref{fig8}), ETS will not be included in our subsequent discussions, with the exception of Fig. \ref{fig8}. {Moreover, we encompass different problem scales such as various numbers of vehicles and different topologies of tasks, aiming to demonstrate the scalability of our methodology. }

\textbf{\textit{1) Performance comparison on CF:}} Fig. \ref{fig4} showcases the average CF value when considering various task types under different VC settings. Our P-HTS outperforms TPTS, DPTS, and RTS in terms of the average CF value because of the optimality of the template $\mathbf{A^{off}}$. The TPTS, DPTS, and RTS methods struggle to obtain all feasible templates, resulting in their higher costs. 

{Apparently, both InstaISS and InstaSISO can achieve a lower practical value of CF than P-HTS. However, these two online methods can incur significant overhead (e.g., decision-making time, also see Fig. \ref{fig9}), despite their ability to attain the optimal solution on the CF. Notably, such methods introduce risks, such as the possibility of a vehicle exiting the studied region due to prolonged decision-making. In this context, an optimal solution does not necessarily equate to an applicable one, particularly when accounting for the dynamic and uncertain nature of the network environment. To this end, our key principle is to introduce a novel and unique stagewise decision-making mechanism where our $\mathbf{A^{off}}$ can alleviate the heavy burden on searching for a proPersolution during online stage. Therefore, as shown in Fig. \ref{fig9}, the running time (RT) of our proposed P-HTS methodology is significantly lower than that of InstaISS and InstaSISO, even by magnitude. Additionally, in Fig. \ref{fig4}, we observe that although the CF value obtained by P-HTS may not be the best, the averaged value of CF over multiple tasks and simulation rounds can be very close to the optimal values achieved by InstaISS and InstaSISO. This demonstrates the superiority of our P-HTS methodology since it offers a good trade-off between the performance on CF and the overhead on decision-making, representing an important reference for future development of dynamic networks with moving nodes such as vehicles. }

To provide a more detailed understanding of practical task scheduling events, we randomly selected 30 simulations (out of 100) with varying VC settings, and Fig. \ref{fig5} portrays the performance on the practical value of CF. In most simulations, P-HTS achieves similar values as InstaISS and InstaSISO, although sometimes, the performance may be sacrificed for time efficiency, which is deemed acceptable in real-world networks. 
 
\textbf{\textit{2) Performance comparison on DEC and TCT:}} {As we consider two significant factors/metrics in the definition of CF in \eqref{eq5}, we have compared performance on both the average DEC and TCT in Fig. \ref{fig6} and Fig. \ref{fig7}, respectively. DEC represents the energy consumption resulting from data exchange in parallel computing, while TCT quantifies the impact of network latency fluctuations and computational performance variations. As shown in Fig. \ref{fig6}, our P-HTS demonstrates a significantly lower DEC as compared to baseline methods such as TPDS, DPTS, and RTS, due to the ability to search for and check across all feasible templates during both offline and online modes. Additionally, P-HTS achieves comparable or superior DEC performance relative to InstaISS and InstaSISO, thanks to the effectiveness of the combination of RA-PilotISS and InstaISS approaches.
In Fig. \ref{fig7}, P-HTS generally outperforms or closely matches the baseline methods in terms of average TCT across most task types and VC settings. While there are instances where P-HTS exhibits a slightly higher TCT, this can be attributed to the inherent risks associated with dynamic networks. Nevertheless, P-HTS ensures reliable task completion, demonstrating commendable performance in both the CF and RT metrics, as shown in Figs. \ref{fig4} and \ref{fig9}. For example, considering Task Type 1, 8 SPs and 17 edges in Fig. \ref{fig7a}, although the average value of TCT of our P-HTS is greater than that of InstaISS and InstaSISO, their performance on RT are far from satisfactory as shown by Fig. \ref{fig9}, imposing challenges in dynamic vehicular network environments.}
\begin{figure}[t!]
	\centering
	\includegraphics[width=0.6 \linewidth]{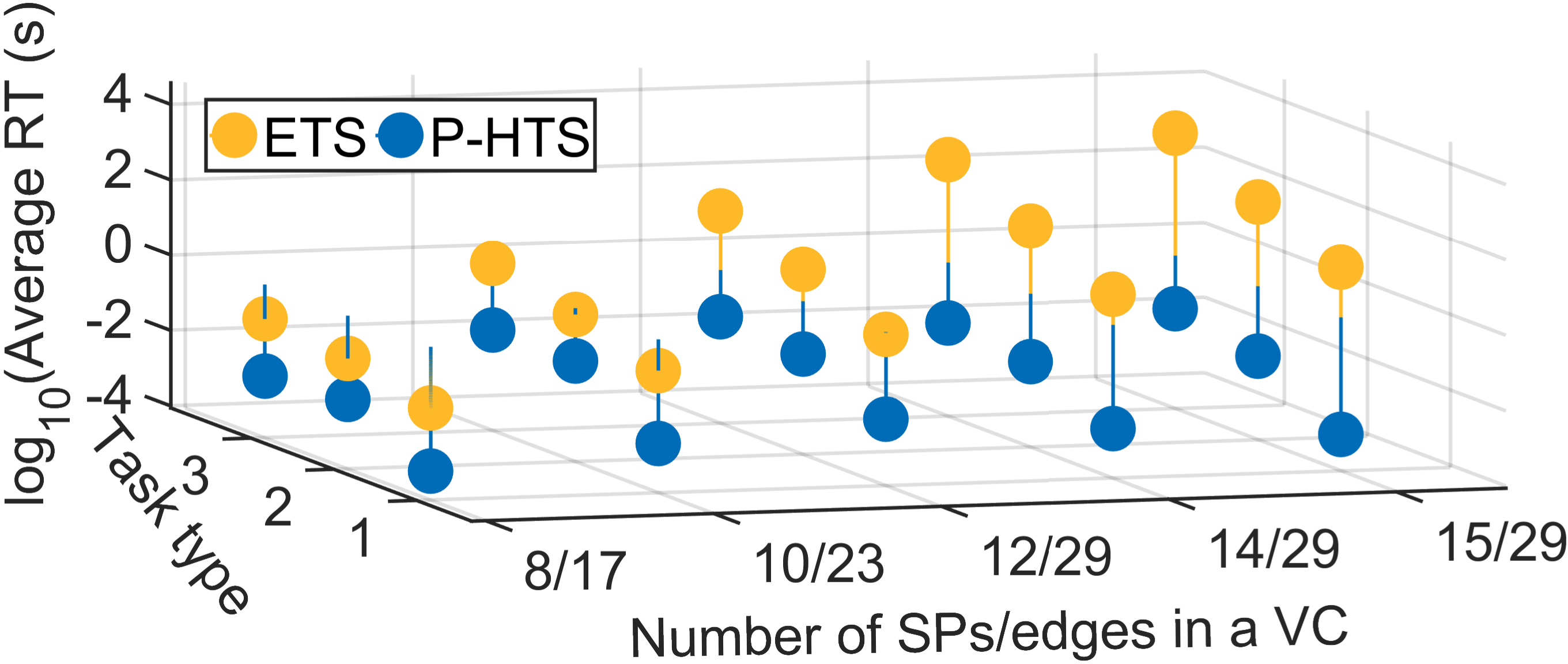}
	\vspace{-0.3 cm}
	\caption{Performance comparison on the average value of RT upon considering various VC settings (e.g., different numbers of SPs and edges among them) and task types.}
	\label{fig8}
	\vspace{-0.7 cm}
\end{figure}

\textbf{\textit{3) Performance comparison on RT:}} 
Time efficiency is crucial for any algorithm deployed over dynamic networks. We thus study the performance in terms of RT in Figs. \ref{fig8}-\ref{fig10}, which demonstrate the time usage on decision-making for task scheduling. A high RT value indicates a lengthy procedure for template searching and optimization, posing challenges in maintaining graph task structures in a dynamic VC where contact duration among SPs is limited. To account for the vast difference in RT values across various methods, especially with complex graph tasks and VC structures, we apply the 10-based logarithm representation in the y-axis of the figures. As shown in Fig. \ref{fig8}, the average RT of ETS increases significantly with the growing number of components, SPs, and edges among SPs, due to the exponential growth of the solution space. For example, for task type 3 and a VC with 14 SPs and 29 edges, the average RT of ETS reaches an unacceptable 178 seconds. Consequently, we do not consider ETS in other performance comparisons.

Fig. \ref{fig9} shows a performance comparison of the average RT, considering different task types and VC settings. Our P-HTS outperforms InstaISS and InstaSISO, making it more suitable for real-world VCs with fast-changing vehicular topologies. The gap between them widens as the graph task complexity and the number of SPs/edges increase, allowing for more templates. In this case, our considered template $\mathbf{A^{off}}$ obtained from RA-PilotISS stands out as it optimizes feasible templates before practical task scheduling, saving more time during online decision-making. Note that TPTS, DPTS, and RTS fail to get all the feasible templates but can provide a quick solution. Nonetheless, our P-HTS performs well with an acceptable RT and significantly better performance on the average value of CF (as shown in Fig. \ref{fig4}) compared to these greedy-based and random-based methods. To analyze the details of RT in each simulation, we sampled 30 cases among the 100 ones as shown in Fig. \ref{fig10}. The curves of our P-HTS remain lower than other methods in most cases, as $\mathbf{A^{off}}$ is often used during practical task scheduling. Although sometimes the proposed TE-InstaISS algorithm as a backup approach can be triggered, such as index 20 in Fig. \ref{fig10b}, the average RT of P-HTS achieves good performance, as discussed earlier.
\begin{table}[htb]	
\scriptsize
	\vspace{-0.3 cm}
	\caption{Performance comparison between InstaISS and P-HTS on task type 1}
	\centering
	\vspace{-0.25 cm} 
	
\setlength{\tabcolsep}{0.1mm}{
	\begin{tabular}{|c|c|c|c|c|c|c|c|}
		\hline
		\multirow{2}{*}{Line} & \multirow{2}{*}{Index} & \multicolumn{3}{c|}{P-HTS} & \multicolumn{3}{c|}{InstaISS} \\
		\cline{3-8}
		&& Template & RT (s) & CF & Template & RT (s) & CF \\
		\hline
		1 & 5 & [$s_3, s_7, s_{10}, s_8, s_{11}$] & 0.00294 & 0.1909 & [$s_4, s_7, s_{10}, s_8, s_{11}$] & 0.08112 & 0.1908 \\
		\hline
		2 & 6 & [$s_{11}, s_7, s_3, s_8, s_4$] & 0.08955 & 0.1897 & [$s_{11}, s_7, s_3, s_8, s_4$] & 0.06484 & 0.1897 \\
		\hline
		3 & 7 & [$s_3, s_7, s_{10}, s_8, s_{11}$] & 0.00016 & 0.1878 & [$s_3, s_7, s_{10}, s_8, s_{11}$] & 0.06118 & 0.1878 \\
		\hline
		4 & 15 & [$s_3, s_7, s_{10}, s_8, s_{11}$] & 0.00004 & 0.1920 & [$s_4, s_7, s_{10}, s_8, s_{11}$] & 0.06775 & 0.1904 \\
		\hline
		5 & 16 & [$s_3, s_7, s_{10}, s_8, s_{11}$] & 0.00005 & 0.1890 & [$s_4, s_7, s_2, s_8, s_3$] & 0.06795 & 0.1875 \\
		\hline
		6 & 17 & [$s_3, s_7, s_{10}, s_8, s_{11}$] & 0.00007 & 0.1873 & [$s_3, s_7, s_{10}, s_8, s_{11}$] & 0.06806 & 0.1873 \\
		\hline
		7 & 25 & [$s_3, s_7, s_{10}, s_8, s_{11}$] & 0.00004 & 0.1892 & [$s_3, s_7, s_{10}, s_8, s_{11}$] & 0.06768 & 0.1892 \\
		\hline
		8 & 26 & [$s_3, s_7, s_{10}, s_8, s_{11}$] & 0.00004 & 0.1965 & [$s_{10}, s_7, s_3, s_{11}, s_8$] & 0.06756 & 0.1898 \\
		\hline
		9 & 27 & [$s_3, s_7, s_{10}, s_8, s_{11}$] & 0.00004 & 0.1932 & [$s_2, s_7, s_4, s_3, s_8$] & 0.06642 & 0.1922 \\
		\hline
	\end{tabular}}
	\label{table2}
	\vspace{-0.2 cm}
\end{table}

\begin{figure*}[htb] \centering    
	\vspace{-0.1 cm} 
	\subfigtopskip=0pt
	\subfigbottomskip=0pt
	\subfigcapskip=0 cm
	\setlength{\abovecaptionskip}{0 cm} 
	\subfigure[] {
		\label{fig9a }     
		\includegraphics[width=0.66 \columnwidth]{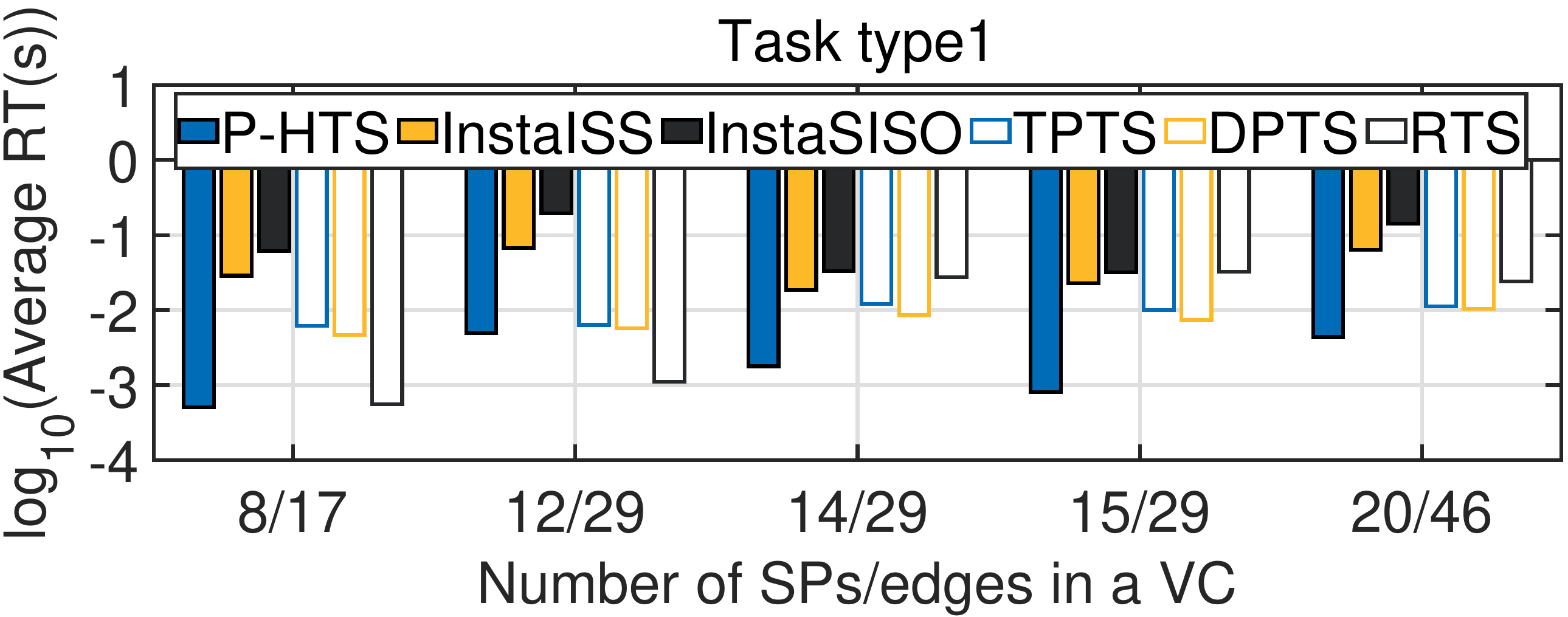}   
	}    \hspace{-3 mm}   
	\subfigure[] { 
		\label{fig9b}     
		\includegraphics[width=0.66\columnwidth]{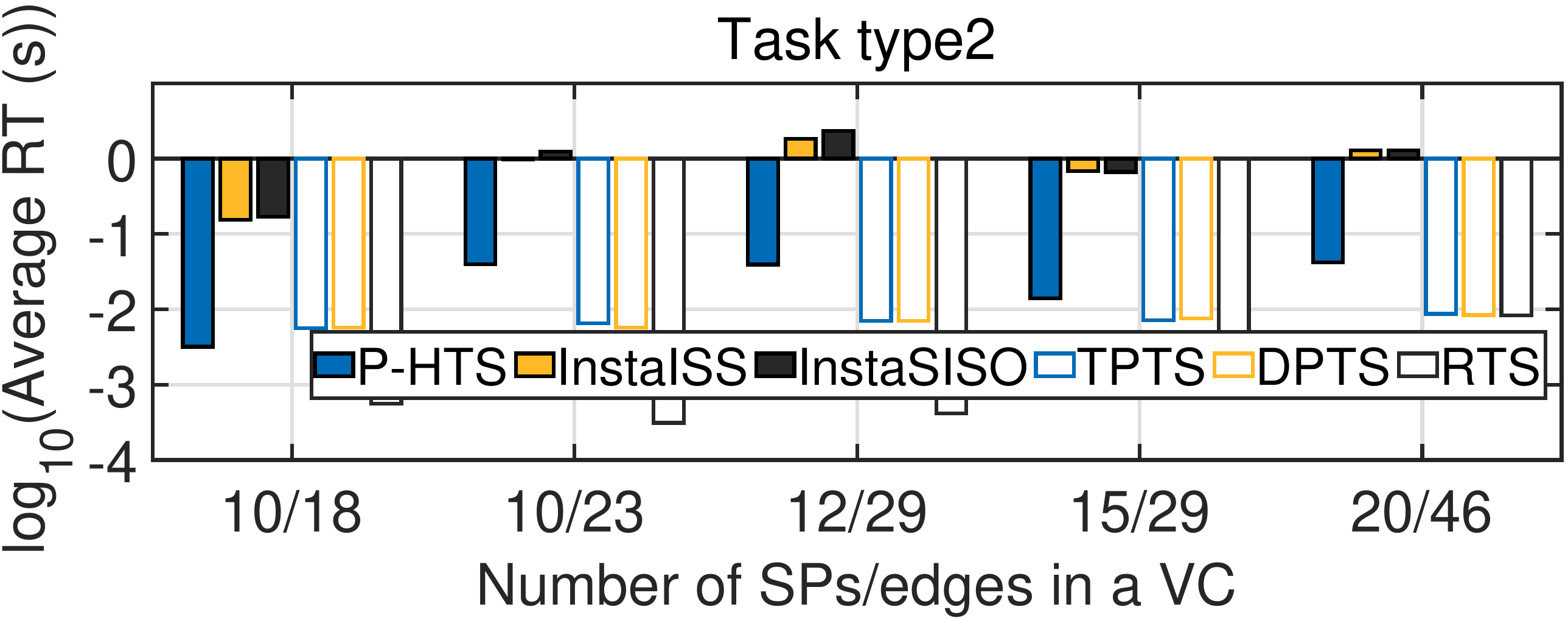}     
	}     \hspace{-3 mm}  
	\subfigure[] {
		\label{fig9c}     
		\includegraphics[width=0.66\columnwidth]{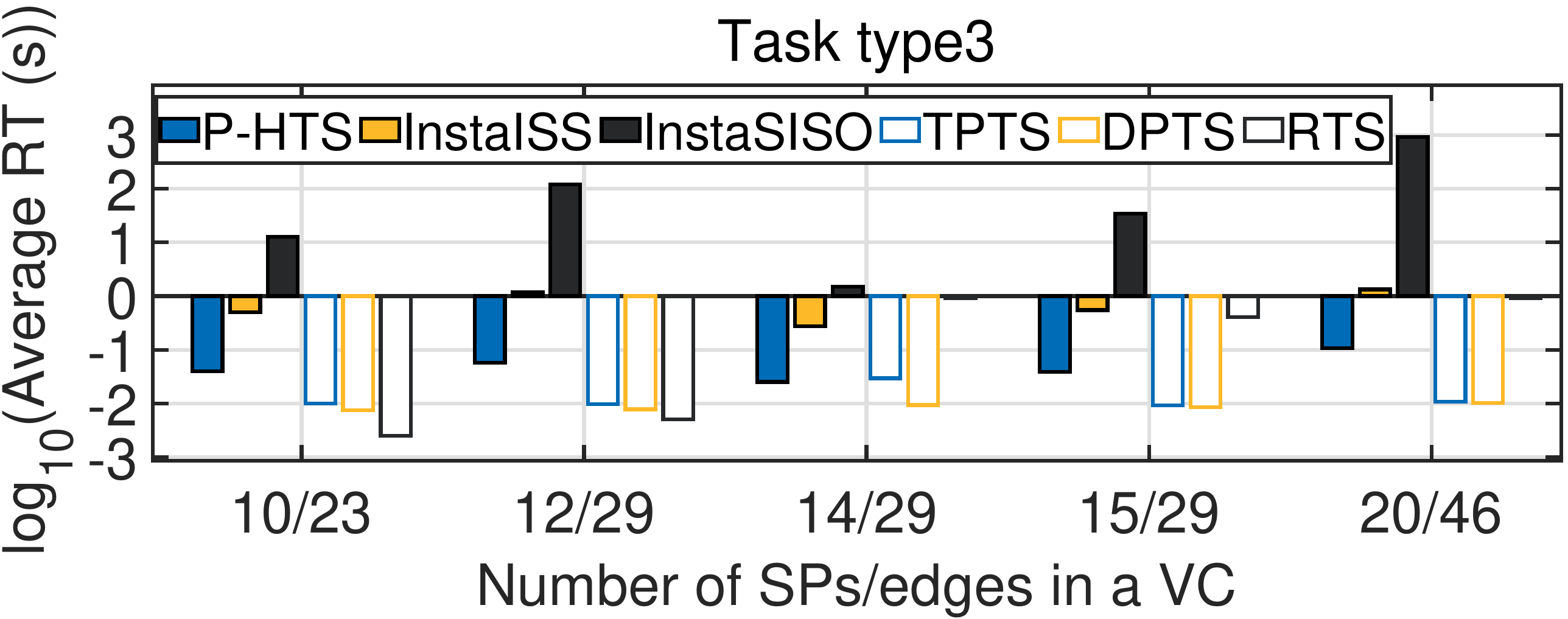}  
	}   		
	\caption{Performance comparison in terms of the average RT under various problem sizes: (a) Task type1; (b) Task type2; (c) Task type3.}      
	\label{fig9}     
\end{figure*}

\begin{figure*}[htb] \centering    
	\vspace{-0.2 cm} 
	\subfigtopskip=1pt
	\subfigbottomskip=1 pt
	\subfigcapskip=0 cm
	\setlength{\abovecaptionskip}{0 cm} 
	\subfigure[] {
		\label{fig10a }     
		\includegraphics[width=0.66 \columnwidth]{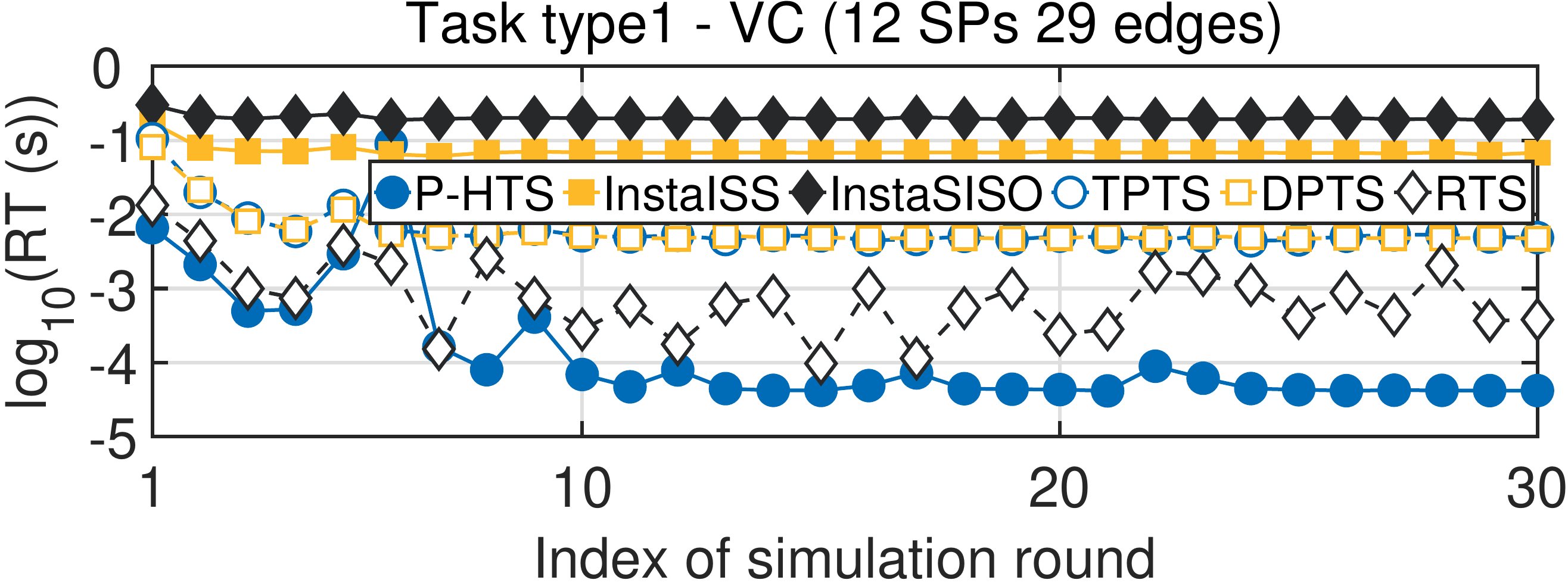}   
	}    \hspace{-3 mm}    
	\subfigure[] { 
		\label{fig10b}     
		\includegraphics[width=0.66 \columnwidth]{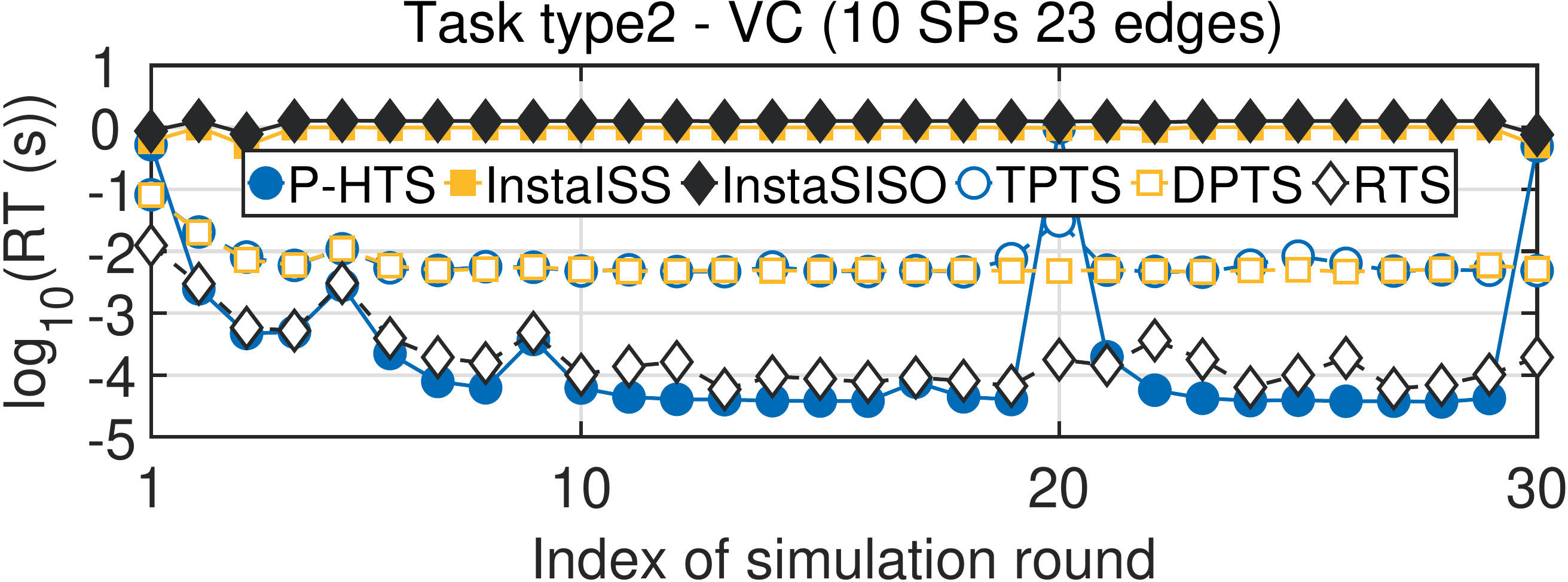}     
	}     \hspace{-3 mm}  
	\subfigure[] {
		\label{fig10c}     
		\includegraphics[width=0.66 \columnwidth]{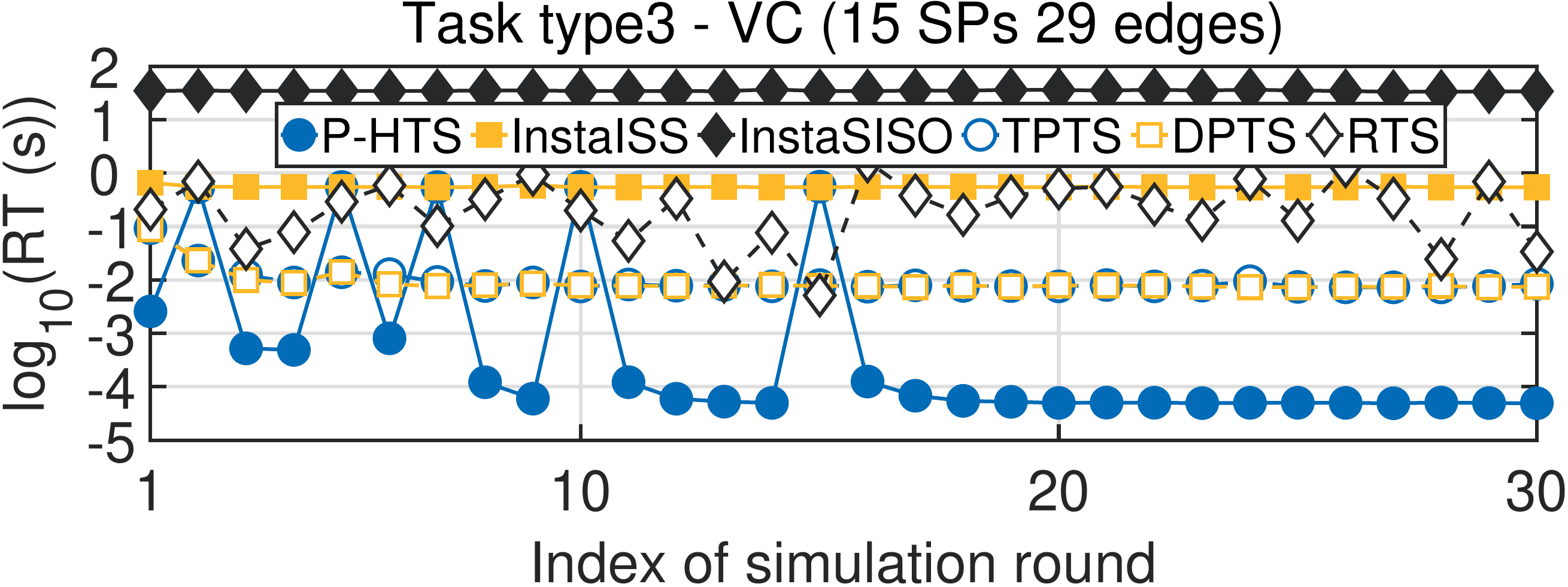}  
	}    		
	\caption{Performance comparison of 30 simulations on the practical value of RT, considering various task types: (a) Task type1; (b) Task type2; (c) Task type3.}      
	\label{fig10}  
	\vspace{-0.45 cm}
\end{figure*}
{
\textbf{\textit{4) Evaluation on larger-scale scenario and highly dynamic vehicular environment:}} To further validate the robustness and scalability of our P-HTS, we conduct another set of experiments regarding the scenario of a rather larger number of SPs and connections between SPs (than the problem sizes shown by the previous figures), and a highly volatile vehicular environment, as illustrated by Fig. \ref{extreme scale} and \ref{highly volite}, respectively. For the former, the VC graph employed in Fig. \ref{extreme scale} consists of 25 SPs and 57 edges. As shown in this figure, our P-HTS maintains its superiority in achieving near-optimal cost function values with lower decision-making time as compared to baseline methods, thus offering a good scalability. 
Then, To simulate highly volatile vehicular environments, we configure the uncertain factors with the following statistical properties:  $t_{m,m^\prime}^{conn}$ is assigned by a mean value of 3 seconds; while the variances are set as 0.002 for $c_{m,m^\prime}^{exch}$, 0.5 Mbps for $r_m$, and 0.2 GHz for $f_m$. These settings are designed to replicate dynamic vehicular network conditions. The results presented in Fig. \ref{highly volite} show that our P-HTS consistently delivers competitive performance, even in such fluctuating environments. Despite the high levels of uncertainty, P-HTS effectively adapts to the changes and ensures optimal or near-optimal task scheduling, further reinforcing the robustness and adaptability of our mechanism in real-world dynamic IoV environments. The above findings collectively highlight the operational resilience, adaptability, and scalability of our P-HTS across diverse settings.}
\begin{figure}[b!] \centering    
	\vspace{-0.6 cm} 
	\subfigtopskip=2pt
	\subfigbottomskip=1pt
	\subfigcapskip=0 cm
	\setlength{\abovecaptionskip}{0 cm} 
	\subfigure[] {
		\label{ExtremeScaleCF}     
		\includegraphics[width=0.48 \columnwidth]{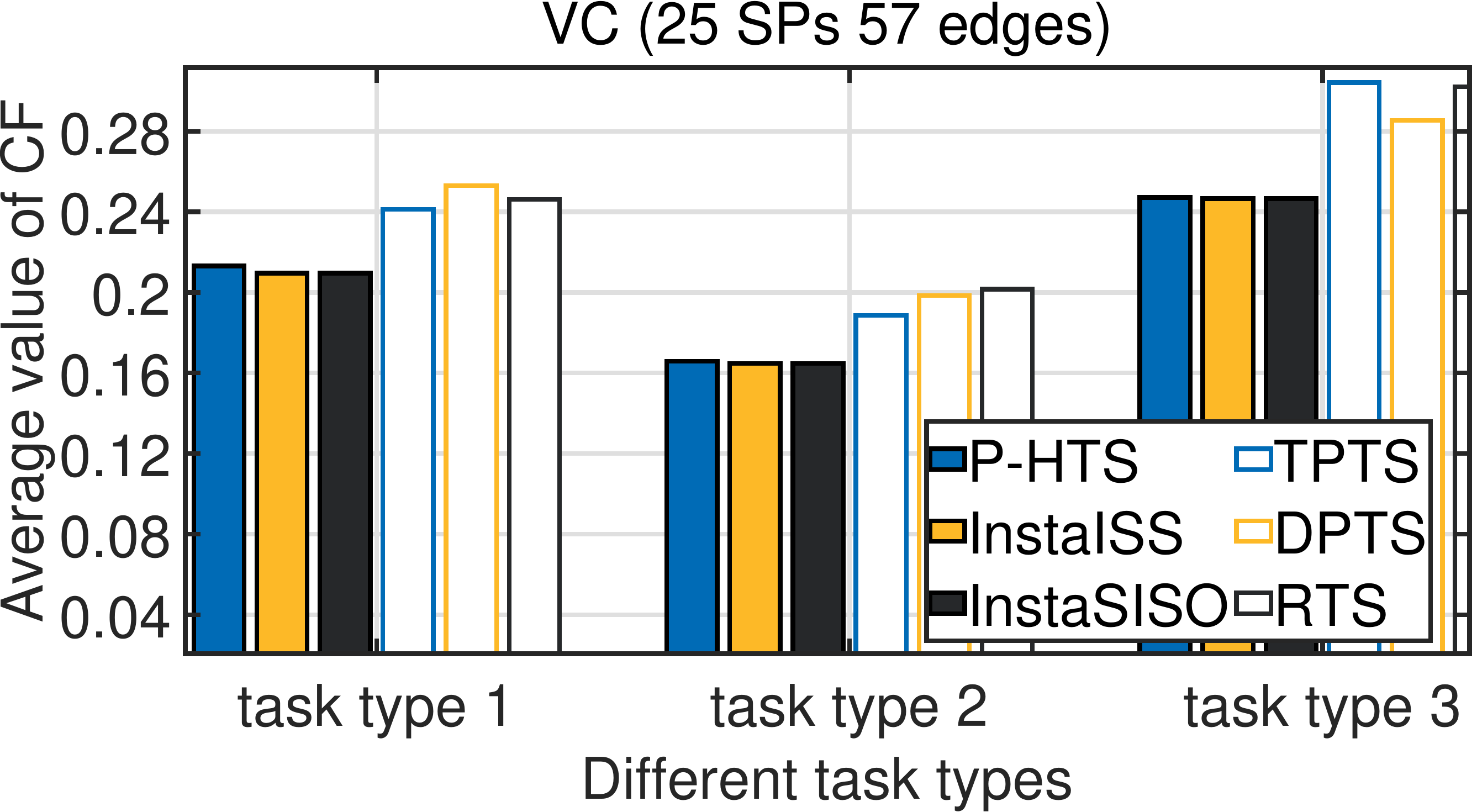}   
	}    \hspace{-3 mm}  
	\subfigure[] { 
		\label{ExtremeScaleRT}     
		\includegraphics[width=0.48 \columnwidth]{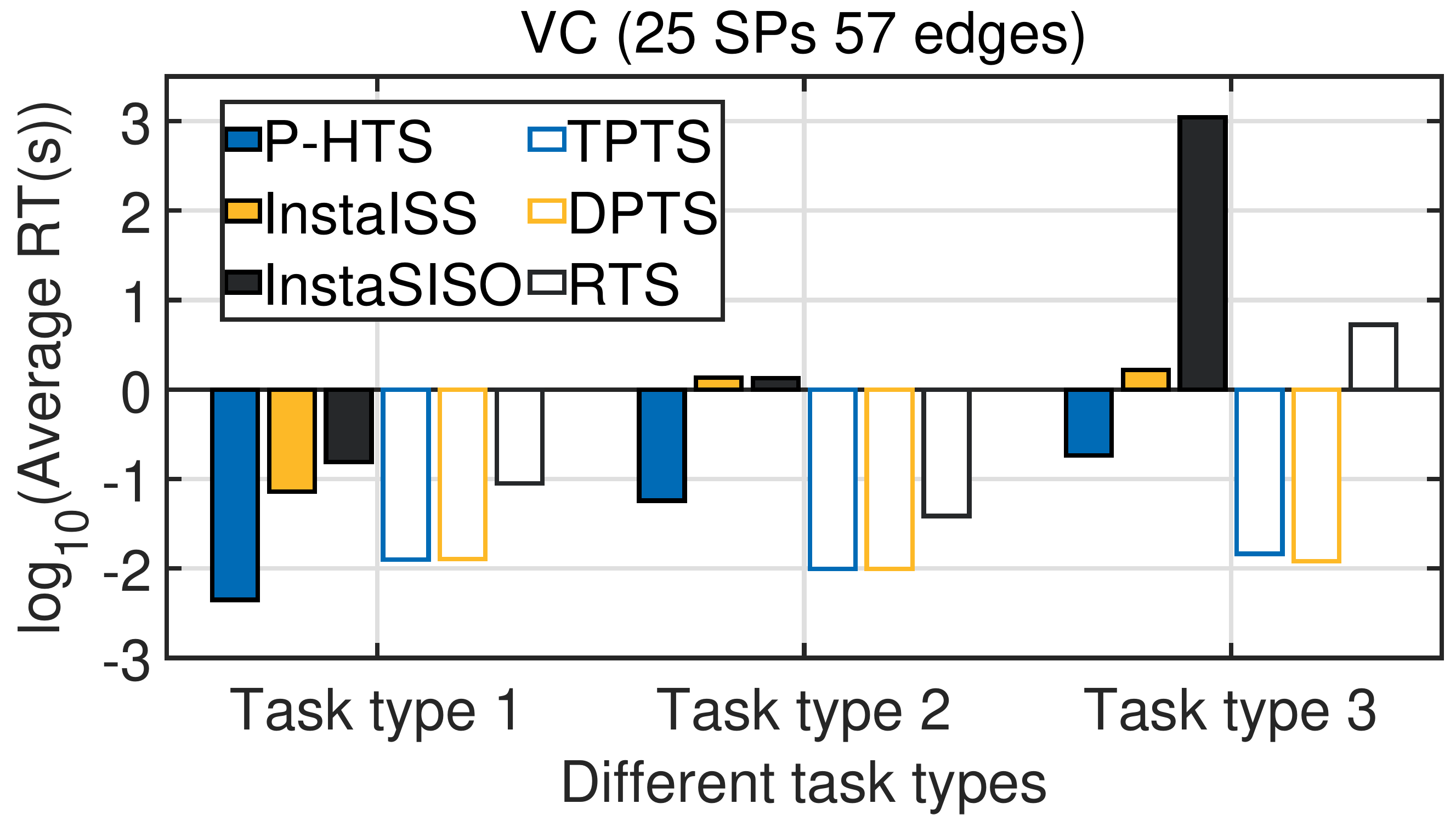}     
	}     \hspace{-3 mm} 
	\caption{{Performance comparison on a large scenario: (a) Value of cost function; (b) Running time.}}      
	\label{extreme scale}     
	\vspace{-0.4 cm}
\end{figure}

\begin{figure}[b!] \centering    
	\vspace{-0.1 cm} 
	\subfigtopskip=2pt
	\subfigbottomskip=1pt
	\subfigcapskip=0 cm
	\setlength{\abovecaptionskip}{0 cm} 
	\subfigure[] {
		\label{HighlyVoliteCF}     
		\includegraphics[width=0.48 \columnwidth]{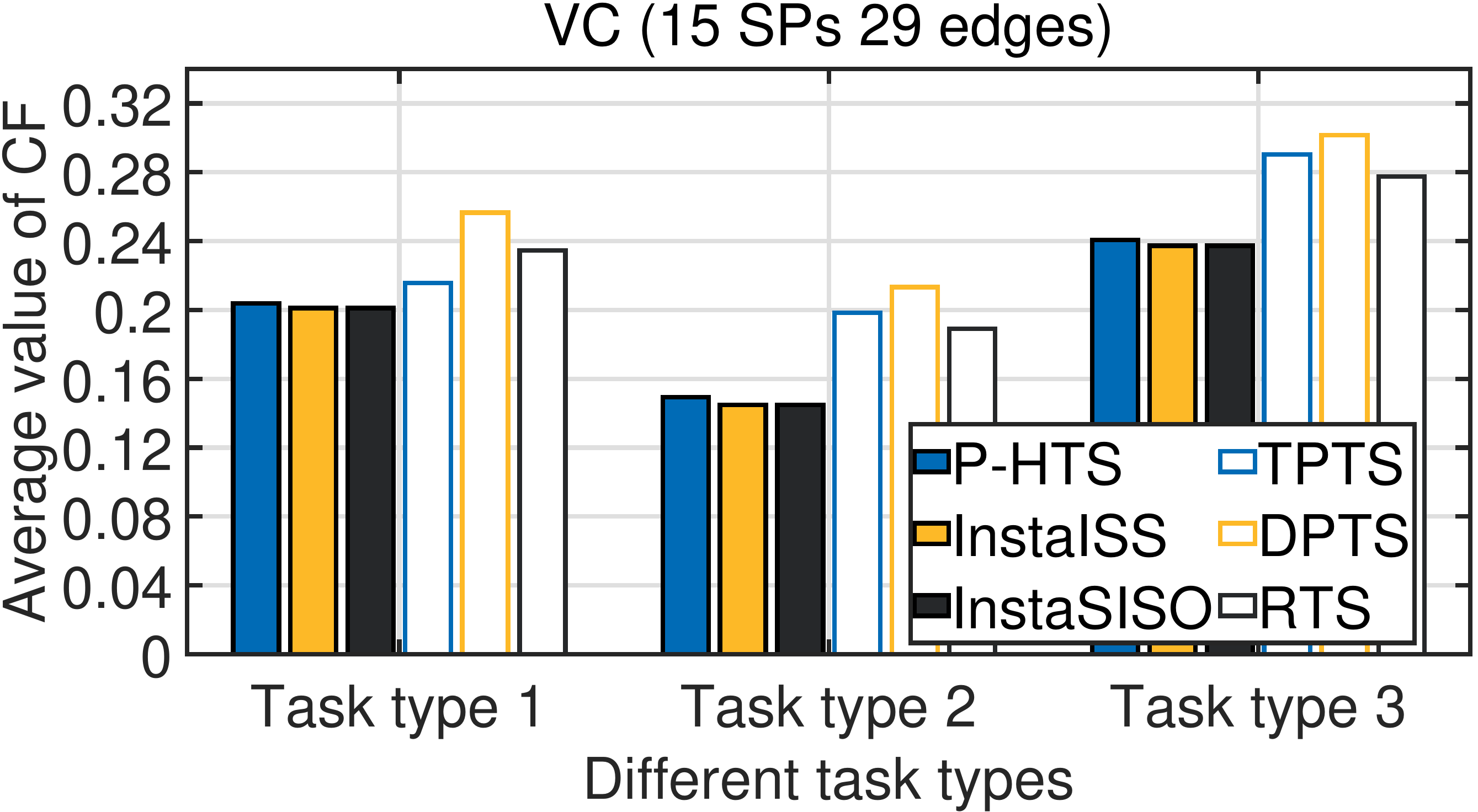}   
	}    \hspace{-3 mm}  
	\subfigure[] { 
		\label{highly volite RF}     
		\includegraphics[width=0.48 \columnwidth]{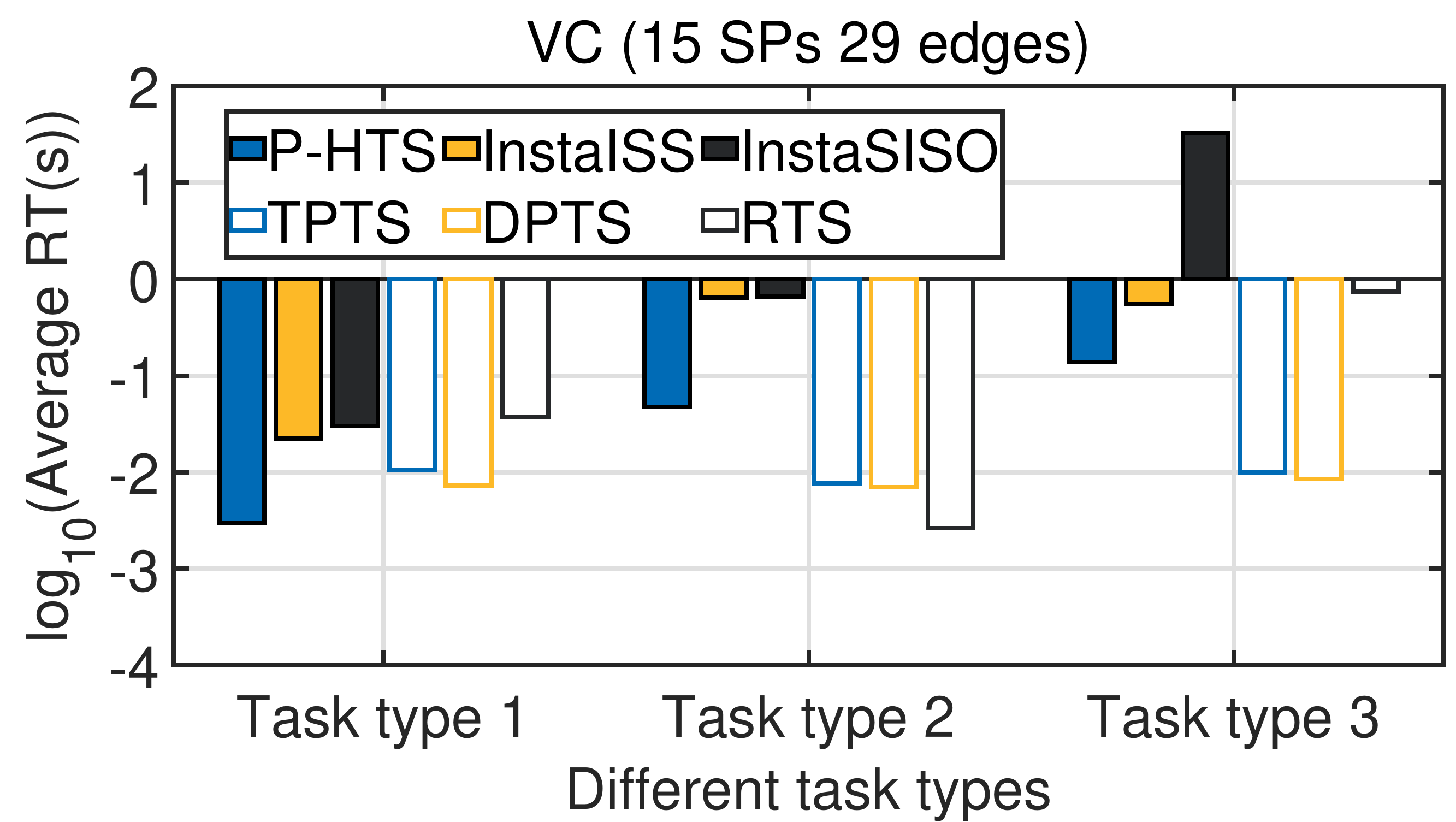}     
	}     \hspace{-3 mm} 
	\caption{{Performance comparison on a highly volatile vehicular environment: (a) Value of cost function; (b) Running time.}}      
	\label{highly volite}     
	\vspace{-0.0 cm}
\end{figure}

To further show the example of what a template looks like, we randomly selected 9 simulations, as displayed in Table \ref{table2}. The task components (type 1) follow the order of $v_1,v_2,v_3,v_4,v_5$ (the same as Fig. \ref{fig3}) and the VC is selected with 12 SPs and 29 edges. Note that template $\mathbf{A^{off}}$ obtained from RA-PilotISS is denoted as $\left\{v_1-s_3, v_2-s_7, v_3-s_{10}, v_4-s_8, v_5-s_{11}\right\}$. As expected, $\mathbf{A^{off}}$ can be used in most simulations (e.g., line 1 and lines 3-9, Table \ref{table2}), which improves time efficiency since our P-HTS does not need to spend time searching for feasible templates during practical task scheduling events. In line 2 (simulation index 6), our proposed TE-InstaISS is triggered due to the failure of $\mathbf{A^{off}}$. The backup template obtained achieves good performance, although a certain period of time is sacrificed. Overall, our P-HTS outperforms InstaISS and achieves a good trade-off between RT and CF in most cases. For example, in line 8 (simulation index 26), our P-HTS achieves over 1600 times faster than InstaISS on RT, while the difference in CF remains within 4\%. The gap in RT becomes dramatically larger when considering more SPs, components, and complicated VC topologies (see Fig. \ref{fig9}).

\begin{figure*}[t!] \centering    
	\vspace{0 cm} 
	\subfigtopskip=2pt
	\subfigbottomskip=1pt
	\subfigcapskip=0 cm
	\setlength{\abovecaptionskip}{0 cm} 
	\subfigure[] {
		\label{fig11a }     
		\includegraphics[width=0.66 \columnwidth]{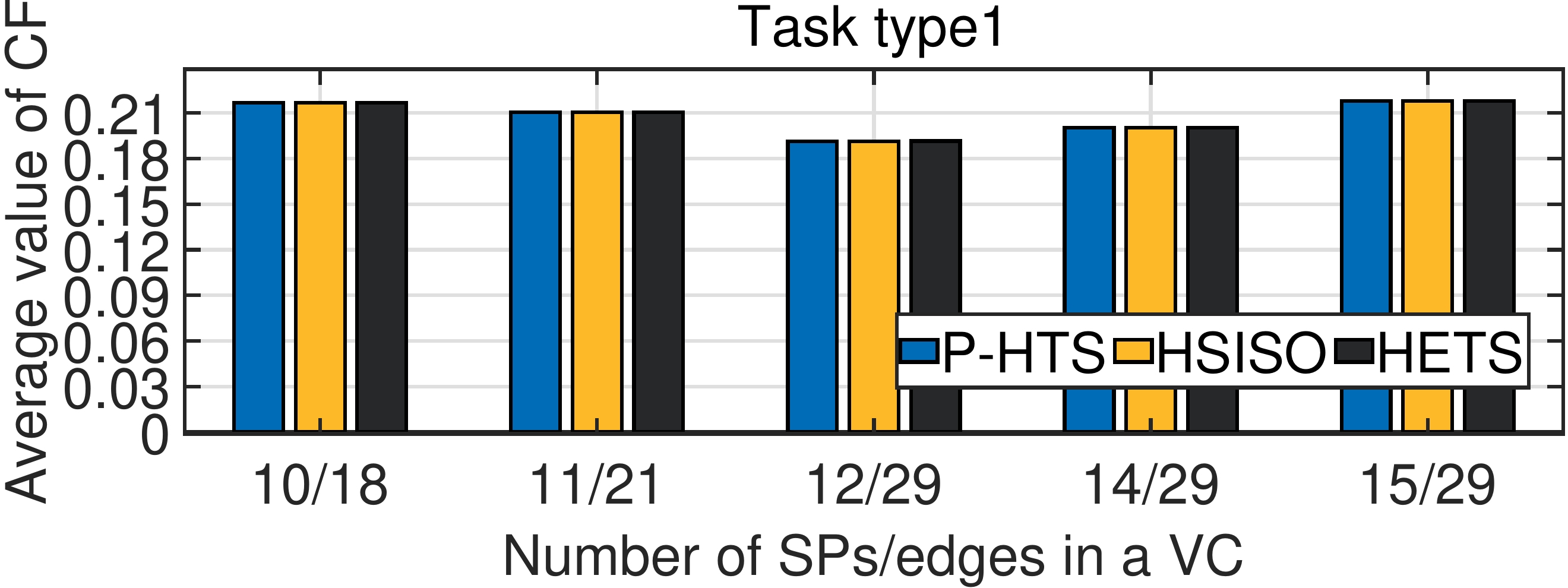}   
	}    \hspace{-3 mm}  
	\subfigure[] { 
		\label{fig11b}     
		\includegraphics[width=0.66\columnwidth]{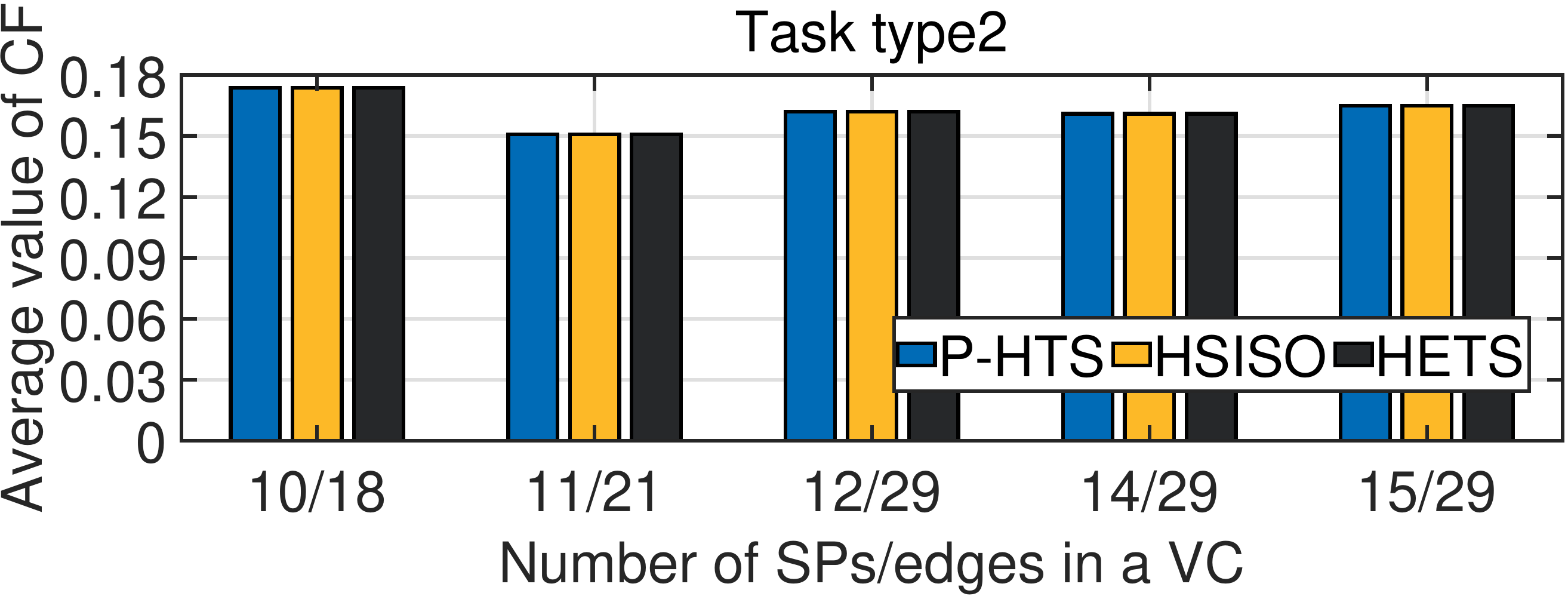}     
	}     \hspace{-3 mm} 
	\subfigure[] { 
		\label{fig11c}     
		\includegraphics[width=0.66\columnwidth]{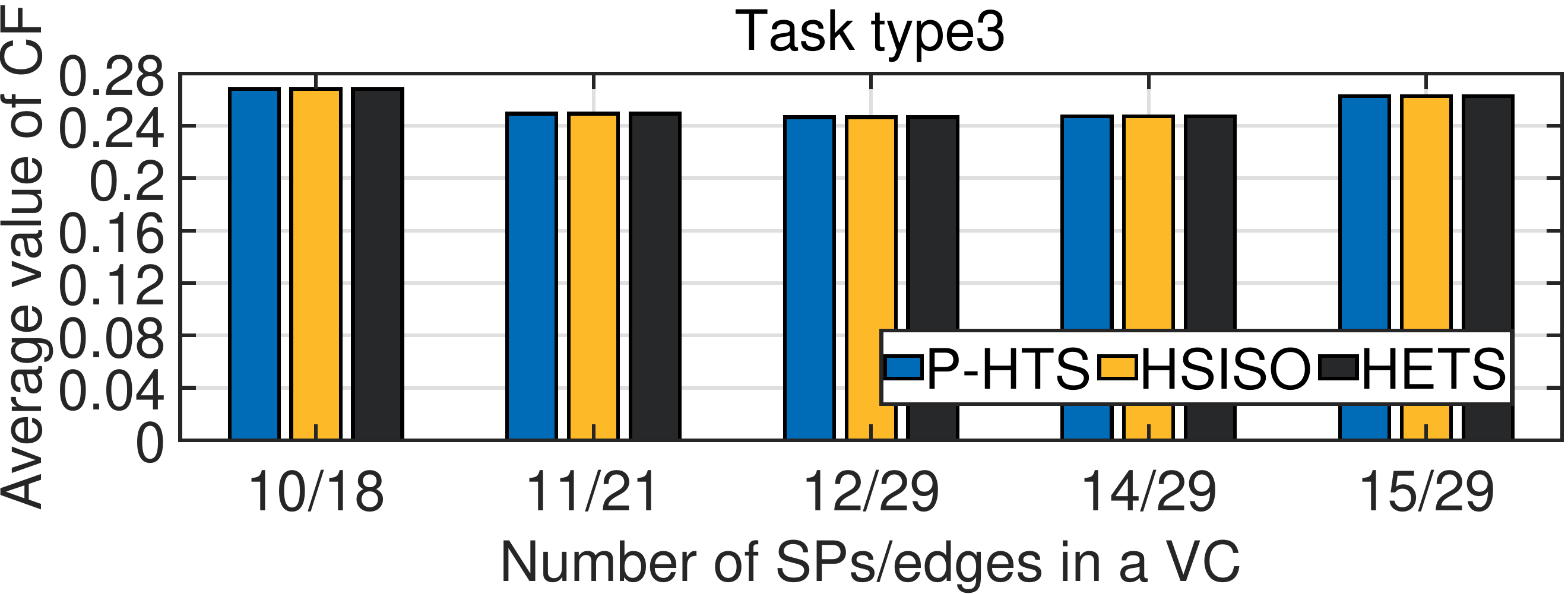}     
	}    
	\caption{Performance comparison in terms of the average value of CF under various problem sizes (among hybrid methods): (a) Task type1; (b) Task type2; (c) Task type3.}      
	\label{fig11}  
	\vspace{-0.3 cm}   
\end{figure*}

\begin{figure*}[t!] \centering    
	\vspace{0 cm} 
	\subfigtopskip=2pt
	\subfigbottomskip=1pt
	\subfigcapskip=0 cm
	\setlength{\abovecaptionskip}{0 cm} 
	\subfigure[] {
		\label{fig12a }     
		\includegraphics[width=0.66 \columnwidth]{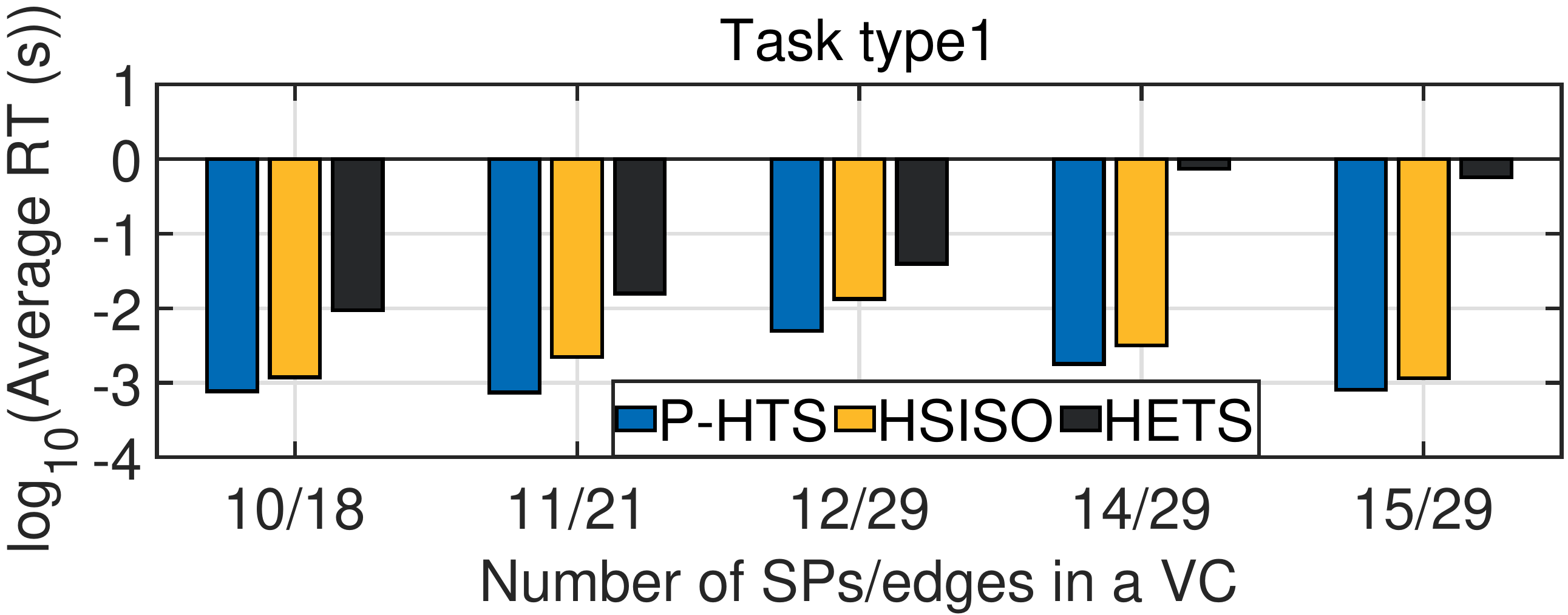}   
	}    \hspace{-3 mm}  
	\subfigure[] { 
		\label{fig12b}     
		\includegraphics[width=0.66 \columnwidth]{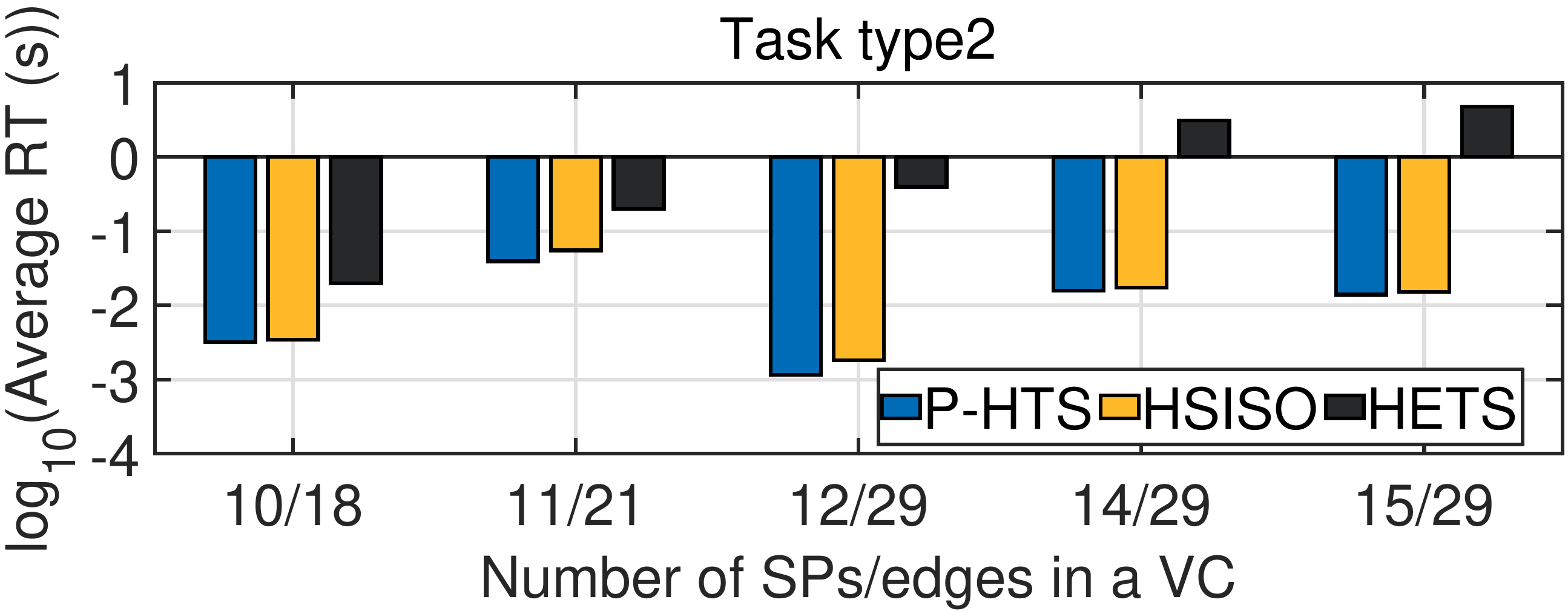}     
	}     \hspace{-3 mm} 
	\subfigure[] { 
		\label{fig12c}     
		\includegraphics[width=0.66 \columnwidth]{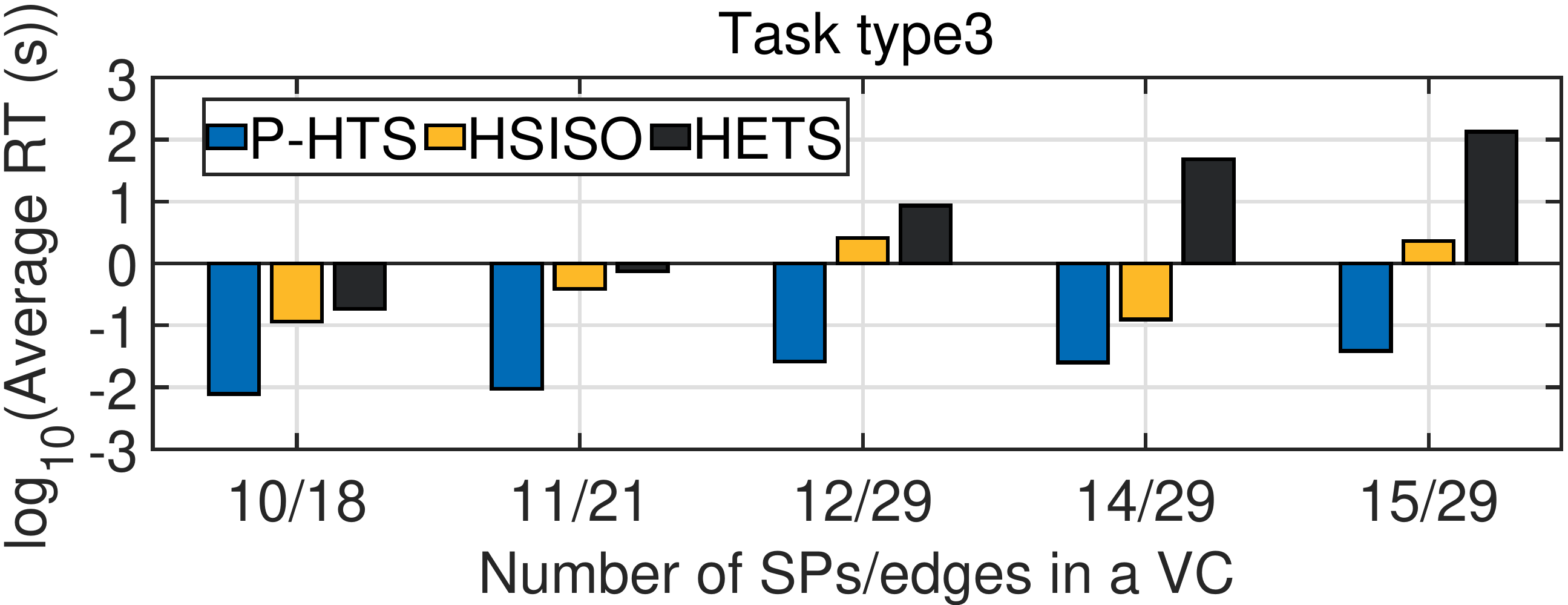}     
	}    
	\caption{Performance comparison in terms of the average RT under various problem sizes (among hybrid methods): (a) Task type1; (b) Task type2; (c) Task type3.}  
	\label{fig12}     
	\vspace{-0.5 cm}
\end{figure*}

\vspace{-0.2cm}
\subsection{Performance Evaluation: Proposed Hybrid Scheduling vs. Other Hybrid Scheduling Methods}
This section presents performance evaluations in terms of CF and RT, considering the following hybrid approaches we developed as baselines for comparison:\\
\textbf{$\bullet$  Hybrid SubISO (HSISO):} We improve SubISO introduced by \cite{abulaish2019subiso} into a hybrid approach by involving both offline and online modes, while applying InstaSISO as the backup plan during practical scheduling events.\\
\textbf{$\bullet$ Hybrid exhaustive template searching (HETS):} We turn ETS into a hybrid approach involving both offline and online modes.

Like P-HTS, both HSISO and HETS have the ability to anticipate and schedule tasks by identifying a template beforehand, while InstaSISO and HTS serve as contingency plans. Figs. \ref{fig11}-\ref{fig12} illustrate the average performance of CF and RT across different task types and SP quantities in a VC. In Fig. \ref{fig11}, the average CF value of P-HTS is compared to that of HSISO and HETS. As shown, the average CF value is consistent across all three methods since P-HTS, HSISO, and HETS can obtain all feasible templates in both offline and online modes. However, Fig. \ref{fig12} highlights the advantages of our P-HTS, which achieves a lower average RT compared to the other methods. Notably, when dealing with more SPs and complex VC structures, HETS experiences a significant increase in RT, while the average RT of P-HTS remains within the range of 0.0001 to 0.1 seconds. For instance, when considering task type 3 and a VC with 15 SPs and 29 edges (see Fig. \ref{fig12c}), the average RT of P-HTS is 0.038 seconds, while HETS takes over 100 seconds.

In summary, our P-HTS offers a reliable and efficient solution for graph task scheduling over dynamic VCs. It exhibits a commendable performance in terms of cost function, task completion time, and data exchange cost. It does so by outperforming the baseline methods in running time, which unequivocally supports the time and energy efficiency in mobile wireless networks. 

\vspace{-0.3cm}
\section{Conclusion}
In this paper, we addressed the challenges of responsive and cost-effective task scheduling in a dynamic VC. Our approach involved modeling both tasks and VCs as undirected weighted graphs, and designing a hybrid graph task scheduling methodology that integrates both offline and online decision-making modes. Specifically, we proposed RA-PilotISS, an offline mechanism that analyzes historical data to identify the template $\mathbf{A^{off}}$ between task components and vehicles, in advance of practical scheduling events. In case that template $\mathbf{A^{off}}$ fails, we also investigate a backup approach TE-InstaISS that can obtain an optimal template $\mathbf{A^{on}}$ under the current network conditions. Our hybrid scheduling mechanism effectively minimizes overhead, including on task completion delay and cost on data exchange, while coping with multiple network uncertainties. We demonstrated the effectiveness of our mechanism through comprehensive experiments. {This work also highlights open directions for future research, such as expanding our methodology to DAG task scheduling, graph task structure preservation by vehicular path planning, and artificial intelligence-promoted risk assessment over large-scale networks and task structures.}


\begin{spacing}{1.0}
	\bibliographystyle{ieeetr}
	\bibliography{reference.bib}

\begin{thebibliography}{10}

\bibitem{liwang2019allocation}
M.~Liwang, S.~Hosseinalipour, Z.~Gao, Y.~Tang, L.~Huang, and H.~Dai, ``Allocation of computation-intensive graph jobs over vehicular clouds in {IoV},'' {\em IEEE Internet Things J.}, vol.~7, no.~1, pp.~311--324, 2019.

\bibitem{luo2021minimizing}
Q.~Luo, C.~Li, T.~H. Luan, and W.~Shi, ``Minimizing the delay and cost of computation offloading for vehicular edge computing,'' {\em IEEE Trans. Serv. Comput.}, vol.~15, no.~5, pp.~2897--2909, 2021.

\bibitem{hu2021efficient}
Z.~Hu, J.~Niu, T.~Ren, B.~Dai, Q.~Li, M.~Xu, and S.~K. Das, ``An efficient online computation offloading approach for large-scale mobile edge computing via deep reinforcement learning,'' {\em IEEE Trans. Serv. Comput.}, vol.~15, no.~2, pp.~669--683, 2021.

\bibitem{mach2017mobile}
P.~Mach and Z.~Becvar, ``Mobile edge computing: A survey on architecture and computation offloading,'' {\em IEEE Commun. Surveys Tut.}, vol.~19, no.~3, pp.~1628--1656, 2017.

\bibitem{rodrigues2016hybrid}
T.~G. Rodrigues, K.~Suto, H.~Nishiyama, and N.~Kato, ``Hybrid method for minimizing service delay in edge cloud computing through {VM} migration and transmission power control,'' {\em IEEE Trans. Comput.}, vol.~66, no.~5, pp.~810--819, 2016.

\bibitem{samie2016computation}
F.~Samie, V.~Tsoutsouras, L.~Bauer, S.~Xydis, D.~Soudris, and J.~Henkel, ``Computation offloading and resource allocation for low-power {IoT} edge devices,'' {\em IEEE world forum Internet Things J. (WF-IoT)}, pp.~7--12, 2016.

\bibitem{liu2021task}
B.~Liu, X.~Xu, L.~Qi, Q.~Ni, and W.~Dou, ``Task scheduling with precedence and placement constraints for resource utilization improvement in multi-user {MEC} environment,'' {\em J. Syst. Architect.}, vol.~114, p.~101970, 2021.

\bibitem{su2024reliable}
S.~Su, P.~Yuan, and Y.~Dai, ``Reliable computation offloading of dag applications in internet of vehicles based on deep reinforcement learning,'' {\em IEEE Transactions on Vehicular Technology}, 2024.

\bibitem{liu2023rfid}
Z.~Liu, M.~Liwang, S.~Hosseinalipour, H.~Dai, Z.~Gao, and L.~Huang, ``{RFID}: towards low latency and reliable {DAG} task scheduling over dynamic vehicular clouds,'' {\em IEEE Trans. Veh. Technol.}, vol.~16, no.~5, pp.~3397--3411, 2023.

\bibitem{waheed2022comprehensive}
A.~Waheed, M.~A. Shah, S.~M. Mohsin, A.~Khan, C.~Maple, S.~Aslam, and S.~Shamshirband, ``A comprehensive review of computing paradigms, enabling computation offloading and task execution in vehicular networks,'' {\em IEEE Access}, vol.~10, pp.~3580--3600, 2022.

\bibitem{liwang2023graph}
M.~Liwang, Z.~Gao, S.~Hosseinalipour, Y.~Su, X.~Wang, and H.~Dai, ``Graph-represented computation-intensive task scheduling over air-ground integrated vehicular networks,'' {\em IEEE Trans. Serv. Comput.}, vol.~16, no.~5, pp.~3397--3411, 2023.

\bibitem{liwang2020multi}
M.~Liwang, Z.~Gao, S.~Hosseinalipour, and H.~Dai, ``Multi-task offloading over vehicular clouds under graph-based representation,'' {\em IEEE Int. Conf. Commun. (ICC)}, pp.~1--7, 2020.

\bibitem{gao2021truthful}
Z.~Gao, M.~Liwang, S.~Hosseinalipour, H.~Dai, and X.~Wang, ``A truthful auction for graph job allocation in vehicular cloud-assisted networks,'' {\em IEEE Trans. Mobile Comput.}, vol.~21, no.~10, pp.~3455--3469, 2022.

\bibitem{hosseinalipour2019power}
S.~Hosseinalipour, A.~Nayak, and H.~Dai, ``Power-aware allocation of graph jobs in geo-distributed cloud networks,'' {\em IEEE Trans. Parallel Distrib. Syst.}, vol.~31, no.~4, pp.~749--765, 2019.

\bibitem{ghaderi2016scheduling}
J.~Ghaderi, S.~Shakkottai, and R.~Srikant, ``Scheduling storms and streams in the cloud,'' {\em ACM Trans. Model. Perform. Eval. Comput. Syst.}, vol.~1, no.~4, pp.~1--28, 2016.

\bibitem{Conte2004}
D.~Conte, P.~Foggia, C.~Sansone, and M.~Vento, ``Thirty years of graph matching in pattern recognition,'' {\em Int. J. Pattern Recogn.}, vol.~18, no.~03, pp.~265--298, 2004.

\bibitem{li2022bi}
T.~Li, N.~Wang, B.~Jiang, and M.~Zhang, ``A bi-objective lane reservation problem considering dynamic traffic flow,'' {\em IEEE Trans. Intell. Transp. Syst.}, vol.~24, no.~1, pp.~367--381, 2022.

\bibitem{liu2020dependency}
Y.~Liu, S.~Wang, Q.~Zhao, S.~Du, A.~Zhou, X.~Ma, and F.~Yang, ``Dependency-aware task scheduling in vehicular edge computing,'' {\em IEEE Internet Things J.}, vol.~7, no.~6, pp.~4961--4971, 2020.

\bibitem{lv2022tbtoa}
X.~Lv, H.~Du, and Q.~Ye, ``{TBTOA}: A {DAG}-based task offloading scheme for mobile edge computing,'' {\em IEEE Int. Conf. Commun (ICC).}, pp.~4607--4612, 2022.

\bibitem{al2020task}
A.~A. Al-Habob, O.~A. Dobre, A.~G. Armada, and S.~Muhaidat, ``Task scheduling for mobile edge computing using genetic algorithm and conflict graphs,'' {\em IEEE Trans. Veh. Technol.}, vol.~69, no.~8, pp.~8805--8819, 2020.

\bibitem{abdisarabshali2023decomposition}
P.~Abdisarabshali, M.~Liwang, A.~Rajabzadeh, M.~Ahmadi, and S.~Hosseinalipour, ``Decomposition theory meets reliability analysis: Processing of computation-intensive dependent tasks over vehicular clouds with dynamic resources,'' {\em IEEE/ACM Trans. Netw.}, 2023.

\bibitem{li2022trade}
R.~Li, C.~S. Lim, M.~E. Rana, and X.~Zhou, ``A trade-off task-offloading scheme in multi-user multi-task mobile edge computing,'' {\em IEEE Access}, vol.~10, pp.~129884--129898, 2022.

\bibitem{alameddine2019dynamic}
H.~A. Alameddine, S.~Sharafeddine, S.~Sebbah, S.~Ayoubi, and C.~Assi, ``Dynamic task offloading and scheduling for low-latency {IoT} services in multi-access edge computing,'' {\em IEEE J. Sel. Areas Commun.}, vol.~37, no.~3, pp.~668--682, 2019.

\bibitem{yue2021todg}
S.~Yue, J.~Ren, N.~Qiao, Y.~Zhang, H.~Jiang, Y.~Zhang, and Y.~Yang, ``{TODG}: Distributed task offloading with delay guarantees for edge computing,'' {\em IEEE Trans. Parallel Distrib. Syst.}, vol.~33, no.~7, pp.~1650--1665, 2021.

\bibitem{tutuncuouglu2022online}
F.~T{\"u}t{\"u}nc{\"u}o{\u{g}}lu, S.~Jo{\v{s}}ilo, and G.~D{\'a}n, ``Online learning for rate-adaptive task offloading under latency constraints in serverless edge computing,'' {\em IEEE/ACM Trans. Netw.}, vol.~31, no.~2, pp.~695--709, 2023.

\bibitem{sun2018cooperative}
F.~Sun, F.~Hou, N.~Cheng, M.~Wang, H.~Zhou, L.~Gui, and X.~Shen, ``Cooperative task scheduling for computation offloading in vehicular cloud,'' {\em IEEE Trans. Veh. Technol.}, vol.~67, no.~11, pp.~11049--11061, 2018.

\bibitem{zhang2022dag}
Z.~Zhang, Q.-S. Hua, X.~Zhang, H.~Jin, and X.~Liao, ``{DAG} scheduling with communication delays based on graph convolutional neural network,'' {\em Wireless Commun. Mobile. Comput.}, vol.~2022, 2022.

\bibitem{shi2016energy}
L.~Shi, Z.~Zhang, and T.~Robertazzi, ``Energy-aware scheduling of embarrassingly parallel jobs and resource allocation in cloud,'' {\em IEEE Trans. Parallel Distrib. Syst.}, vol.~28, no.~6, pp.~1607--1620, 2016.

\bibitem{liwang2020energy}
M.~Liwang, Z.~Gao, S.~Hosseinalipour, H.~Dai, and X.~Wang, ``Energy-aware allocation of graph jobs in vehicular cloud computing-enabled software-defined,'' {\em IEEE Int. Conf. Comput. Commun. Workshops. (INFOCOM WKSHPS)}, pp.~604--609, 2020.

\bibitem{zhang2022task}
D.~Zhang, L.~Cao, H.~Zhu, T.~Zhang, J.~Du, and K.~Jiang, ``Task offloading method of edge computing in internet of vehicles based on deep reinforcement learning,'' {\em Cluster Comput.}, vol.~25, no.~2, pp.~1175--1187, 2022.

\bibitem{hu2019task}
Y.~Hu, T.~Cui, X.~Huang, and Q.~Chen, ``Task offloading based on {L}yapunov optimization for {MEC}-assisted platooning,'' {\em IEEE Int. Conf. Wireless Commun. Signal Process. (WCSP)}, pp.~1--5, 2019.

\bibitem{Josilo2020}
S.~Jo{\v{s}}ilo and G.~D{\'a}n, ``Computation offloading scheduling for periodic tasks in mobile edge computing,'' {\em IEEE/ACM Trans. Netw.}, vol.~28, no.~2, pp.~667--680, 2020.

\bibitem{dey2016vehicle}
K.~C. Dey, A.~Rayamajhi, M.~Chowdhury, P.~Bhavsar, and J.~Martin, ``Vehicle-to-vehicle {(V2V)} and vehicle-to-infrastructure (v2i) communication in a heterogeneous wireless network--performance evaluation,'' {\em Trans. Res. Part C: Emerg. Technol.}, vol.~68, pp.~168--184, 2016.

\bibitem{hou2020reliable}
X.~Hou, Z.~Ren, J.~Wang, W.~Cheng, Y.~Ren, K.-C. Chen, and H.~Zhang, ``Reliable computation offloading for edge-computing-enabled software-defined {IoV},'' {\em IEEE Internet Things J.}, vol.~7, no.~8, pp.~7097--7111, 2020.

\bibitem{deb2020deft}
P.~K. Deb, C.~Roy, A.~Roy, and S.~Misra, ``{DEFT}: Decentralized multiuser computation offloading in a fog-enabled {IoV} environment,'' {\em IEEE Trans. Veh. Technol.}, vol.~69, no.~12, pp.~15978--15987, 2020.

\bibitem{pham2019coalitional}
Q.-V. Pham, H.~T. Nguyen, Z.~Han, and W.-J. Hwang, ``Coalitional games for computation offloading in {NOMA}-enabled multi-access edge computing,'' {\em IEEE Trans. Veh. Technol.}, vol.~69, no.~2, pp.~1982--1993, 2019.

\bibitem{huang2012dynamic}
D.~Huang, P.~Wang, and D.~Niyato, ``A dynamic offloading algorithm for mobile computing,'' {\em IEEE Trans. Wireless Commun.}, vol.~11, no.~6, pp.~1991--1995, 2012.

\bibitem{jovsilo2018joint}
S.~Jo{\v{s}}ilo and G.~D{\'a}n, ``Joint allocation of computing and wireless resources to autonomous devices in mobile edge computing,'' {\em Proc. WKSHPS. Mobile Edge Commun.}, pp.~13--18, 2018.

\bibitem{west2001introduction}
D.~B. West, ``Introduction to graph theory,'' {\em Upper Saddle River: Prentice hall.}, vol.~2, 2001.

\bibitem{ghimire2021dynamic}
B.~Ghimire and D.~B. Rawat, ``Dynamic clustering in {IoV} using behavioral parameters and contention window adaptation,'' {\em IEEE Trans. Veh. Technol.}, vol.~71, no.~2, pp.~2031--2040, 2021.

\bibitem{Huang2021}
X.~Huang, R.~Yu, D.~Ye, L.~Shu, and S.~Xie, ``Efficient workload allocation and user-centric utility maximization for task scheduling in collaborative vehicular edge computing,'' {\em IEEE Trans. Veh. Technol.}, vol.~70, no.~4, pp.~3773--3787, 2021.

\bibitem{zhan2020deep}
W.~Zhan, C.~Luo, J.~Wang, C.~Wang, G.~Min, H.~Duan, and Q.~Zhu, ``Deep-reinforcement-learning-based offloading scheduling for vehicular edge computing,'' {\em IEEE Internet Things J.}, vol.~7, no.~6, pp.~5449--5465, 2020.

\bibitem{Chen2022}
C.~Chen, Y.~Zeng, H.~Li, Y.~Liu, and S.~Wan, ``A multihop task offloading decision model in {MEC}-enabled internet of vehicles,'' {\em IEEE Internet Things J.}, vol.~10, no.~4, pp.~3215--3230, 2022.

\bibitem{luo2020hfel}
S.~Luo, X.~Chen, Q.~Wu, Z.~Zhou, and S.~Yu, ``{HFEL}: Joint edge association and resource allocation for cost-efficient hierarchical federated edge learning,'' {\em IEEE Trans. Wireless Commun.}, vol.~19, no.~10, pp.~6535--6548, 2020.

\bibitem{abulaish2019subiso}
M.~Abulaish, Z.~A. Ansari, {\em et~al.}, ``{SubISO}: a scalable and novel approach for subgraph isomorphism search in large graph,'' {\em IEEE Int. Conf. Commun. Syst. Netw. (COMSNETS)}, pp.~102--109, 2019.

\end{thebibliography}
\end{spacing}

\clearpage
\begin{appendices}
\section{Derivation of $\overline{\mathbbm{T}}\left(\mathbf{A}\right)$}
Recalling the previous sections, our model adopts an approximation of $\overline{\mathbbm{T}}\left(\mathbf{A}\right)$. This part describes the complexity on obtaining its actual value, which relies heavily on the number of components in a task. To do so, we show two cases of $\left|\bm{V}^{\bm{task}}\right|=2$ and $\left|\bm{V}^{\bm{task}}\right|=3$, as typical examples.
First, according to \eqref{eq2}, we can describe the expected value of $\mathbbm{T}\left(\mathbf{A}\right)$ as the following (14):
	\begin{equation}
		\small
	\setlength{\abovedisplayskip}{3pt}
	\setlength{\belowdisplayskip}{3pt}
		\begin{aligned}
			\overline{\mathbbm{T}}\left(\mathbf{A}\right)
			&=\mathrm{E}[\mathrm{max}[\alpha_{n,m}\mathbbm{t}_{n,m}^{sum}]]_{1\le n\le\left|\bm{V}^{\bm{task}}\right|,1\le m\le\left|\bm{V}^{\bm{serv}}\right|} \\ 
		\end{aligned}
		\label{eq27} 
	\end{equation}
	Considering $\left|\bm{V}^{\bm{task}}\right|=2$, where $\alpha_{n,m} = 1$ and $\alpha_{n',m'} = 1$, we then have $\overline{\mathbbm{T}}\left(\mathbf{A}\right)$ as:
	\begin{equation}
		\setlength{\abovedisplayskip}{3pt}
		\setlength{\belowdisplayskip}{3pt}
		\small
		\begin{aligned}
			\overline{\mathbbm{T}}\left(\mathbf{A}\right)
			&=\mathrm{E}\left[\mathrm{max}\left(\mathbbm{t}_{n,m}^{sum},\mathbbm{t}_{n',m'}^{sum}\right)\right] \\ 
		\end{aligned}
		\label{eq28} 
	\end{equation}
	Due to (16) given below,
	\begin{equation}
		\setlength{\abovedisplayskip}{3pt}
		\setlength{\belowdisplayskip}{3pt}
		\small
		\begin{aligned}
			\mathrm{max}(\mathbbm{t}_{n,m}^{sum},\mathbbm{t}_{n',m'}^{sum}) = \frac{1}{2}(\mathbbm{t}_{n,m}^{sum}+\mathbbm{t}_{n',m'}^{sum}+|\mathbbm{t}_{n,m}^{sum}-\mathbbm{t}_{n',m'}^{sum}|),
		\end{aligned}
		\label{eq29} 
	\end{equation}
	we can transform \eqref{eq28} as:
	\begin{equation}
		\setlength{\abovedisplayskip}{3pt}
		\setlength{\belowdisplayskip}{3pt}
		\small
		\begin{aligned}
			\overline{\mathbbm{T}}\left(\mathbf{A}\right)
			&=\mathrm{E}\left[\frac{1}{2}\left(\mathbbm{t}_{n,m}^{sum}+\mathbbm{t}_{n',m'}^{sum}+\left|\mathbbm{t}_{n,m}^{sum}-\mathbbm{t}_{n',m'}^{sum}\right|\right)\right] \\ 
			&=\frac{1}{2}\mathrm{E}\left(\mathbbm{t}_{n,m}^{sum}\right)+\frac{1}{2}\mathrm{E}\left(\mathbbm{t}_{n',m'}^{sum}\right)+\frac{1}{2}\mathrm{E}\left(\left|\mathbbm{t}_{n,m}^{sum}-\mathbbm{t}_{n',m'}^{sum}\right|\right).
		\end{aligned}
		\label{eq30} 
	\end{equation}
	Similarly, for $\left|\bm{V}^{\bm{task}}\right|=3$, where $\alpha_{n,m} =\alpha_{n',m'} =\alpha_{n'',m''} = 1$, we have (18). Apparently, it is challenging to get the close form of $\overline{\mathbbm{T}}\left(\mathbf{A}\right)$, especially upon having $\left|\bm{V}^{\bm{task}}\right|\ge4$. To facilitate analysis, we simply use the maximum value of the mean of $\mathbbm{t}_{n,m}^{sum}$ instead of the intractable expectation of $\mathbbm{T}\left(\mathbf{A}\right)$, given in \eqref{eq32}.
	\begin{equation}
		\setlength{\abovedisplayskip}{3pt}
		\setlength{\belowdisplayskip}{3pt}
		\small
		\begin{aligned}
			\overline{\mathbbm{T}}\left(\mathbf{A}\right)
			&=\mathrm{E}\left[\mathrm{max}\left(\mathbbm{t}_{n,m}^{sum},\mathbbm{t}_{n',m'}^{sum},\mathbbm{t}_{n'',m''}^{sum}\right)\right] \\
			&=\mathrm{E}\left[\mathrm{max}\left(\mathrm{max}\left(\mathbbm{t}_{n,m}^{sum},\mathbbm{t}_{n',m'}^{sum}\right),\mathrm{max}\left(\mathbbm{t}_{n',m'}^{sum},\mathbbm{t}_{n'',m''}^{sum}\right)\right)\right]\\
			&=\mathrm{E}[\mathrm{max}(\frac{1}{2}(\mathbbm{t}_{n,m}^{sum}+\mathbbm{t}_{n',m'}^{sum}+|\mathbbm{t}_{n,m}^{sum}-\mathbbm{t}_{n',m'}^{sum}|),\\
			& \ \ \ \ \ \frac{1}{2}(\mathbbm{t}_{n,m}^{sum}+\mathbbm{t}_{n',m'}^{sum}+|\mathbbm{t}_{n,m}^{sum}-\mathbbm{t}_{n',m'}^{sum}|))] \\ 
			&=\frac{1}{2}\mathrm{E}[\frac{1}{2}(\mathbbm{t}_{n,m}^{sum}+\mathbbm{t}_{n'',m''}^{sum}+2\mathbbm{t}_{n',m'}^{sum}+|\mathbbm{t}_{n,m}^{sum}-\mathbbm{t}_{n',m'}^{sum}|\\
			&\ \ \ \
			+|\mathbbm{t}_{n'',m''}^{sum}-\mathbbm{t}_{n',m'}^{sum}|)+|\mathbbm{t}_{n,m}^{sum}-\mathbbm{t}_{n'',m''}^{sum}\\
			&\ \ \ \ +|\mathbbm{t}_{n,m}^{sum}-\mathbbm{t}_{n',m'}^{sum}|-|\mathbbm{t}_{n'',m''}^{sum}-\mathbbm{t}_{n',m'}^{sum}||]\\~&=\frac{1}{4}\mathrm{E}(\mathbbm{t}_{n,m}^{sum})+\frac{1}{4}\mathrm{E}(\mathbbm{t}_{n'',m''}^{sum})+\frac{1}{2}\mathrm{E}(\mathbbm{t}_{n',m'}^{sum})\\
			&\ \ \ \
			+\frac{1}{4}\mathrm{E}(|\mathbbm{t}_{n,m}^{sum}-\mathbbm{t}_{n',m'}^{sum}|)+\frac{1}{4}\mathrm{E}(|\mathbbm{t}_{n'',m''}^{sum}-\mathbbm{t}_{n',m'}^{sum}|)\\
			&\ \ \ \
			+\frac{1}{2}\mathrm{E}(|\mathbbm{t}_{n,m}^{sum}-\mathbbm{t}_{n'',m''}^{sum}+|\mathbbm{t}_{n,m}^{sum}-\mathbbm{t}_{n',m'}^{sum}|\\
			&\ \ \ \ -|\mathbbm{t}_{n'',m''}^{sum}-\mathbbm{t}_{n',m'}^{sum}||)
		\end{aligned}
		\label{eq31} 
	\end{equation}
	
	\begin{equation}
		\setlength{\abovedisplayskip}{3pt}
		\setlength{\belowdisplayskip}{3pt}
		\small
		\begin{aligned}
				\overline{\mathbbm{T}}\left(\mathbf{A}\right)
				&=\mathrm{E}[\mathrm{max}[\alpha_{n,m}\mathbbm{t}_{n,m}^{sum}]]_{1\le n\le\left|\bm{V}^{\bm{task}}\right|,1\le m\le\left|\bm{V}^{\bm{serv}}\right|} \\ 
				&\approx\mathrm{max}{\left[\alpha_{n,m}\overline{\mathbbm{t}_{n,m}^{sum}}\right]}_{1\le n\le\left|\bm{V}^{\bm{task}}\right|,1\le m\le\left|\bm{V}^{\bm{serv}}\right|}
		\end{aligned}
		\label{eq32} 
	\end{equation}
	
\section{Derivation of Risk \eqref{eq10}}
\normalfont
We consider the risk $R^{time}_{n,m}$ given in \eqref{eq10} under $\alpha_{n,m}=1$.	
According to \eqref{eq1} and \eqref{eq8}, we have
\begin{equation}
	\setlength{\abovedisplayskip}{3pt}
	\setlength{\belowdisplayskip}{3pt}
	\mathbbm{t}_{n,m}^{sum} = \frac{q_n}{f_m} + \frac{d_n}{r_m} = \tau_m^{comp} q_n + \tau_m^{comm} d_n,
	\label{eq14} 
	\small
\end{equation}
where $\tau_m^{comp}=\frac{1}{f_m}$ and $\tau_m^{comm}=\frac{1}{r_m}$. Suppose that $f_m$ and $r_m$ are independent and continuous random variables with probability density functions (PDFs) $\phi_{f_m}$ and $\phi_{r_m}$, respectively. According to the transformation theorem of random variables with continuous distributions, we can obtain the PDFs of $\tau_m^{comp}$ and $\tau_m^{comm}$ ($\phi_{\tau_m^{comp}}$ and $\phi_{\tau_m^{comm}}$) according to $\phi_{f_m}$ and $\phi_{r_m}$, as given by
\begin{equation}
	\setlength{\abovedisplayskip}{3pt}
	\phi_{\tau_m^{comp}}(\tau_m^{comp}) = \phi_{f_m}\left(f_m\right) \left|\frac{d{f_m}}{d{\tau_m^{comp}}}\right|=\frac{1}{{\left(\tau_m^{comp}\right)}^2}\phi_{f_m}\left(\frac{1}{\tau_m^{comp}}\right),
	\label{eq15} 
	\footnotesize
\end{equation}
\begin{equation}
	\phi_{\tau_m^{comm}}(\tau_m^{comm}) = \phi_{r_m}\left(r_m\right) \left|\frac{d{r_m}}{d{\tau_m^{comm}}}\right|=\frac{1}{{\left(\tau_m^{comm}\right)}^2}\phi_{r_m}\left(\frac{1}{\tau_m^{comm}}\right)
	\label{eq16} 
	\footnotesize
\end{equation}
The cumulative distribution function (CDF) of $\mathbbm{t}_{n,m}^{sum}$, denoted as $\Phi_{\mathbbm{t}_{n,m}^{sum}}$, is defined by:
\begin{equation}
	\setlength{\abovedisplayskip}{3pt}
	\setlength{\belowdisplayskip}{3pt}
	\Phi_{\mathbbm{t}_{n,m}^{sum}}(t)=\text{Pr}\left(\mathbbm{t}_{n,m}^{sum} \le t\right)
	=\text{Pr}\left(\tau_m^{comp} q_n+\tau_m^{comm}d_n\le t\right)
	\label{eq17} 
	\small
\end{equation}
Since $\tau_m^{comp}$ and $\tau_m^{comm}$ are independent, their joint probability density function refers to as the product of the individual density functions, as shown by
\begin{equation}
	\setlength{\abovedisplayskip}{3pt}
	\setlength{\belowdisplayskip}{3pt}
	\phi_{\tau_m^{comp}, \tau_m^{comm}}(\tau_m^{comp}, \tau_m^{comm})= \phi_{\tau_m^{comp}}(\tau_m^{comp}) \phi_{\tau_m^{comm}}(\tau_m^{comm})
	\label{eq19} 
	\small
\end{equation}
Accordingly, we can calculate $\Phi_{\mathbbm{t}_{n,m}^{sum}}$ as \eqref{eq20}.
\begin{figure*}
	\hrulefill
	\begin{equation}
		\setlength{\abovedisplayskip}{3pt}
		\setlength{\belowdisplayskip}{3pt}
		\begin{aligned}
			&\Phi_{\mathbbm{t}_{n,m}^{sum}}(\mathbbm{t}_{n,m}^{sum}\le t)
			=\text{Pr}(\tau_m^{comp}q_n+\tau_m^{comm}d_n\le t) \\
			&=\iint_{\tau_m^{comp} q_n+\tau_m^{comm} d_n\le t}
			{\phi_{\tau_m^{comp}, \tau_m^{comm}}(\tau_m^{comp}, \tau_m^{comm})d\tau_m^{comp}d\tau_m^{comm}} \\
			&=\iint_{\tau_m^{comp} q_n\le t-\tau_m^{comm} d_n}{\phi_{\tau_m^{comp}, \tau_m^{comm}}(\tau_m^{comp}, \tau_m^{comm})d\tau_m^{comp}d\tau_m^{comm}} \\
			&=\int_{-\infty}^{+\infty}\int_{-\infty}^{\left(t-\tau_m^{comm} d_n\right)/q_n}{\phi_{\tau_m^{comp}, \tau_m^{comm}}(\tau_m^{comp}, \tau_m^{comm})d\tau_m^{comp}d\tau_m^{comm}} \\
			&=\int_{-\infty}^{+\infty}{\phi_{\tau_m^{comm}}}(\tau_m^{comm})\int_{-\infty}^{(t-\tau_m^{comm} d_n)/q_n}{\phi_{\tau_m^{comp}}(\tau_m^{comp})d\tau_m^{comp}d\tau_m^{comm}}
			\small
		\end{aligned}
		\small
		\label{eq20}
	\end{equation}
\end{figure*}

The inner integral with respect to $\tau_m^{comp}$ is given by
\begin{equation}
	\setlength{\abovedisplayskip}{3pt}
	\setlength{\belowdisplayskip}{3pt}
	\int_{-\infty}^{(t-\tau_m^{comm} d_n)/q_n}{\phi_{\tau_m^{comp}}(\tau_m^{comp})d\tau_m^{comp}},
	\label{eq21} 
	\small
\end{equation}
where the integral starts from $-\infty$ to a certain constant, which thus evaluates the CDF of $\tau_m^{comp}$ at the point $(t-\tau_m^{comm} d_n)/q_n$, (denote as $\Phi_{\tau_m^{comp}}[(t-\tau_m^{comm}d_n)/q_n]$).

Therefore, we have the following \eqref{eq22}.
\begin{figure*}
	\begin{equation}	
		\setlength{\abovedisplayskip}{3pt}
		\setlength{\belowdisplayskip}{3pt}
		\Phi_{\mathbbm{t}_{n,m}^{sum}}(\mathbbm{t}_{n,m}^{sum}\leq t) =\int_{-\infty}^{+\infty} \phi_{\tau_m^{comm}}(\tau_m^{comm}) \Phi_{\tau_m^{comp}}\left[\frac{t-\tau_m^{comm} d_n}{q_n}\right] \, d\tau_m^{comm}	
		\label{eq22} 
		\small
	\end{equation}
\end{figure*}

We differentiate this expression with respect to $t$ to obtain the PDF of $\mathbbm{t}_{n,m}^{sum}$:
\begin{equation}
	\setlength{\abovedisplayskip}{3pt}
	\setlength{\belowdisplayskip}{3pt}
	\small
	\begin{aligned}
		\label{eq23} 	
		&\phi_{\mathbbm{t}_{n,m}^{sum}}=\frac{d}{d\mathbbm{t}_{n,m}^{sum}}\Phi_{\mathbbm{t}_{n,m}^{sum}}(\mathbbm{t}_{n,m}^{sum}\le t) \\
		&= \frac{d}{d\mathbbm{t}_{n,m}^{sum}}\int_{-\infty}^{+\infty} \phi_{\tau_m^{comm}}(\tau_m^{comm})  \Phi_{\tau_m^{comp}}\left[\frac{t-\tau_m^{comm} d_n}{q_n}\right] \, d\tau_m^{comm} \\
		&=\int_{-\infty}^{+\infty} \phi_{\tau_m^{comm}}(\tau_m^{comm}) \frac{d}{d\mathbbm{t}_{n,m}^{sum}}\Phi_{\tau_m^{comp}}\left[\frac{t-\tau_m^{comm} d_n}{q_n}\right] d\tau_m^{comm} \\
		&=\int_{-\infty}^{+\infty} \phi_{\tau_m^{comm}}(\tau_m^{comm}) \phi_{\tau_m^{comp}}\left[\frac{\mathbbm{t}_{n,m}^{sum}-\tau_m^{comm} d_n}{q_n}\right]  \frac{1}{q_n} \, d\tau_m^{comm},
	\end{aligned}
\end{equation}
where $\phi_{\tau_m^{comp}}$ is the PDF of $\tau_m^{comp}$.
Then, $R^{time}_{n,m}$ can be represented as the above equation \eqref{eq24}.
\begin{figure*}
	\begin{equation}
		\small
		\begin{aligned}
			&\text{Pr}(\mathbbm{t}_{n,m}^{sum}>t_n^{{max}})=\int_{t_n^{{max}}}^{+\infty}{\phi_{\mathbbm{t}_{n,m}^{sum}}(\mathbbm{t}_{n,m}^{sum})d\mathbbm{t}_{n,m}^{sum}} \\
			&=\int_{t_n^{{max}}}^{+\infty}\int_{-\infty}^{+\infty} \phi_{\tau_m^{comm}}(\tau_m^{comm}) \phi_{\tau_m^{comp}}\left[\frac{\mathbbm{t}_{n,m}^{sum}-\tau_m^{comm} d_n}{q_n}\right]  \frac{1}{q_n} \, d\tau_m^{comm}  d\mathbbm{t}_{n,m}^{sum}
			\label{eq24} 
			\small	
		\end{aligned}
	\end{equation}
	\hrulefill
\end{figure*}

According to \eqref{eq15} and \eqref{eq16}, we can rewrite \eqref{eq24} as the following:
\begin{equation}
	\setlength{\abovedisplayskip}{3pt}
	\setlength{\belowdisplayskip}{3pt}
	\small
	\begin{aligned}	
		&\text{Pr}(\mathbbm{t}_{n,m}^{sum}>t_n^{{max}}) \\
		&=\int_{t_n^{{max}}}^{+\infty}\int_{-\infty}^{+\infty} {\left(\tau_m^{comm}\right)}^2 \phi_{r_m}\left(\frac{1}{\tau_m^{comm}}\right) \\
		&\phi_{f_m}\left(\frac{q_n}{\mathbbm{t}_{n,m}^{sum}-\tau_m^{comm} d_n}\right) \frac{q_n}{{(\mathbbm{t}_{n,m}^{sum}-\tau_m^{comm} d_n)}^2} d\tau_m^{comm}  d\mathbbm{t}_{n,m}^{sum} 
		\label{eq25} 
		\small
	\end{aligned}
\end{equation}

\section{Derivation of Risk \eqref{eq11}}
\normalfont
For risk $R^{struc}_{m,m^\prime}$ given in \eqref{eq11}, we consider the situation where $\beta_{m,m^\prime}=1$. Assume that $t_{m,m^\prime}^{conn}$ follows a continuous distribution, and its probability density function is denoted by $\phi_{t_{m,m\prime}^{conn}}$, we have
\begin{equation}
	\setlength{\abovedisplayskip}{3pt}
	\setlength{\belowdisplayskip}{3pt}
	\label{eq26}
	\small
	R^{struc}_{m,m^\prime}=\mathrm{Pr}\left({t_{m,m^\prime}^{conn}}< w_{n,n^\prime}^{task}\right)=\int_{-\infty}^{w_{n,n\prime}^{task}}{\phi_{t_{m,m\prime}^{conn}}(t_{m,m\prime}^{conn})dt_{m,m\prime}^{conn}}
\end{equation}

\section{Details of TE-InstaISS}
	\normalfont
	Review the previous Sec. 3.3, TE-InstaISS is introduced as a backup approach detailed by Algorithm 7 when $\mathbf{A^{off}}$ is unavailable, which involves a series of steps similar to RA-PilotISS (detailed by Algorithms 8-11):
	\begin{algorithm}[htb]
	\setstretch{0.6}
	\small
			{
				\caption{Time-efficient instantaneous isomorphic subgraph searching (TE-InstaISS)}
				\KwIn{$\bm{G^{task}}$, $\bm{G^{serv}}$, $\bm{r}$, $\bm{f}$, $\bm{W^{serv}}$, $\bm{c}$}
				\KwOut{Template $\mathbf{A^{on}}$}
				$\widetilde{{v}_n},\mathrm{ASP}\left(\widetilde{v_n} \right)\leftarrow{{\mathrm{PivotCSelectOnline}}}$ \% Algorithm 8 \\
				${SG}_{all}\gets\emptyset$ \\
				\ForEach{$\widetilde{s_m}\ \in\ {\mathrm{ASP}\left(\widetilde{v_n}\right)}$}
				{${Candi}\gets {\rm{RegionExploreOnline}}$ \% Algorithm 9 \\
					\If{${Candi}\neq\emptyset$}
					{
						${SG}_m\gets {\rm{SubGSearchOnline}}$ \% Algorithm 10
					}
					${SG}_{all}\gets{SG}_{all}\cup{SG}_m$ \\
				}
				$\mathbf{A^{on}} \gets {\rm{OptTSearchOnline}}$ \% Algorithm 11 \\
			}
		\label{Alg7}
	\end{algorithm}
	
	\noindent$\bullet$ \textbf{Step 1.} Pivot component selection: To facilitate our proposed TE-InstaISS, we start with selecting a pivot component ($\widetilde{{v}_n}$) through the PivotCSelectOnline function (Algorithm 8), along with its corresponding alternative SPs ($\mathrm{ASP}\left(\widetilde{v_n}\right)$). Algorithm 8 employs constraint \eqref{C4} and considers the degree and neighborhood information of all components to determine the alternative SP set $\mathrm{ASP}\left(v_n\right)$ (lines 1-6). Additionally, the function calculates the eccentricity $\mathrm{Ecc}\left(v_n\right)$ for all components (line 7). Finally, this algorithm selects the pivot component $\widetilde{v_n}$ by applying line 8, which is the component with the minimum value of ${\left|\mathrm{ASP}\left(v_n\right)\right|}\times{\mathrm{Ecc}\left(v_n\right)}$.
	\begin{algorithm}[htb]
	\setstretch{0.6}
		\small
		\caption{PivotCSelectOnline}\label{algorithm}
		\KwIn{$\bm{G^{task}}$, $\bm{G^{serv}}$, $\bm{r}$, $\bm{f}$, $\bm{W^{serv}}$}
		\KwOut{The pivot component $\widetilde{v_n}$, its alternative SPs $\mathrm{ASP}\left(\widetilde{v_n}\right)$, and eccentricity $\mathrm{Ecc}\left({v_n}\right)$}
		$\mathcal{L} \gets $  List of ordered pairs $\left< v_n,\left|\mathrm{ASP}\left(v_n\right)\right|\right>$ \hspace{6em} 
		\% $\left|\mathrm{ASP}\left(v_n\right)\right|$ is the number of alternative SPs\\
		\ForEach{$v_n \in \bm{V^{task}}$}
		{\ForEach{$s_m \in \bm{V^{serv}}$}
			{
				\If
				{
					$\mathbbm{t}_{n,m}^{sum}\leq t_{n}^{max} \ \mathrm{and} \  \mathrm{Deg}\left(v_n\right)\le \mathrm{Deg}\left(s_m\right) \  \mathrm{and} \   
					{\mathrm{MNdeg}{\left(v_n\right)}} \le {\mathrm{MNdeg}{\left(s_m\right)}}$
				}
				{
					$\mathrm{ASP}\left(v_n\right)\gets\mathrm{ASP}\left(v_n\right) \ \cup \ s_m$
				}
			}
			$\mathcal{L}(n) \gets \left<v_n,\left|\mathrm{ASP}\left(v_n\right)\right|\right>$ \\
			$\mathrm{Ecc}\left({v_n}\right) \gets$ Calculate the eccentricity of ${v_n}$ \\
		}
		$\widetilde{v_n}\gets v_n \ \text{with the minimum value of} ~ {\left|\mathrm{ASP}\left(v_n\right)\right|\times \mathrm{Ecc}\left(v_n\right)}$ \\
		return $\widetilde{v_n}, \mathrm{ASP}\left(\widetilde{v_n}\right),\mathrm{Ecc}\left({\widetilde{v_n}}\right)$
		\label{Alg8}
	\end{algorithm}
	
	\noindent$\bullet$ \textbf{Step 2.} Subgraph search: By utilizing functions RegionExploreOnline (Algorithm 9), as well as SubGSearchOnline (Algorithm 10) and thoroughly examining the SPs in set $\mathrm{ASP}\left(\widetilde{v_n}\right)$, we can obtain all feasible isomorphic subgraphs. It is worth noting that the SPs within $\mathrm{ASP}\left(\widetilde{v_n}\right)$ are appropriately labeled as $\widetilde{s_m}$ to distinguish them.

\begin{algorithm}[h!]
	\small
	\setstretch{0.6}
	\caption{RegionExploreOnline}
	\KwIn{$\bm{G^{task}}$, $\bm{G^{serv}}$, $\widetilde{v_n}$, $\widetilde{s_m}$, $\bm{r}$, $\bm{f}$, $\bm{W^{serv}}$}
	\KwOut{$Candi$: SPs in $\bm{G^{sub}}$ that can be mapped to task components}
	$\bm{G^{sub}}$\ $\leftarrow$\ {Subgraph with $\widetilde{s_m}$ as center and $\mathrm{Ecc}\left(\widetilde{v_n}\right)$ as the radius in VC graph $\bm{{G}^{serv}}$} \\
	\If{$\left|\bm{V^{sub}}\right|<\left|\bm{V^{task}}\right|$}
	{return $\emptyset$}
	$Candi \gets \emptyset$ \\
	\ForEach{$v_n\in \bm{V^{task}}$}
	{$\mathrm{ASP}\left(v_n\right) \gets \emptyset$ \\
		\ForEach{$s_m\in\bm{V^{sub}}$}
		{\If
			{
				$\mathrm{Dst}(v_n,\widetilde{v_n}) \geq \mathrm{Dst}(s_m,\widetilde{s_m})$
			}
			{
				\If
				{
					$\mathbbm{t}_{n,m}^{sum} \leq t_{n}^{max}\ \mathrm{and} \  \mathrm{Deg}\left(v_n\right)\le \mathrm{Deg}\left(s_m\right) \  \mathrm{and} \   
					{\mathrm{MNdeg}{\left(v_n\right)}} \le {\mathrm{MNdeg}{\left(s_m\right)}}$
				}
				{
					$\mathrm{ASP}\left(v_n\right)\gets\mathrm{ASP}\left(v_n\right) \ \cup \ s_m$
				}
			}	
		}
		
	}
	$Candi \gets Candi \ \cup\ \left<v_n,\mathrm{ASP}\left(v_n\right)\right>$ \\
	return $Candi$
	\label{Alg9}
\end{algorithm}
	\begin{algorithm}[htb]
	\setstretch{0.6}
		\small
		\caption{SubGSearchOnline} \label{algorithm}
		\KwIn{$\bm{G^{task}}$, $\bm{G^{serv}}$,\ $Candi$, $\bm{r}$, $\bm{f}$, $\bm{W^{serv}}$}
		\KwOut{${{SG}_m}$: The set of isomorphic subgraphs from\ \ $Candi$}
		${{SG}_m}$\ $\leftarrow$\ All combinations of candidate SPs for task components from\ $Candi$ \\
		\ForEach{${{G}_i}\in{{SG}_m}$}
		{
			$\mathbf{B}_i \gets$  Profile of $\beta_{m,m^\prime}$ associated with ${G}_i$ \\
			\ForEach{$e_{n,n^\prime}^{task}\in\bm{E^{task}} \ \ \mathrm{and} \ \ \beta_{m,m^\prime}=1$}
			{
				\If
				{
					${e_{m,m^\prime}^{serv} \notin \bm{E^{serv}}} \ \ \mathrm{or} \ \  w_{n,n^\prime}^{task}> t_{m,{m}^\prime}^{conn}$ 
				}
				
				{
					Delete ${{G}_i}$ from ${{SG}_m}$
				}
			}
		}
		return ${{SG}_m}$
		\label{Alg10}
	\end{algorithm}

	\begin{algorithm}[htb]
		\small
		\setstretch{0.6}
		\caption{OptTSelectOnline} \label{algorithm}
		\KwIn{$\bm{G^{task}}$, $\bm{G^{serv}}$, ${{SG}_{all}}$, ${\bm{r}}$, ${\bm{f}}$, ${\bm{W^{serv}}}$, ${\bm{c}}$}
		\KwOut{Template $\mathbf{A^{on}}$}
		$\mathcal{F}_{min} \gets $ Initialize the minimum value of cost function with a large value\\
		\ForEach{${{G}_i }\in {{SG}_{all}}$}
		{
			$\mathbf{A}_i,\mathbf{B}_i$ $\gets$  Scheduling template from subgraph ${G}_i$ \\
			$\mathcal{F}(\mathbf{A}_i,\mathbf{B}_i)$ $\gets$ The\ value\ of\ cost\ function\ on\ template $\mathbf{A}_i,\mathbf{B}_i$ \\
			\If{$\mathcal{F}(\mathbf{A}_i,\mathbf{B}_i) < \mathcal{F}_{min}$}
			{
				$\mathbf{A^{on}} \gets \mathbf{A}_i$, $\mathcal{F}_{min} \gets \mathcal{F}(\mathbf{A}_i,\mathbf{B}_i)$
			}
		}
		return $\mathbf{A^{on}}$
		\label{Alg11}
	\end{algorithm}

Particularly, in Algorithm 9, each SP $\widetilde{s_m}\in\mathrm{ASP}\left(\widetilde{v_n}\right)$ undergoes a process where a subgraph, denoted as $\bm{G^{sub}}$, is extracted from $\bm{G^{serv}}$. In this subgraph, $\widetilde{s_m}$ is regarded as the center point and $\mathrm{Ecc}\left(\widetilde{v_n}\right)$ as the radius. The radius indicates the longest path from $\widetilde{s_m}$ to other SPs (line 1). For instance, if $\mathrm{Ecc}\left(\widetilde{v_n}\right)$ is equal to 3, $\bm{G^{sub}}$ will encompass all the SPs within 3 hops from $\widetilde{s_m}$ in $\bm{G^{serv}}$. If the SPs in $\bm{G^{sub}}$ are insufficient to support component processing, an empty set is returned (lines 2-3), indicating that no isomorphic subgraph can be found in $\bm{G^{sub}}$. If there are enough SPs, Algorithm 9 then examines the degree and neighborhood details of all the task components and searches alternative SPs for them under $\bm{G^{sub}}$. These alternative SPs are saved in set $Candi$ (lines 4-12).

Afterwards, Algorithm 10 first identifies all isomorphic subgraphs in $\bm{G^{sub}}$ that have the same structure as task graph $\bm{G^{task}}$. This operation can be implemented by exploring alternative SPs of different components in $Candi$ (line 1) to find all possible isomorphic subgraphs. Next, lines 2-6 eliminate subgraphs that fail to meet constraint \eqref{C5}. Finally, Algorithm 10 saves the set of isomorphic subgraphs in $SG_m$.

\noindent$\bullet$ \textbf{Step 3.} Determination of template $\mathbf{A^{on}}$: Once all isomorphic subgraphs have been obtained, referred to as set ${{SG}_{all}}$, TE-InstaISS utilizes OptTSelectOnline function (as outlined in Algorithm 11) to identify the optimal template. This involves calculating the practical value of $\mathcal{F}(\mathbf{A,B})$ over various templates, with the optimal template $\mathbf{A^{on}}$ being selected based on the minimum value of the cost function.

\vfill
\end{appendices}

\end{document}